\documentclass[showkeys,floatfix,noshowpacs, preprintnumbers, aps,superscriptaddress,pra,10pt,twocolumn,longbibliography,nofootinbib]{revtex4-2}

\usepackage{comment}
\usepackage{graphicx}
\usepackage{tikz}
\usepackage{tikz-cd}
\usepackage{csquotes}
\usetikzlibrary{decorations.markings,positioning}
\usetikzlibrary{shadows.blur}

\usepackage{physics}
\usepackage{braket}
\usepackage{float}  
\usepackage[english]{babel}
\usepackage[utf8]{inputenc}
\usepackage[normalem]{ulem}

\usepackage[colorlinks=true,linktocpage=true]{hyperref}
\hypersetup{
    allcolors  = {blue},
}
\usepackage{booktabs} 

\usepackage{amsmath}
\usepackage{amssymb} 
\usepackage{bbold}
\usepackage{amsthm}

\usepackage{xfrac}
\usepackage{mathtools}
\usepackage{leftindex}

\usepackage{standalone}

\usepackage{thmtools}
\usepackage{thm-restate}

\usepackage{listings}
\usepackage{enumitem}   

\usepackage[page,toc,titletoc,title]{appendix}

\usepackage{footmisc}
\usepackage{subcaption}

\usepackage{microtype} 
\usepackage[capitalize]{cleveref} 

\DeclareUnicodeCharacter{0301}{*************************************}

\usepackage{nomencl}
\makenomenclature

\usepackage{soul}

\theoremstyle{definition}

\DeclareMathOperator{\arccosh}{arccosh}

\usepackage{mathrsfs}

\usepackage{todonotes}

\usepackage{soul}

\begin{document}

\title{Particle detector in a position-superposed black hole spacetime}
\author{Laurens Walleghem}\thanks{\newline Email: \href{laurens.walleghem@york.ac.uk}{laurens.walleghem@york.ac.uk}}
\affiliation{Department of Mathematics, University of York, Heslington, York YO10 5DD, United Kingdom}
\affiliation{International Iberian Nanotechnology Laboratory (INL), Av. Mestre Jos\'{e} Veiga, 4715-330 Braga, Portugal}
\author{Carlo Cepollaro}\thanks{\newline Email: \href{carlo.cepollaro@oeaw.ac.at}{carlo.cepollaro@oeaw.ac.at}}
\affiliation{University of Vienna, Faculty of Physics, Vienna Center for Quantum Science and
Technology, Boltzmanngasse 5, 1090 Vienna, Austria}
\affiliation{Institute for Quantum Optics and Quantum Information (IQOQI), Austrian
Academy of Sciences, Boltzmanngasse 3, A-1090 Vienna, Austria}
\affiliation{University of Vienna, Vienna Doctoral School in Physics, Boltzmanngasse 5, 1090
Vienna, Austria}

\begin{abstract}
    We calculate the response of an Unruh--DeWitt detector in a 2+1d spacetime that contains a BTZ black hole in a superposition of locations as viewed in a reference system associated with the detector. Upon performing a Quantum Reference Frame (QRF) transformation, this can also be seen as a detector in a superposition of locations in a single classical black hole spacetime. 
    We use this to derive the form of the interaction of the detector and scalar field in such a superposition of spacetimes, ignoring backreaction.
    We define a measurement whose outcome probabilities contain a nonclassical contribution that would be absent for a black hole described by a classical mixture of positions. Finally, we compare our results with a previously studied setup involving a mass-superposed black hole by Foo et al. [Phys. Rev. Lett. 129, 181301 (2022)], and highlight a key difference. 
    We show analytically how this difference arises and find singularities in the mass-superposed spectrum probed by the detector.
\end{abstract}

\maketitle

\section{Introduction}
Black holes serve as a prime example of a natural setting where quantum theory intersects with general relativity. Ideally, a full theory of quantum gravity is needed to completely describe the behavior of black holes as specific cases within a broader framework. This approach, often referred to as \enquote{top-down}, has so far struggled to yield concrete predictions due to the challenges in making calculations with existing models.

Crucially, one of the most intriguing aspects of black holes—their emission of Hawking radiation—was predicted using a different, \enquote{bottom-up} approach. Without relying on a full theory of quantum gravity, but instead using only established principles from general relativity and quantum theory, Hawking predicted that black holes emit (approximately) thermal radiation~\cite{hawking1975particle}. This discovery led to fundamental puzzles in theoretical physics, such as the black hole information puzzle~\cite{hawking1976breakdown,harlow2016jerusalem,Polchinski_2016}, and the cloning~\cite{susskind1993stretched,hayden2007black} and firewall paradoxes~\cite{page1993average,page1993information,braunstein2007quantum,braunstein2013better,almheiri2013black}.

Other examples of \enquote{bottom-up} approaches include theoretical proposals to study quantum-gravitational effects that would be absent if gravity was classical, such as table-top quantum gravity experiments~\cite{bose2017spin,marletto2017gravitationally,bose2022mechanism,tino2021testing,marshman2020locality,moore2021searching,margalit2021realization,Bengyat_2024,tobar2024detecting}, quantum-information protocols probing quantum features of black holes~\cite{howl2022quantum}, or the study of quantum superpositions of spacetime geometries, including effective metrics sourced by matter in quantum superposition~\cite{foo2023quantum,foo2022quantum,suryaatmadja2023signaturesrotatingblackholes,foo_schrodingers_2021,Foo_2022,akil2022semiclassical}.
In the last framework, Quantum Reference Frames (QRFs)~\cite{giacomini2019quantum,de_la_Hamette_2023,de_la_Hamette_2020,Vanrietvelde_2020,delahamette2021perspectiveneutralapproachquantumframe,cepollaro2024entanglement,Loveridge_2018,carette2024operationalquantumreferenceframe,fewster2024quantumreferenceframesmeasurement,kabel2024identification,Cepollaro_2024,zych2018relativityquantumsuperpositions,zych2018quantum,kabel2023quantum,goeller2022diffeomorphisminvariantobservablesdynamicalframes,Araujo_Regado_2025} have caught recent attention, as a tool to study quantum gravitational phenomena: when spacetime exhibits non-classical behavior, classical coordinates may no longer provide an adequate description of the physical situation, necessitating the use of a quantum framework for reference frames. 
Moreover, quantum reference frames have also been linked to asymptotic symmetries~\cite{bondi1962gravitational,sachs1962gravitational,compere2019advancedlecturesgeneralrelativity,compere_poincare_2020}, memory~\cite{Zeldovich:1974gvh,christodoulou1991nonlinear,thorne1992gravitational,Bieri_2024,maibach2026balancefluxlawsgeneral}, and soft and edge modes~\cite{Freidel_2020,Donnelly_2015,Donnelly_2016,Speranza_2018} in Refs.~\cite{kabel2023quantum,Araujo_Regado_2025,Carrozza_2022,gomes2025boundariesframesissuephysical}, which endow asymptotically flat spacetimes with soft hair through infrared physics~\cite{strominger2018lecturesinfraredstructuregravity}, argued to play an important role for the black hole information puzzle, for example~\cite{hawking2016soft,Donnay_2016,carney_infrared_2017,strominger2017blackholeinformationrevisited}.

An operational way to analyze the behavior of black
holes and Hawking radiation is by probing the radiation with a particle detector, such as the Unruh--DeWitt detector~\cite{birrell1984quantum}. 
This approach has recently regained much attention: extensive work has been done on the response of detectors
around black holes~\cite{Ng_2014,bhattacharya2024probinghiddentopologyquantum,Smith_2014,Barbado_2011,Hodgkinson_2012,hodgkinson2013particledetectorscurvedspacetime,paczos2023hawking,henderson2018harvesting,Henderson_2020,Henderson_2022,juarez2022quantum,Ju_rez_Aubry_2014,Lifschytz_1994} and in AdS spacetime~\cite{jennings2010response}, infalling detectors crossing the black hole horizon~\cite{preciado2024more,Ng_2022,wang2024singular,spadafora2024deepknottedblackhole,Barbado_2016}, detectors in a superposed trajectory~\cite{Foo_2020,PhysRevResearch.3.043056,foo2021entanglement}, detectors in superposed quotiented Minkowski~\cite{foo2023quantum} and Rindler spacetimes~\cite{goel2024accelerateddetectorsuperposedspacetime}, in superposed BTZ black holes~\cite{foo2022quantum,suryaatmadja2023signaturesrotatingblackholes}, superposed detectors in de Sitter and Schwarzschild spacetime~\cite{niermann2024particle,paczos2023hawking}, entanglement harvesting~\cite{Stritzelberger_2021,Hotta_2020,Stritzelberger_2020,Cong_2020,Ng_2018,henderson2018harvesting,Gallock_Yoshimura_2021,Mendez_Avalos_2022,Tjoa_2020,robbins2020entanglementamplificationrotatingblack,chakraborty2024entanglementharvestingquantumsuperposed,dubey2025harvesting,Membrere_2023,Barman_2023,wang2025harvestinginformationhorizon}, acceleration- and gravity-induced transparency effects~\cite{vsoda2022acceleration,pan2024gravity}, anti-Unruh~\cite{brenna2016anti,garay2016thermalization} and anti-Hawking~\cite{Henderson_2020,CAMPOS2021136198} phenomena, covariantly smeared detectors in spacetime~\cite{Mart_n_Mart_nez_2020}, etc. 
Furthermore, more complex models for detectors have been proposed in ~\cite{giacomini2022second,wood2022quantized,sudhir2021unruh,gale2023relativistic}, used for example for investigating the Unruh effect~\cite{barbado2020unruh,sudhir2021unruh}.

In this paper, we use quantum reference frame techniques to describe a BTZ black hole in superposition of different locations as viewed from a reference frame associated with the detector, by studying the response of an Unruh--DeWitt detector in such a superposition of spacetimes.
We do so without relying on the details of a full theory of quantum gravity, but showing that this physical situation can be equivalently (and more conveniently) described in a different reference frame as a detector in a superposition of different locations in a fully classical gravitational field. 
In this frame, we can calculate the detector's evolution without involving any quantum property of gravity. Next, we perform a second QRF transformation to return to the original frame, where the black hole is in superposition, and examine the resulting dynamics. 
Our findings demonstrate that a joint measurement of the detector and the black hole's position can reveal effects of the superposition, manifesting as interference fringes. 

Our methods and setup are inspired by Ref.~\cite{foo2022quantum}, which studies a BTZ black hole in superposition of different masses. 
By contrast, we place the black hole in a superposition of locations. 
Crucially, we find similar but different results: the resulting interference pattern is smooth - showing no sharp peaks - unlike the pronounced peaks presented in Ref.~\cite{foo2022quantum}. We analytically identify the origin of this difference by studying the spectrum probed by the detector in the cases of mass and position superposition. This further supports the interpretation in Ref.~\cite{foo2022quantum} that such peaks are a consequence of black hole mass quantization~\cite{bekenstein2020quantum,hod1998bohr,Bekenstein_1995,bekenstein1998quantumblackholesatoms,deppe2024echoesbeyonddetectinggravitational}.

The rest of this work is organized as follows. In \Cref{sec:setup} we introduce the necessary concepts for our calculations: we provide a short pedagogical introduction to the BTZ black hole, Unruh--DeWitt detectors and quantum reference frames. The reader who is already familiar with these concepts can directly read \Cref{sec:main_results}, where we describe the physical scenario we consider, and present our theoretical, numerical results and corresponding analytical expressions. Finally, we conclude in \Cref{sec:discussion}.

\section{A pedagogical introduction to BTZ black holes, Unruh--DeWitt detectors, and Quantum Reference Frames}\label{sec:setup}

\subsection{BTZ Black hole} \label{sec:BTZ}
The BTZ black hole~\cite{Banados_1992,Carlip_1995,carlip2023quantumgravity21dimensions} is a 2+1-dimensional black hole solution to Einstein's equation with negative cosmological constant $\Lambda = -1/l^2$, asymptotically anti-de Sitter ($AdS_3$) with radius $l$. It is particularly interesting as its field correlation functions have closed-form analytic expressions~\cite{Lifschytz_1994,Carlip_1995,Hodgkinson_2012,carlip2003quantum,foo2022quantum,Smith_2014,henderson2018harvesting}, making theoretical explorations more tractable.

We consider here a non-rotating and uncharged BTZ black hole, with metric
\begin{equation} \label{eq:BTZ_metric}
    \mathrm{d} s^2=-f(r) \, \mathrm{d} t^2+f(r)^{-1} \mathrm{~d} r^2+r^2 \mathrm{~d} \varphi^2,
\end{equation} with $f(r)=r^2/l^2-M$ and coordinate range $\sqrt{M}l < r < \infty,-\infty < t < \infty, \varphi \in [0,2\pi)$. 

\begin{figure}[h]  \centering   \tikzset{every picture/.style={line width=0.75pt}} 

\begin{tikzpicture}[x=0.75pt,y=0.75pt,yscale=-1,xscale=1]

\draw   (83.97,124) -- (188,124) -- (188,228.03) -- (83.97,228.03) -- cycle ;
\draw    (83.97,124) -- (188,228.03) ;
\draw    (83.97,228.03) -- (188,124) ;
\draw    (83.97,122.5) -- (188,122.5)(83.97,125.5) -- (188,125.5) ;
\draw    (83.97,226.53) -- (188,226.53)(83.97,229.53) -- (188,229.53) ;

\draw (139.08,152.19) node [anchor=north west][inner sep=0.75pt]  [font=\scriptsize,rotate=-315] [align=left] {$\displaystyle r=r_{h}$};
\draw (109.1,129.91) node [anchor=north west][inner sep=0.75pt]  [font=\scriptsize,rotate=-45] [align=left] {$\displaystyle r=r_{h}$};
\draw (126.22,106.54) node [anchor=north west][inner sep=0.75pt]  [font=\scriptsize] [align=left] {$\displaystyle r=0$};
\draw (124.22,232.54) node [anchor=north west][inner sep=0.75pt]  [font=\scriptsize] [align=left] {$\displaystyle r=0$};
\draw (190.22,174.54) node [anchor=north west][inner sep=0.75pt]  [font=\scriptsize] [align=left] {$\displaystyle r=\infty $};
\draw (49.22,174.54) node [anchor=north west][inner sep=0.75pt]  [font=\scriptsize] [align=left] {$\displaystyle r=\infty $};

\end{tikzpicture} \caption{The (maximally) extended Penrose diagram of the BTZ black hole~\cite{Carlip_2005}.} \label{fig:BTZ_Penrose}
\end{figure}
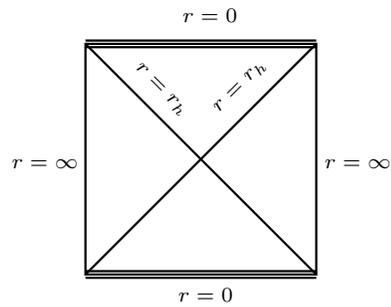

The black hole has an event horizon at $r_h=\sqrt{M}l$, with maximally extended Penrose diagram as in \cref{fig:BTZ_Penrose}. Notably, being a 2+1-dimensional gravitational spacetime, the BTZ black hole does not admit a Newtonian limit~\cite{DESER1984220,barrow1986three,cornish1991gravitation,Carlip_1995}, but other limits such as the Newton--Hooke one can be considered~\cite{Bacry:1968zf,Gibbons_2003,Bizo__2018,Pal_2025}.

The conformal boundary at infinity is timelike, so that information can enter and leave at spatial infinity. Therefore, boundary conditions must be imposed for a field theory on the BTZ spacetime. 
In general, 2+1d gravity has no propagating degrees of freedom, but it is due to this boundary that nontrivial dynamics can occur, and so-called `boundary gravitons' can appear in a quantum theory of gravity~\cite{carlip2003quantum,carlip2023quantumgravity21dimensions,Carlip_2005}.
The BTZ black hole can be obtained from the universal covering space of $AdS_3$~\cite{Kraus2008} by means of the quotient under $\Gamma_M:\varphi \rightarrow \varphi+2 \pi \sqrt{M}$, see Supplemental Material of Ref.~\cite{foo2022quantum}.
This fact will allow for correlation functions of quantum fields in BTZ black hole spacetime to be calculated precisely using correlations in $AdS_3$. 
The temperature assigned to a BTZ black hole equals $T_0  = \sqrt{M}/(2\pi l) $~\cite{Carlip_1995}, and indeed a static 
detector around a BTZ black hole detects a red- or blueshifted local Tolman temperature~\cite{Santiago_2019,Lima_2019} $T_0/\sqrt{-g_{00}}$ in the Hartle--Hawking state~\cite{Hodgkinson_2012,hodgkinson2013particledetectorscurvedspacetime}.

A test particle at fixed radius $R$ in space moving only in time has proper time that relates to coordinate time $t$ as $g_{00} (dt/d\tau)^2=-1$ so \begin{equation} \label{eq:tau_to_t}
    \tau = \sqrt{-g_{00}} t = \sqrt{f(R)} t =: \gamma(R) t = \sqrt{\frac{R^2}{l^2}-M} t,
\end{equation} with $\gamma(R)$ the redshift factor. Thus, when we consider two clocks at fixed, but different radii $R_1$ and $R_2$ from the black hole, there will be a relative time dilation between them. 

We introduce a particle field that interacts with the gravitational field of the black hole, and later we will measure the particle field with an Unruh--DeWitt detector. Specifically, we consider a massless scalar field $\phi(x)$ conformally coupled to the metric, with vacuum analogous to the Hartle--Hawking state~\cite{Carlip_1995,Lifschytz_1994}.

The measurement probabilities of the Unruh--DeWitt detector, as explained below, depend on the two-point correlator of the scalar field in the BTZ black hole background. Such correlator can be calculated from the $AdS_3$ correlators (see~\cite{foo2022quantum,Hodgkinson_2012,Lifschytz_1994,hodgkinson2013particledetectorscurvedspacetime} for details):
\begin{equation} \label{eq:W_BTZ_from_image_sum}
    W_{\mathrm{BTZ}}\left(\mathrm{x}, \mathrm{x}^{\prime}\right)=\sum_n \Upsilon^n W_{AdS}\left(\mathrm{x}, \Gamma^n \mathrm{x}^{\prime}\right)
\end{equation} with $\Upsilon=+1,-1$ for an untwisted and twisted field~\cite{Lifschytz_1994}. We will consider the untwisted case $\Upsilon=1$ unless stated otherwise.
The correlation function in the scalar vacuum of $AdS_3$ is given by \begin{equation} \label{eq:BTZ_correlation}
    \begin{split}
        W_{\mathrm{AdS}}\left(x, x^{\prime}\right)=\frac{1}{4 \pi l \sqrt{2}}\left[\frac{1}{\sqrt{\rho\left(x, x^{\prime}\right)}}-\frac{\zeta}{\sqrt{\rho\left(x, x^{\prime}\right)+2}}\right],
    \end{split}
\end{equation} with $\rho(x,x')$ the squared geodesic distance in the 2+2 embedding space of $AdS_3$, and $\zeta \in\{-1,0,1\}$ specifies the boundary conditions at asymptotic infinity: Neumann, transparent or Dirichlet~\cite{carlip2003quantum,Lifschytz_1994,Shiraishi_1994}. 
Unless stated otherwise, we use transparent boundary conditions, $\zeta=0$. When viewing $AdS$ as conformally equivalent to half of an Einstein static universe, Dirichlet and Neumann boundary conditions reflect waves at the conformal boundary out of phase and in phase, respectively. Transparent boundary conditions instead correspond to quantizing a massless conformally coupled field on the full Einstein static universe, allowing it to pass through the $AdS$ boundary rather than being reflected~\cite{avis1978quantum}. Thus, energy and angular momentum can flow across conformal infinity, leading to a prescription of an effective two-time Cauchy surface for the field on $AdS$~\cite{avis1978quantum}.
The corresponding implications for the eternal BTZ spacetime are unclear and lie beyond our scope. Nevertheless, the BTZ two-point function can be constructed from the $AdS_3$ one through the image sum in \cref{eq:W_BTZ_from_image_sum}. More general Robin boundary conditions have also been considered~\cite{Bussola_2017,CAMPOS2021136198}; see Appendix~\ref{sec:app_W_BTZ} for further details.

\subsection{Unruh--DeWitt detectors} \label{sec:UDW}
In flat spacetime, the usual definition of a particle relies on excitations of fields, decomposed into positive and negative frequency modes --- a decomposition that is possible when a global timelike Killing vector field exists.
On a general curved spacetime, nevertheless, this straightforward definition is not possible, due to the lack of a preferred notion of time. A possible way of defining particles in a more operational way was introduced by Unruh~\cite{unruh1976notes}, suggesting that \enquote{a particle is what a particle detector detects}. 

Unruh modeled a particle detector as a scalar field in a cavity, and showed what we now call the Unruh effect~\cite{Crispino_2008}: an accelerated detector in Minkowski spacetime registers a thermal bath of particles. DeWitt~\cite{dewitt1979} proposed a simplified detector model, replacing the scalar field in a cavity by a simple two-level system, obtaining the Unruh--DeWitt detector~\cite{birrell1984quantum,martin2013wavepacket}.

Specifically, an Unruh--DeWitt (UDW) detector consists of a detector modeled as a quantum system following a path $x(\tau)$ in spacetime that interacts linearly with a scalar field.\footnote{Extensions to detectors for fermionic or other boson fields exist~\cite{hummer2016renorm,ali2021unruhdewittdetectorresponsescomplex}, as well as explicit models for $n$-level detectors~\cite{Guedes_2024,lima2023unruh}, detectors with quantized center of mass~\cite{Stritzelberger_2020,sudhir2021unruh}, second-quantized detectors~\cite{giacomini2022second,gale2023relativistic} and detector models derived relativistic-covariantly using the influence functional \cite{torres24particle}. For simplicity, we will use a 2-level detector model.} The spacetime path\footnote{Usually the spacetime path is some classical trajectory on a background spacetime; later, we will allow the detector in a superposition of paths (meaning the path is an operator with state highly peaked around two classical trajectories).} can be freely chosen, geodesic or non-geodesic driven by some external force. The detector has internal degrees of freedom, typically taken to be an $n$-level system with energy gaps, that \enquote{click} when a particle is measured, 
and external degrees of freedom, corresponding to the position of the detector.  It interacts with the particle field through a linear coupling, with the Schrödinger-picture interaction Hamiltonian in the detector's rest frame being 
\begin{equation} \label{eq:H_int}
 \hat H_\text{int}^\tau = \lambda \eta(\tau)  \hat \mu \otimes \phi(\hat{\mathbf{x}}),
\end{equation} with $\hat{\mathbf{x}}$ the operator corresponding to the spatial position of the detector and $\hat \mu$ the monopole moment which has off-diagonal elements in the detector's energy eigenbasis so that this interaction allows the detector to switch between its internal energy levels. Here $\hat H^\tau_\text{int}$ is written with respect to the proper time of the detector; choosing a different time $t$ to describe the evolution  yields the transformed Hamiltonian~\cite{Mart_n_Mart_nez_2020} \begin{equation} 
\hat H^t_\text{int} = \frac{d\tau}{dt} \hat H^\tau_\text{int}. \end{equation}
The coupling strength $\lambda$ will be assumed to be small, and $\eta(\tau)$ is a switching (-on and -off) function of the detector, which we will take to be a Gaussian $\eta(\tau) = e^{-\frac{\tau^2}{2\sigma^2}}$ with width $\sigma$ much smaller than the timescale $\Delta \tau$ of the experiment. We say the detector has detected a particle when the detector transitions to a higher internal energy level. An example of a detector could be an atom as a detector for particles of the electromagnetic field.
 
We will perform our calculations in the interaction picture, in which the UDW interaction Hamiltonian becomes~\cite{Stritzelberger_2020,gale2023relativistic,sudhir2021unruh,wood2022quantized} (with pre-subscript $I$ for interaction picture) 
\begin{equation} \label{eq:H_int_I_pic}
    \leftindex_I {\hat H}^\tau_\text{int}(\tau)= \lambda \eta(\tau) \int d \nu( \mathbf{x}) \ket{\mathbf{x}(\tau)} \bra{\mathbf{x}(\tau)}_D \otimes \hat \mu(\tau) \otimes \phi(x(\tau)),
\end{equation} with $\ket{\mathbf{x}(\tau)}_D$ the spatial position of the detector,  $\hat \mu(\tau)= e^{i \hat H_0 \tau}\, \hat \mu \, e^{-i\hat H_0\tau}$ with $\hat H_0$ the detector's monopole moment Hamiltonian so that $\hat H_0 \ket{E}_\mu = E \ket{E}_\mu$. Here $\phi(x(\tau))$ is the scalar field evaluated at the detector's spacetime path $x(\tau)$ with $\tau$ the detector's proper time.
We denote states of the detector's position and internal degrees of freedom by $\ket{\mathbf{x}}_D,\ket{E}_\mu$, respectively.

This detector can be used to measure properties of the scalar field to gain information on the semiclassical gravitational field.
Let us consider an initial state preparation $\ket{\psi(\tau_i)}=|\mathbf{x}_i\rangle_D |E_0\rangle_\mu |0\rangle_\phi$, i.e. with the detector in its ground state at some initial position $\mathbf{x}_i$ and field $\phi$ in the Hartle--Hawking vacuum state $\ket{0}_\phi$. The detector \enquote{clicks} when a transition to another internal state $\ket{E}_\mu$ happens, due to the interaction with the particle field. This leads to a transition amplitude (see Appendix~\ref{sec:app_interaction} for details)
\begin{equation} \label{eq:single_detector_amplitude}
   \begin{split}
       \braket{\vartheta,E,\mathbf{x}_f|\psi(\tau_f)} &= -i \lambda \langle E |\mu |E_0\rangle_\mu e^{i\Phi}  \times \\ &\times \int_{\tau_i}^{\tau_f} d\tau \eta(\tau) e^{i(E-E_0)\tau} \langle \vartheta| \phi(x(\tau)) |0\rangle_{\phi},
   \end{split}
\end{equation} 
with $\Phi$ a potential phase factor, $\ket{\vartheta}$ a final scalar field state that we will sum over and later we will take $\tau_i = - \tau_f$. 
The probability of the transition $E_0 \rightarrow E$ happening, independently of the final state of the field and for a final position of the detector $\mathbf{x}_f$, is
\begin{align} \label{eq:transition_prob}
    P_{\mathbf{x}_f} &= \int d\vartheta \left | \braket{\vartheta,E,\mathbf{x}_f|\psi(\tau_f)}\right|^2 \nonumber \\
    &=  |\langle E | \mu |E_0\rangle|^2 \lambda^2  \mathcal{F}(E-E_0),
\end{align}
with the detector response function given by \begin{equation}
\begin{split}
    \mathcal{F}(\Omega) =& \int_{\tau_i}^{\tau_f} d\tau \int_{\tau_i}^{\tau_f}  d\tau' \eta(\tau) \eta(\tau') e^{-i \Omega(\tau-\tau')} \\& \langle 0 | \phi(x(\tau)) \phi(x(\tau')) |0\rangle \\ &= \int_{\tau_i}^{\tau_f} d\tau \int_{\tau_i}^{\tau_f} d\tau' \eta(\tau) \eta(\tau') e^{-i \Omega (\tau-\tau')} \\ & W_{BTZ}(x(\tau), x(\tau'))  . 
 \end{split}
\end{equation} 
Here $W_{BTZ}(x,x')$ is the scalar field two-point correlator in a BTZ black hole background.

In \Cref{sec:main_results} we will consider the detector trajectory where the detector is held fixed at its initial position ($\mathbf{x}_f=\mathbf{x}_i$), and will denote the transition probabilities as $P_{\mathbf{x}_i}$.

\subsection{Quantum Reference Frames} \label{sec:QRFS}
When describing physical systems, we typically do so relative to a chosen reference frame. A compelling question emerges when considering the possibility that the reference itself exhibits quantum properties. This idea has sparked significant research into quantum reference frames \cite{giacomini2019quantum,de_la_Hamette_2023,de_la_Hamette_2020,Vanrietvelde_2020,delahamette2021perspectiveneutralapproachquantumframe,Loveridge_2018,carette2024operationalquantumreferenceframe,fewster2024quantumreferenceframesmeasurement,cepollaro2024entanglement,kabel2024identification,Cepollaro_2024}, which investigate how the physical world is described when taking the reference frame to be a quantum system. Notably, it has been shown that fundamental quantum features, such as superposition and entanglement, depend on the chosen frame~\cite{giacomini2019quantum,cepollaro2024entanglement}: a system that appears to be in superposition with respect to one quantum reference frame may be localized in another. This idea has been especially valuable in the study of quantum-gravitational phenomena, as it allows for a description within a reference frame where gravity is entirely classical. As a result, it offers insights into such phenomena without requiring a complete theory of quantum gravity.

To introduce these ideas, consider three classical systems, $B$, $D$, and $F$, modeled as point particles on a Cartesian line. In this setting, a change of reference system is simply a translation of the spatial origin.
From the reference frame of $B$, we say that $D$ and $F$ are at position $\mathbf{x}_D$ and $\mathbf{x}_F$ respectively. On the other hand, when taking $D$ as a reference, one assigns positions $\mathbf{q}_B=-\mathbf{x}_D$ and $\mathbf{q}_F=\mathbf{x}_F-\mathbf{x}_D$ to $B$ and $F$ respectively. 
Can we extend this notion to quantum systems? Is it possible to define a transformation between \emph{quantum} reference frames?

\begin{figure}[h]  \centering   \includegraphics[width=0.4\textwidth]{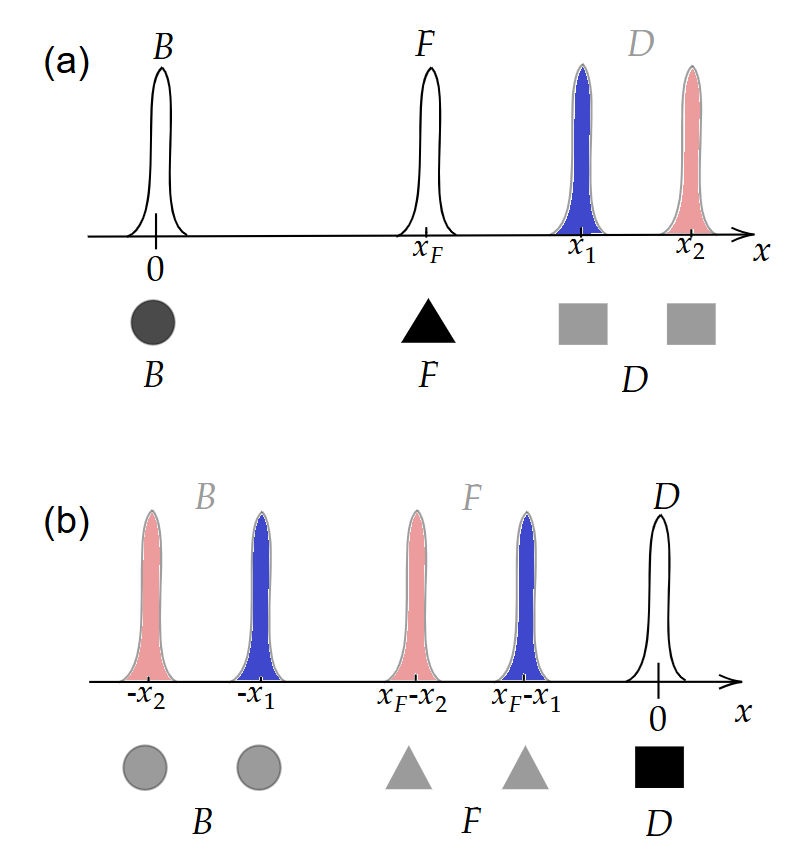} \caption{(a): Systems $B$ and $F$, near a position-superposed system $D$, which under a QRF transformation can be mapped to (b), where $B$ and $F$ are in indefinite, entangled locations but $D$ has a definite single-valued location. Adapted from Ref.~\cite{giacomini2019quantum}.} \label{fig:QRF_BH}
\end{figure}

If the three systems are now quantum, it could be that $D$ is in a superposition state of two locations $\mathbf{x}_1$ and $\mathbf{x}_2$. How can we transform to the reference frame of $D$ in this scenario? A natural idea is to extend the classical transformations by linearity, namely applying the classical transformation for each branch of the superposition.
The previous situation, from the reference frame of $B$, is described by 
\begin{equation}
    \ket{\psi}^{(B)}=\frac{1}{\sqrt{2}}\bigg( \ket{\mathbf{x}_1 }_{\!D} + \ket{\mathbf{x}_2}_{\!D} \bigg) \ket{\mathbf{x}_F}_{F}.
\end{equation} 
A QRF transformation, i.e. a coherent translation controlled on system $D$'s position, is implemented through the unitary transformation 
\begin{equation}\label{eq:QRF_transformation}
    \hat S^{(B)\to(D)} = \int d\mathbf{x} \,\ket{-\mathbf{x}}_{B}\bra{\mathbf{x}}_D \otimes \hat T_F^\dagger(\mathbf{x}),
\end{equation} with $\hat T_F$ the translation operator on the Hilbert space of the system $F$.
The state in the perspective of $D$ is then
\begin{align}
    \ket{\psi}^{(D)} &=  S^{(B)\to(D)} \ket{\psi}^{(B)} \nonumber \\
    &= \frac{\ket{-\mathbf{x}_1}_B\ket{\mathbf{x}_F-\mathbf{x}_1}_F + \ket{-\mathbf{x}_2}_B\ket{\mathbf{x}_F-\mathbf{x}_2}_F}{\sqrt{2}}.
\end{align} 
This describes the same physical setup but when choosing the origin to be system $D$'s location\footnote{Sometimes the origin state is also written explicitly, i.e. for example $\ket{\mathbf{0}}_D$ in this case.}, see also \cref{fig:QRF_BH}. 

The analogous QRF map $ \hat S^{(B)\to(D)} $ used later in this work in the BTZ spacetime is a branch-dependent change of reference-field coordinates, rather than a translation symmetry or isometry of the BTZ geometry.

Consider now the situation where not only we want to (spatially) translate the physical description, but also change the time coordinate we use in Schrödinger's equation. To do so, the (classical) reference frame change involves two steps: we first apply the translation, and then transform the Hamiltonian to account for the new time parameter. Under this change of classical reference frame, the Hamiltonian transforms as
\begin{equation}
    \hat H^{(D)} = \frac{dt}{d\tau} \hat T \hat H^{(B)} \hat T^\dagger, 
\end{equation}
where $\hat T$ implements the classical spatial translation, $\tau$ is the new time parameter and $t$ is the old one. Applied to our case, if we choose $t$ to be the coordinate time, and $\tau$ to be the proper time in the detector frame, we obtain
\begin{align}
    \tau &= \gamma(\textbf{x}_D) \,t, \\
    \gamma(\textbf{x}_D)&=\sqrt{-g_{00}(\textbf{x}_D)} = \sqrt{\frac{\textbf{x}_D^2}{l^2}-M}.
\end{align}
We now generalize this transformation to superposition of translations, as we have explained before. The resulting Hamiltonian is (see~\cite{giacomini2022second} for details)
\begin{equation}
    \hat H^{(D)} =  \hat S^{(B) \to (D)} \gamma^{-1} (\hat{\textbf{x}}_D) \hat H^{(B)} \hat S^{\dagger (B) \to (D)}
\end{equation}
The reparametrization of time is now promoted to an operator: when $D$ is in superposition with respect to $B$, their proper times are in superposition with respect to each other.

In the rest of the paper we apply these ideas to the standard Unruh--DeWitt detector scenario described in the previous section, involving a detector $D$, a black hole $B$ and a scalar field $\phi$. The interaction depends on the field only through the local operator $\phi(\hat x_D)$, namely the field evaluated at the detector position. Since $\phi$ is a scalar, the change of frame introduces no additional internal or tensorial transformation on the field degrees of freedom. Its only effect is the corresponding change of the argument at which the field is evaluated, which is already accounted for by the transformation of the detector position. Therefore, the QRF transformation acts non-trivially only on the relative position degrees of freedom of $B$ and $D$, while acting as the identity on the field Hilbert space:
\begin{equation}\label{eq:QRF_transformation_field}
    \hat S^{(B)\to(D)} = \int d\nu(\mathbf{x}) \,\ket{-\mathbf{x}}_{B}\bra{\mathbf{x}}_D \otimes \mathbb{1}_\phi.
\end{equation}
In the next section we describe this set-up in more detail. 
A QRF treatment of this set-up and transformation using quantum reference fields, which models the operator $\hat{\mathbf{x}}$ more explicitly as a relational radial coordinate distance between references frames associated with the black hole and detector, can be found in Appendix~\ref{app:new_reference_fields}. 

\section{Main results: Response of a detector near a position-superposed black hole} \label{sec:main_results}
\subsection{Setup} \label{sec:setup_scenario}

\begin{figure}[h]  \centering   \tikzset{every picture/.style={line width=0.75pt}} 

\begin{tikzpicture}[x=0.75pt,y=0.75pt,yscale=-1,xscale=1]

\draw [color={rgb, 255:red, 74; green, 74; blue, 74 }  ,draw opacity=0.95 ]   (207.5,142.06) -- (371.48,140.02) ;
\draw [color={rgb, 255:red, 74; green, 74; blue, 74 }  ,draw opacity=0.95 ]   (207.5,142.06) -- (416.5,138.52) ;
\draw  [color={rgb, 255:red, 0; green, 0; blue, 0 }  ,draw opacity=1 ][fill={rgb, 255:red, 184; green, 210; blue, 235 }  ,fill opacity=1 ] (85.73,147.02) .. controls (85.73,137.49) and (93.46,129.77) .. (102.98,129.77) .. controls (112.51,129.77) and (120.23,137.49) .. (120.23,147.02) .. controls (120.23,156.54) and (112.51,164.27) .. (102.98,164.27) .. controls (93.46,164.27) and (85.73,156.54) .. (85.73,147.02) -- cycle ;
\draw  [fill={rgb, 255:red, 184; green, 210; blue, 235 }  ,fill opacity=1 ] (127.5,130.03) -- (176,130.03) -- (176,163) -- (127.5,163) -- cycle ;
\draw  [color={rgb, 255:red, 0; green, 0; blue, 0 }  ,draw opacity=1 ][fill={rgb, 255:red, 255; green, 255; blue, 255 }  ,fill opacity=1 ] (90,147.02) .. controls (90,139.84) and (95.81,134.03) .. (102.98,134.03) .. controls (110.16,134.03) and (115.97,139.84) .. (115.97,147.02) .. controls (115.97,154.19) and (110.16,160) .. (102.98,160) .. controls (95.81,160) and (90,154.19) .. (90,147.02) -- cycle ;
\draw    (106.5,138.03) -- (102.98,147.02) ;
\draw    (107.5,153.03) -- (102.98,147.02) ;
\draw    (157.5,145.03) -- (152.5,153.03) ;
\draw  [fill={rgb, 255:red, 255; green, 255; blue, 255 }  ,fill opacity=1 ] (138.5,153.03) .. controls (138.5,153.03) and (138.5,153.03) .. (138.5,153.03) .. controls (138.5,144.2) and (144.77,137.05) .. (152.5,137.05) .. controls (160.23,137.05) and (166.5,144.2) .. (166.5,153.03) -- cycle ;
\draw    (157.5,145.03) -- (152.5,153.03) ;
\draw  [fill={rgb, 255:red, 74; green, 74; blue, 74 }  ,fill opacity=1 ][blur shadow={shadow xshift=0pt,shadow yshift=0pt, shadow blur radius=14.25pt, shadow blur steps=19 ,shadow opacity=100}] (353.5,140.02) .. controls (353.5,130.08) and (361.55,122.03) .. (371.48,122.03) .. controls (381.42,122.03) and (389.47,130.08) .. (389.47,140.02) .. controls (389.47,149.95) and (381.42,158) .. (371.48,158) .. controls (361.55,158) and (353.5,149.95) .. (353.5,140.02) -- cycle ;
\draw  [fill={rgb, 255:red, 74; green, 74; blue, 74 }  ,fill opacity=1 ][blur shadow={shadow xshift=0pt,shadow yshift=0pt, shadow blur radius=14.25pt, shadow blur steps=19 ,shadow opacity=100}] (416.5,138.52) .. controls (416.5,128.31) and (424.78,120.03) .. (434.98,120.03) .. controls (445.19,120.03) and (453.47,128.31) .. (453.47,138.52) .. controls (453.47,148.72) and (445.19,157) .. (434.98,157) .. controls (424.78,157) and (416.5,148.72) .. (416.5,138.52) -- cycle ;
\draw  [fill={rgb, 255:red, 0; green, 0; blue, 0 }  ,fill opacity=1 ] (207.5,142.06) .. controls (207.5,137.23) and (211.41,133.32) .. (216.24,133.32) .. controls (221.07,133.32) and (224.98,137.23) .. (224.98,142.06) .. controls (224.98,146.89) and (221.07,150.8) .. (216.24,150.8) .. controls (211.41,150.8) and (207.5,146.89) .. (207.5,142.06) -- cycle ;
\draw [color={rgb, 255:red, 245; green, 166; blue, 35 }  ,draw opacity=1 ]   (157.5,145.03) .. controls (159.21,143.41) and (160.87,143.45) .. (162.5,145.16) .. controls (164.13,146.87) and (165.79,146.91) .. (167.5,145.29) .. controls (169.21,143.67) and (170.88,143.72) .. (172.49,145.43) .. controls (174.12,147.14) and (175.78,147.18) .. (177.49,145.56) .. controls (179.2,143.94) and (180.86,143.98) .. (182.49,145.69) .. controls (184.12,147.4) and (185.78,147.44) .. (187.49,145.82) .. controls (189.2,144.2) and (190.86,144.24) .. (192.49,145.95) -- (195.5,146.03) -- (195.5,146.03) ;
\draw [color={rgb, 255:red, 255; green, 255; blue, 255 }  ,draw opacity=1 ]   (342,214) -- (425.5,224.03) ;

\draw (75,95) node [anchor=north west][inner sep=0.75pt]   [align=left] {detector and clock};
\draw (350,69) node [anchor=north west][inner sep=0.75pt]   [align=left] {position-superposed \\ \ \ \ \ \ \ black hole};
\draw (361,167) node [anchor=north west][inner sep=0.75pt]  [font=\large,color={rgb, 255:red, 74; green, 74; blue, 74 }  ,opacity=0.95 ] [align=left] {$\displaystyle R_{1}$};
\draw (427.47,167) node [anchor=north west][inner sep=0.75pt]  [font=\large,color={rgb, 255:red, 74; green, 74; blue, 74 }  ,opacity=0.95 ] [align=left] {$\displaystyle R_{2}$};
\draw (366,131) node [anchor=north west][inner sep=0.75pt]  [font=\large,color={rgb, 255:red, 255; green, 255; blue, 255 }  ,opacity=1 ] [align=left] {$\displaystyle 1$};
\draw (430,131) node [anchor=north west][inner sep=0.75pt]  [font=\large,color={rgb, 255:red, 255; green, 255; blue, 255 }  ,opacity=1 ] [align=left] {$\displaystyle 2$};

\end{tikzpicture} \caption{A detector with a clock is used to measure scalar particles created some fixed distance away from a position-superposed black hole. Relative to the detector, the black hole is in a superposition of radial distances $R_1$ and $R_2$.} \label{fig:setup2}
\end{figure}
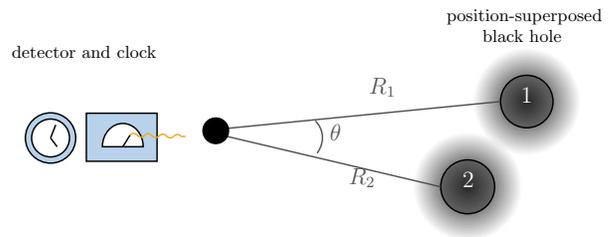

\textit{Setup.} Consider the following scenario, illustrated in \cref{fig:setup2}. A particle detector is used to measure a massless conformally coupled scalar field near a 2+1d BTZ black hole that can be interpreted as being in a superposition of locations as viewed from a reference frame associated with the detector. Such a black hole can, for example, result from the collapse of a shell in a superposition of two different positions in this frame. The detector comes with a clock and is held stationary rather than freely falling. The detector is turned on for a finite proper time interval according to its clock. 

How does the detector respond to this situation? Can it witness quantum effects of the superposed spacetime?
\vspace{0.5cm}

\textit{Approximations.} 
In our calculations, we neglect the gravitational backreaction of the scalar field and detector, as well as reaction effects of the detector's clicking and we assume the two configurations of the black hole we consider are sufficiently distinguishable to be associated with orthogonal states. We also assume the black hole metric to be static, i.e. we neglect potential evaporation. As introduced in \Cref{sec:UDW}, we work with a point-like detector, i.e. we assume the detector to be highly localized in each branch of the evolution. Moreover, we also neglect a potential conjugate variable to the black hole position~\cite{Eyheralde_2017,Eyheralde_2020,louko_hamiltonian_1998}.

\vspace{0.5cm}

\textit{Using QRFs.} In order to model the setup described above, we cannot employ the usual treatment of BTZ black holes and Unruh--DeWitt detectors: since the black hole is in superposition of locations in the detector's reference frame, the evolution of the particle field will be in a quantum superposition of spacetimes, and the usual description of QFT on curved spacetime does not suffice to describe this situation.

Despite not subscribing to any specific theory of quantum gravity, we assume here that any such theory must have a semiclassical limit, where perfectly distinguishable gravitational states are orthogonal, and such that in each branch classical general relativity applies. We assign to each of these states a vector on a Hilbert space $\mathcal{H}_B = L^2(\mathbb{R})$, labeled by the position of the black hole with respect to the detector. This can be seen as the relational radial coordinate distance between the black hole and detector, as we discuss in more detail using quantum reference fields in Appendix~\ref{app:new_reference_fields}.

Under these assumptions, we can model the physical setup in the perspective of the detector, before it is switched on, as 
\begin{equation} \label{eq:initial_state_BH_superp}
    \ket{\psi(\tau_i)}^{(D)} = \frac{\ket{\mathbf{x}_1 }_{\!B} + \ket{\mathbf{x}_2}_{\!B} }{\sqrt{2}} \ket{0}_\phi \ket{E_0}_{\mu}.
\end{equation}
Here $\ket{\mathbf{x}_{1/2} }_{\!B}$ are the two orthogonal semiclassical states of the black hole, corresponding to two different coordinate positions relative to the detector with $\ket{0}_\phi$ the vacuum state of the scalar field in the gravitational BTZ background, 
evolved according to standard QFT on curved spacetime, and $\ket{E_0}_{\mu}$ represents the internal degrees of freedom of the detector before it is switched on.

In order to model the response of the detector, we need to define the Hamiltonian under which it evolves in this situation. A possible solution is to postulate it, as was done in Ref.~\cite{foo2022quantum}. We follow here an alternative approach, using QRF transformations to derive it.

As described in the previous section, the physical setup of the black hole in superposition of locations with respect to a detector can be described in the reference frame of the black hole as the detector being in superposition of locations and the black hole being localized (cf.~\cref{fig:QRF_BH}). In other words, by applying the QRF change defined in Eq.~\eqref{eq:QRF_transformation_field}, we obtain the state
\begin{align}
    \ket{\psi(t_i)}^{(B)} =  \frac{\ket{-\mathbf{x}_1}_D + \ket{-\mathbf{x}_2}_D}{\sqrt{2}} \ket{0}_\phi \ket{E_0}_\mu.
\end{align}
where we choose the initial coordinate time to be $t_i=\tau_i=0$.
In this reference frame, there is no source of gravitational field in quantum superposition, and hence we can employ the usual tools of QFT in curved spacetime. Specifically, the UDW interaction for a localized black hole and quantum UDW detector has the following form with respect to the coordinate time~\cite{giacomini2022second}: 
\begin{equation}  \label{eq:H_int_time_changes}
    \hat H^{(B),t}_{\text{int}} = \gamma(\hat{\textbf{x}}_D) \lambda\, \eta(\gamma(\hat{\textbf{x}}_D)t)\,\mu \otimes \phi(\hat{\mathbf{x}}_D).
\end{equation} 
Hence we can perform the QRF transformation~\cref{eq:QRF_transformation_field} to obtain the Hamiltonian in the detector's frame with respect to its proper time as
\begin{equation} \label{eq:H_int_QRF_transformed}
    \hat H_\text{int}^{(D),\tau}=\lambda \, \eta(\tau) \mu \,  \otimes \phi(-\hat{ \textbf{x}}_B),
\end{equation} see also Appendix~\ref{app:H_int_superp_BH}.
This Hamiltonian is analogous to the one that was postulated in Ref.~\cite{foo2022quantum}, in the case of a mass-superposed black hole.

\subsection{Results} \label{sec:new_results}

\textit{Measuring the final state.} We now have a full description of the physical situation: given the initial state and interaction Hamiltonian in \cref{eq:H_int_QRF_transformed,eq:initial_state_BH_superp}, we can calculate the final state of the system for a transition to energy level $E$ and final scalar field state $\vartheta$ (which we will sum over), to first order in the interaction coupling. The resulting state is simply a linear superposition, i.e. 
\begin{equation} \label{eq:transition_amplitude_superp}
    \langle{E,\vartheta}|{\psi(\tau_f)}^{(D)}\rangle = \frac{\ket{\mathbf{x}_1}_B \Theta_1 +e^{i \Delta \Phi} \ket{\mathbf{x}_2}_B \Theta_2}{\sqrt{2}},
\end{equation}
where $\Theta_{1/2}$ are the transition amplitudes defined in \cref{eq:single_detector_amplitude} for detector position at the origin and black hole at position $\mathbf{x}_{1/2}$. 

We set the relative phase $\Delta\Phi$ to zero. Such a phase may arise from the gravitational dynamics of either the black hole or the apparatus holding the detector in place, both of which we neglect here. If $\Delta\Phi$ is non-negligible, it can nevertheless be absorbed into the definition of the measurement introduced below; see Appendix~\ref{sec:app_interaction} for details.

In order to witness interference effects, we must lose which-way information. For this reason, a simple measurement on the internal transition of the detector from $E_0$ to $E$ does not suffice, as it would produce a mixed signal $(P_1+P_2)/2$ with no interference effect. Instead, we measure a transition to an energy $\ket{E}$, together with a measurement on the BH position that ignores which-way (or which-branch) information\footnote{If we model the physical reference fields of Appendix~\ref{app:new_reference_fields} as well, we would have to do a joint measurement that ignores which-branch information by including all systems $S$ that (might) contain such which-branch information $S_i$, including the physical reference fields, and, in fact, also the device holding the detector at fixed radial location (which is not a geodesic), meaning $\ket{\pm} =1/\sqrt{2} ( \ket{\mathbf{x}_1}_B \ket{S_1}_S \pm \ket{\mathbf{x}_2}_B \ket{S_2}_S)$. If we did not do so, systems holding which-branch information might decohere the superposition leading to suppression of the off-diagonal term $P_{12}$.} $\ket{\pm}_B = \frac{\ket{\mathbf{x}_1}_B \pm \ket{\mathbf{x}_2}_B}{\sqrt{2}}$.
This is the standard measurement scheme in interferometry: the two branches are recombined with a beam splitter, and the output ports are then measured. Although such a protocol is likely to remain a thought experiment for a cosmological black hole, it nevertheless provides a useful conceptual probe, and an experimental realization in analogue-gravity systems~\cite{barcelo2005analogue,braunstein2023analogue,unruh1981experimental,garay2000sonic,weinfurtner2011measurement} may be possible. The measurement probabilities are
\begin{equation}\label{eq:probabilities}
    P_\pm = \frac{1}{4} \big( P_1 + P_2 \pm 2 P_{12} \big),
\end{equation}
where $P_{1/2}$ are the probabilities defined in Eq.~\eqref{eq:transition_prob} for a detector in a single black hole spacetime (but with the black hole at a different location $\mathbf{x}_1,\mathbf{x}_2$ relative to the detector). $P_{12}$ is an interference term obtained as $P_{12} = \text{Re}[F_1^* F_2]$, with $F_i$ the amplitudes of Eq.~\eqref{eq:single_detector_amplitude}, now summed over the final field states, for the black hole at positions $\mathbf{x}_1,\mathbf{x}_2$, that is 
\begin{align} \label{eq:interference_term}
    P_{12} = &|\langle E | \mu |E_0\rangle|^2 {\lambda^2} \nonumber \\ & \int_{\tau_i}^{\tau_f} d\tau d\tau'  \eta(\tau) \eta(\tau') e^{-i\Delta E(\tau-\tau')} W^{(12)}_{BTZ}(x(\tau),x(\tau'))
\end{align} with $\Delta E = E-E_0$ and $W^{(12)}_{BTZ}$ the two-point function \begin{equation}
    W^{(12)}_{BTZ}(x,x') = \langle 0 |_\phi \phi(x) \phi(x') |0 \rangle_\phi,
\end{equation} with $x,x'$ on the worldline of the detector at fixed spatial position $\mathbf{x}_1,\mathbf{x}_2$, respectively. 

We therefore find an interference term that is a genuine quantum effect extracted from the Unruh--DeWitt detector due to the black hole being in a quantum superposition. 

\subsection{Numerical analysis} \label{sec:results}

\begin{figure}[h]  \centering   \includegraphics[width=0.45\textwidth]{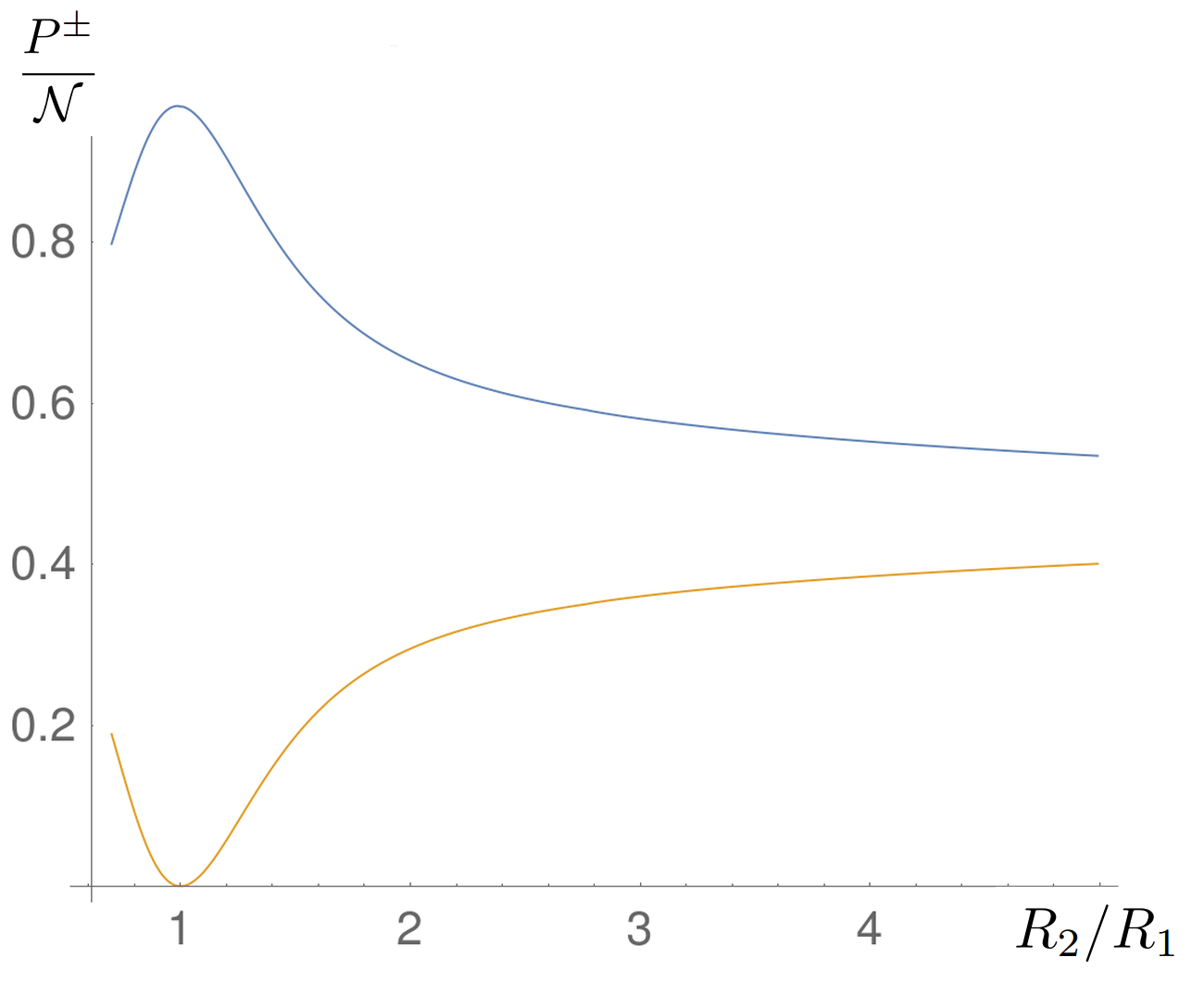} \caption{Measurement probabilities $P_{E}^{\pm}/\mathcal{N}$ with $\mathcal{N}=\sigma \lambda^2/2 |\langle E |\mu |E_0\rangle|^2$ for a position-superposed black hole as a function of ${R_2/R_1}$. Here $P^+_E/\mathcal{N}$ is in blue and $P^-_E/\mathcal{N}$ in orange.
Parameters used: $l/\sigma = 5,\sigma=1$, $\tau_f = 10 \sigma = -\tau_i$ in the detector's own clock time, and $R_1/\sigma = 2 r_h,Ml^2 =4,E-E_0 = 0.00016$.} \label{fig:Fig1own}
\end{figure}

In \cref{fig:Fig1own} we plot the measurement probabilities given in Eq.~\eqref{eq:probabilities}, rescaled by a normalization factor $\mathcal{N}=\sigma \lambda^2/2 |\langle E |\mu |E_0\rangle|^2$ that depends on details of the detector's internal structure, as a function of the ratio $R_2/R_1$ of the radial coordinate distances between the detector and black hole in the branches of the superposition. 

We see here the effect of the interference term, which is proportional to the difference between the two graphs. The interference is peaked around $R_1=R_2$, where the (coordinate) distance between the black hole and detector is the same in the two branches, and decreases with increasing ratio and as the distances between the detector and black hole increase in both branches, an effect in line with the decay of the conformally coupled scalar field two-point function with distance, i.e. due to $W^{(12)}_{BTZ}((t,R_1),(t',R_2))$ in \cref{eq:interference_term}.
However, right at and very near to $R_1 = R_2$, this calculation cannot be trusted, as then the black hole is not in a superposition of locations (as seen from the detector). Indeed, we have $\ket{+}_B = 2/\sqrt{2}\ket{x_1}, \ket{-} = 0$ meaning that the $\ket{\pm}$ measurement is ill-defined and the assumed initial state in \cref{eq:initial_state_BH_superp} with $\mathbf{x}_1 = \mathbf{x}_2$ is not normalized, such that the calculation of the interference probabilities \cref{eq:probabilities} cannot be trusted there. 

\begin{figure}[h]  \centering   \includegraphics[width=0.45\textwidth]{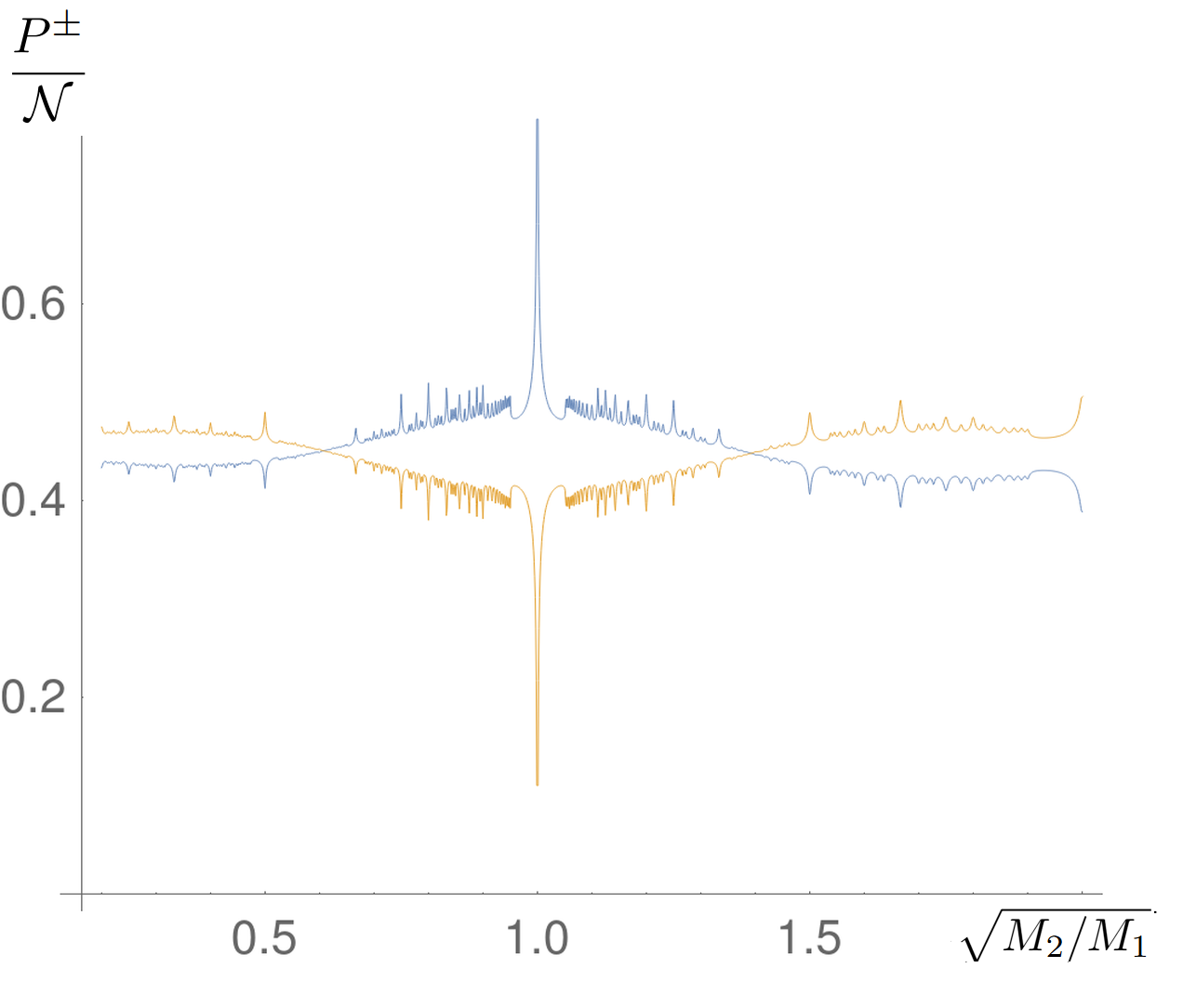} \caption{Results analogous to Fig.~1 of Ref.~\cite{foo2022quantum}: Measurement probabilities $P_{E}^{\pm}/\mathcal{N}$ with $\mathcal{N}=\sigma \lambda^2/2 |\langle E |\mu |E_0\rangle|^2$ for a mass-superposed black hole as a function of $\sqrt{M_2/M_1}$. Here $P^+_E/\mathcal{N}$ is in blue and $P^-_E/\mathcal{N}$ in orange. 
Parameters used: $l/\sigma = 5,\sigma=1$, $t_f = 5 \sigma = -t_i,R/\sigma = 25,M_1 l^2 =4, E-E_0 = \Omega = 0.01 M_1$ (with $t_f,t_i$ coordinate start and end time). Note that Ref.~\cite{foo2022quantum} included an extra phase $\Delta \Phi = (\sqrt{M_1}-\sqrt{M_2})\Delta t$ in the interference term, i.e. $\pm 2 \cos(\Delta \Phi) P_{12}$ in $P^\pm$~\cite{foo2022quantum}, a convention we follow here.} \label{fig:Fig1Foocompare}
\end{figure}

Furthermore, it is meaningful to compare our results with those of Ref.~\cite{foo2022quantum}, where a mass-superposed black hole was considered, instead of the position-superposed one we consider. In particular, we are comparing the measurement probabilities in Eq.~\eqref{eq:probabilities}, with the measurement probabilities of Eq.~(11) of Ref.~\cite{foo2022quantum}, that are analogous to the setup studied here, but when considering a black hole in superposition of masses. We plot their probabilities in \cref{fig:Fig1Foocompare}, reproducing Fig.~1 of Ref.~\cite{foo2022quantum}. 
The authors argue that the sharp peaks that they observe are a consequence of having a superposition of different masses, and in particular interpret it in light of black hole mass quantization~\cite{bekenstein2020quantum,hod1998bohr,Bekenstein_1995,bekenstein1998quantumblackholesatoms,deppe2024echoesbeyonddetectinggravitational}. Our results further support this claim: when the black hole is in superposition of locations, but with single-valued mass, the sharp peaks are indeed absent. In the next subsection we will show in detail how this phenomenon arises from the spectrum probed by the detector.

\subsection{Mass- vs. position-superposition} \label{sec:analytical}
In this subsection we derive an analytical expression for the spectrum probed by the detector in the case of a mass- and position-superposed black hole. We use it to identify the origin of the peaks that were observed numerically in the detector response for a mass-superposed black hole in~\cite{foo2022quantum} (see~\cref{fig:Fig1Foocompare}). We find that the peaks can be traced to a coherent enhancement of the image sum at rational square-root mass ratios, and that additional singularities of the spectrum arise at certain rational mass ratios and corresponding position values in the mass-superposed case. Concretely, we will consider a generic mass- and position-superposed black hole, with position and mass $R_1,M_1$ and $R_2,M_2$ in the two branches of the superposition, respectively. 
Details can be found in Appendix \ref{sec:analytical_appendix}.

\vspace{0.5cm}

\textit{Single-spacetime detector response.} In a non-superposed spacetime, the detector response from \cref{eq:transition_prob} can be expressed as~\footnote{See Ref.~\cite{Fewster_2016} for a more precise treatment.}
\begin{equation} \label{eq:Fewster_eq_chi_hatW}
    \mathcal{F}(E)=\frac{1}{2 \pi} \int_{-\infty}^{\infty} d \omega \,|\widehat{\eta}({\scriptstyle\frac{\omega}{\gamma}})|^2 \,\widehat{W}_{\scriptstyle BTZ}(E \gamma +\omega),
\end{equation} with time integration limits pushed to $\pm \infty$ and the Fourier transform defined as
\begin{equation}
    \widehat{f}(\omega)=\int_{-\infty}^{\infty} d s f(s) \, \mathrm{e}^{-\mathrm{i} \omega s}.
\end{equation}
An analytical expression for $\widehat{W}_{BTZ}(K)$ can be found~\cite{Lifschytz_1994}, see Appendix \ref{sec:analytical_appendix} for details.

\textit{Interference term in detector response.} The interference term $\mathcal{F}_{12}(E)$, for both mass- and position-superposed black holes now, can be written in a similar form as \cref{eq:Fewster_eq_chi_hatW}, obtaining \begin{equation}
\begin{split}
    \mathcal{F}_{12}(E) = \frac{1}{2\pi} \int_{-\infty}^{+ \infty}  \mathrm{d}\omega  \hspace{0.05cm} \widehat{\eta}({\scriptstyle\frac{\omega}{\widetilde{\gamma_1}}})^* \hspace{0.05cm} \widehat{\eta}\left({\scriptstyle\frac{\omega}{\widetilde{\gamma_2}} + E [\frac{\widetilde{\gamma_1}}{\widetilde{\gamma_2}} -1] } \right) \\ \hspace{0.05cm} \widehat{W}^{(12)}_{BTZ}\left(\omega+\widetilde{\gamma_1} E\right). 
\end{split}
\end{equation}
Following Ref.~\cite{foo2022quantum}, we consider AdS time coordinates $\overline{t}$ related to the BTZ coordinates $t$ as $\overline{t}=t \sqrt{M}$ (see Supplemental Material of Ref.~\cite{foo2022quantum}, for instance).
Using the formulas in Appendix B of Ref.~\cite{Lifschytz_1994}, an analytical expression for $\widehat{W}^{(12)}_{BTZ}$ can be found, with the additional subtlety that for $\sqrt{M_1/M_2} \in \mathbb{Q} \backslash \{1\}$ and $R_1 /\sqrt{M_1} = R_2 /\sqrt{M_2}$ an additional pole arises, giving a singular term $\textsf{sing.}$:
\begin{equation} \label{eq:widehatW_12_BTZ_K}
    \begin{split}
        \widehat{W}^{(12)}_{BTZ}(K) = \textsf{sing.} + \frac{1}{2  \sqrt{\widetilde{\gamma_1} \widetilde{\gamma_2}}} \frac{1}{\sum_k \Upsilon^{2k}} \hspace{0.07cm} \frac{1}{e^{K/T_H^{AdS}}+1} \\ \sum_{m,n} \bigg[ P_{\frac{i K}{2 \pi T_H^{AdS}}-\frac{1}{2}}\big(\beta_{mn}^{(12)} \big) - \zeta  P_{\frac{i K}{2 \pi T_H^{AdS}}-\frac{1}{2}}\big(\beta_{mn}^{'(12)}\big) \bigg],
    \end{split}
\end{equation} 
with $T_H^{AdS}$ the Hawking temperature of the AdS–Rindler horizon with respect to AdS–Rindler time, $T_H^{AdS} = 1/(2 \pi l), \widetilde{\gamma_i}=\gamma(R_i,M_i)/\sqrt{M_i}$, and $\beta_{nm}^{(12)},\beta_{nm}^{'(12)}$ factors that depend on $M_1,M_2,R_1,R_2$. Here $P_\nu$ is a Legendre function of the first kind.
If $\sqrt{M_1/M_2} \notin \mathbb{Q} \backslash \{1\}$, then we have $\textsf{sing.} = 0$, else it gives the following contribution if also $R_1 = R_2 \sqrt{M_1/M_2}$: 
\begin{equation} \label{eq:sing.}
\begin{split}
    \textsf{sing.} = & \frac{1}{4  \sqrt{\widetilde{\gamma_1}\widetilde{\gamma_2} p q}} \\  &\text{ if } \sqrt{\frac{M_1}{M_2}} = \frac{p}{q} \in \mathbb{Q} \backslash \{1\} \text{ and } \frac{R_1}{R_2} =  \sqrt{\frac{M_1}{M_2}}    
\end{split}
\end{equation} with $\text{gcd}(p,q)=1$. Here we used a natural cut-off procedure for the infinite sums $\sum_{m,n}$ as suggested by the group structures that define the BTZ spacetime through the periodic identification of AdS--Rindler space, see also Appendix \ref{sec:analytical_appendix}. 

For other values of $R_1,R_2$, one must resort to the `regular part' of $\widehat{W}^{(12)}_{BTZ}(K)$ of \Cref{eq:widehatW_12_BTZ_K}. For rational $\sqrt{M_1/M_2}=\frac{p}{q}$ and $R_1/R_2$ not this same rational number but $R_1,R_2$ fixed, $\beta_{mn}^{(12)}$ has the same value for terms in the sum over $m,n$ satisfying $n/m=\sqrt{M_1/M_2}$, and thus add up coherently. In addition, for these coherently repeated terms, when $R_1,R_2$ are both large compared to their horizon radii, $\beta_{mn}^{(12)} \rightarrow 1+$, the Legendre function $P_{-\frac{1}{2}-iK...}$ remains close to 1 until large $|K|$-values before decaying, leading to sharper peaks.  
On the other hand, for irrational $\sqrt{M_1/M_2}$, for large $|K|$ oscillatory behavior (decaying with $|K|$) of the Legendre function dephases the sum leading to cancellations on average. For near-rational irrational $\sqrt{M_1/M_2}$ cut-off dependent peaks may occur at large $R_1,R_2$, but only to a limited extent, as determined by the irrational offset, with more averaged cancellations for large $|K|$.
Additionally, large positive $K$-values are exponentially suppressed by the thermal factor, and thus main contributions to (peaked) behavior come from large negative $K$-values. 
This leads to the peaks in the interfered response function in Ref.~\cite{foo2022quantum}, as pictured in \Cref{fig:Fig1Foocompare}.

\section{Discussion and outlook} \label{sec:discussion}

We have analyzed the transition probability of a static detector in a position-superposed BTZ black hole spacetime, where our measurement loses which-way information on the position of the black hole. To do so, we employed a quantum reference frame transformation, showing that the physical situation is equivalent to a position-superposed Unruh--DeWitt detector evolving in a classical spacetime. As a result, we find a `nonclassical' term that would be absent for a black hole in a probabilistic mixture of locations.

The same setup but for a mass-superposed BTZ black hole was considered by Foo et al. in Ref.~\cite{foo2022quantum}, where the interferometric measurement was performed upon superposed mass states for the black hole, and sharp peaks reminiscent of black hole quantization were found.
In our setup we find no such sharp peaks, providing further evidence that those peaks are a consequence of black hole mass quantization.
Moreover, we have provided analytical expressions for the spectrum probed by an Unruh--DeWitt detector in the case of a mass- and position-superposed BTZ black hole. 
These expressions show that the sharp peaks originate from a coherent enhancement of the image sum at rational square-root mass ratios, and that at certain position ratios an additional pole singularity appears at rational square-root mass ratios.

\vspace{0.5cm}

Building on our results, and more broadly on the study of detectors in superposed spacetimes, several promising avenues for future research emerge. A natural direction for future work is to investigate the physical regimes in which our approximations break down. For instance, we have assumed that the two branches of the superposition are classically perfectly distinguishable, so that the corresponding black hole states can be treated as orthogonal. This assumption is expected to fail when the two superposed locations become arbitrarily close. A first step toward addressing this regime could be to apply linearized quantum gravity, along the lines of Refs.~\cite{chen2023quantum,chen2024quantum}. In addition, we might relax the assumption of gravity being sourced only by position. In the more general situation in which gravity is sourced by both position and momentum, a more refined treatment of semiclassical states is required, for example in terms of gravitational coherent states~\cite{carney_dressed_2018,ashtekar2015geometryphysicsnullinfinity,ashtekar_null_2018,Prabhu_2024,prabhu_infrared_2024,strominger2018lecturesinfraredstructuregravity,Thiemann_2001,klauder1985coherent,zhang1990coherent,Kiefer:2004xyv,hartle1983wave}.

Furthermore, another interesting possibility would be to investigate coherences in a detector with many energy levels near a superposed BTZ black hole, as was done for the Schwarzschild black hole in \cite{paczos2023hawking}.
Moreover, we have investigated the first nontrivial perturbative order in the Unruh--DeWitt interaction, but a more complete understanding of this interaction requires the investigation of higher-order terms~\cite{bradler2018unitaryevolutionpairunruhdewitt,han2025reliablequantummasterequation} and nonperturbative treatments~\cite{Brown_2013,Polo_G_mez_2024}.

As the quantum BTZ black hole has an effective description as a two-dimensional conformal field theory (CFT) at infinity that can be coupled to classical matter~\cite{Carlip_2014,Carlip_2005,emparan1998quantization}, it would be interesting to investigate the results regarding transition amplitudes and probabilities of detectors in superposed BTZ black hole spacetimes in the related CFT.
One would expect similar effects as for the position- and mass-superposed BTZ black holes in other superposed spacetimes, such as superposed Schwarzschild black holes for example, and how the spectra of peaks in solvable 1+1d (black hole) spacetimes~\cite{mertens2023solvable} relate to CFTs on the boundary in AdS/CFT~\cite{maldacena1999large}. 
Our analytical results suggest that this could be achieved by probing the spectrum of interfered two-point functions between superposed spacetimes, and looking for additional Hadamard-like singularities~\cite{Kay:1988mu,wald1994quantum,Radzikowski:1996pa,Fewster_2013} in these interfered two-point functions that relate properties such as black hole mass with the quantized angular structure of the quantized field solutions.
In general, calculating Wightman functions in such superpositions might be challenging, but branch-dependent coordinate transformations, a form of a quantum reference frame transformation, or embeddings in higher-dimensional (Minkowski) spacetime may offer potential routes. 
An example of the latter is Schwarzschild in 6d Minkowski spacetime~\cite{fronsdal1959completion} in the GEMS (Global Embedding in Minkowski Spacetime) paradigm~\cite{russo2008accelerating,deser1997accelerated,deser1998equivalence,deser1999mapping,sheykin2019global,sheykin2021global}, even though it is not always clear when (in which cases) exactly this mapping works~\cite{Paston_2014}. 

\vspace{0.5cm}

Finally, this work combines aspects of quantum reference frames, quantum gravity and detector models, and thus works toward an operational understanding of quantum theories of gravity. 
This approach allows to critically assess some common assumptions of black hole paradoxes~\cite{hawking1976breakdown,harlow2016jerusalem,Polchinski_2016,raju2022lessons}, especially when combining the viewpoints of different observers such as the cloning~\cite{susskind1994gedanken,hayden2007black} and firewall paradoxes~\cite{almheiri2013black,braunstein2007quantum,braunstein2013better,harlow2016jerusalem}.
Regarding the black hole information puzzle~\cite{hawking1976breakdown}, expectations of unitarity come from AdS/CFT~\cite{maldacena1999large} and recent path integral calculations, making use of so-called entanglement islands and replica wormholes~\cite{penington2020entanglement,almheiri2020page,almheiri2019entropy,wang2024entanglement,Hartman_2020,wang2021page,Hashimoto_2020,Balasubramanian_2021,almheiri2020entanglementislandsinhigher,kibe2022holographic,chen2020quantumextremalislandseasyI,chen2020quantumextremalislandseasyII,bousso_islands_2023,almheiri2023islandsoutsidehorizon,Krishnan_2021,Van_Raamsdonk_2021,geng2021information,geng2021informationparadoxandits,geng2026seeingpagecurvesislands,geng2022inconsistency,raju2022failure,geng2025revisitingrecentprogresskarchrandall,geng2025makingcasemassiveislands,Geng_2020,geng2026mechanisminformationencodingislands,Uhlemann2022information,Uhlemann_2021}. 
These path integral calculations reproduce the unitary Page curve~\cite{page1993information,page2013time} for the radiation, but a precise mechanism and physical understanding remain lacking.
An interesting development which may be important for the mechanism of information transfer is a black hole's soft hair~\cite{hawking2016soft,Donnay_2016,carney_infrared_2017,strominger2017blackholeinformationrevisited}, closely related to quantum reference frames as shown in Refs.~\cite{kabel2023quantum,Araujo_Regado_2025,Carrozza_2022,gomes2025boundariesframesissuephysical}.

Standard assumptions of black hole paradoxes featuring infalling observers, such as the cloning and firewall paradoxes, are subject to several subtleties, see for example Sec. VI of Ref.~\cite{walleghem2026wignersfriendsblackhole}.
First, an old black hole has a nonstationary and, expectedly, a nontrivial superposed quantum geometry~\cite{flanagan_infrared_2021,page1980is,Bao_2018,calmet2022brief,akil2025quantumsuperpositionblackhole,pasterski_hps_2021,Hutchinson_2016}, which an infalling observer will interact with. 
Another issue is the gauge invariance of observables and, relatedly, the definition of a radiation subsystem (inside the bulk)~\cite{Uhlemann2022information,JACOBSON_2013,Jacobson_2019,raju2022failure}. 
Observables for infalling observers can be made gauge-invariant by considering relational observables or dressings~\cite{Rovelli:1990ph,Brown_1995,Giddings:2019hjc,Giddings:2022hba,Giddings:2025xym,Giddings:2025bkp,Donnelly:2016rvo,Donnelly:2018nbv,Giddings_2006,goeller2022diffeomorphisminvariantobservablesdynamicalframes,Jacobson_2019,Giddings:2024qcf,JACOBSON_2013,deboer2022frontiersquantumgravityshared}, which can be seen as a use of quantum reference frames~\cite{goeller2022diffeomorphisminvariantobservablesdynamicalframes,fewster2024quantumreferenceframesmeasurement}, with nontrivial implications for spacetime locality.

Explicit calculations using quantum reference frames and detectors probing the radiation can further quantify these subtleties. 
For example, claims concerning the entanglement of subsystems can be probed operationally with entanglement harvesting~\cite{Stritzelberger_2021,Hotta_2020,Stritzelberger_2020,Cong_2020,Ng_2018,henderson2018harvesting,Gallock_Yoshimura_2021,Mendez_Avalos_2022,Tjoa_2020,robbins2020entanglementamplificationrotatingblack,chakraborty2024entanglementharvestingquantumsuperposed,dubey2025harvesting,Membrere_2023,Barman_2023,wang2025harvestinginformationhorizon}, including claims of early and late radiation entanglement, and which observers would detect high entanglement across the horizon and in radiation from entangled black holes~\cite{Maldacena_2013,susskind2014ereprghzconsistencyquantum,Susskind_2016}.
\vspace{0.5cm}
\section*{Acknowledgements}
\vspace{-0.5cm}
LW thanks David Maibach for many interesting discussions on black holes and information. 
LW thanks Bernard Kay, Benito A. Juárez-Aubry and Claudio Dappiaggi for interesting discussions on detectors, the Unruh and Hawking effect. CC thanks Luis Barbado and Flaminia Giacomini for useful comments on an early version of the draft. LW thanks Harkan Kirk-Karakaya for an introduction to the University of York's (UoY) research server.
The UoY research and Viking cluster was used during this project, which is a high performance compute facility provided by the University of York. We are grateful for computational support from the University of York, IT Services and the Research IT team.
LW acknowledges the use of the AI tools Claude and ChatGPT for making numerical code more efficient and improving language. 
LW acknowledges support from the United Kingdom Engineering and Physical Sciences Research Council (EPSRC) DTP Studentship (grant number EP/W524657/1). LW also thanks the International Iberian Nanotechnology Laboratory -- INL in Braga, Portugal and the Quantum and Linear-Optical Computation (QLOC) group for the kind hospitality. This research was funded in whole or in part by
the Austrian Science Fund (FWF) [10.55776/F71] and
[10.55776/COE1]. This publication was made possible through the
financial support of WOST (WithOutSpaceTime) grant
from the John Templeton Foundation. The opinions expressed in this publication are those of the authors and
do not necessarily reflect the views of the John Templeton Foundation.

\bibliography{refs}

@article{foo2022quantum,
  title={Quantum signatures of black hole mass superpositions},
  author={Foo, Joshua and Arabaci, Cemile Senem and Zych, Magdalena and Mann, Robert B},
  journal={Phys. Rev. Lett.},
  volume={129},
  number={18},
  pages={181301},
  year={2022},
  publisher={APS},
 url ={https://doi.org/10.1103/PhysRevLett.129.181301}
}

@article{giesel2025linkingedgemodesgeometrical,
  title = {Linking edge modes and geometrical clocks in linearized gravity},
  author = {Giesel, Kristina and Kabel, Viktoria and Wieland, Wolfgang},
  journal = {Phys. Rev. D},
  volume = {112},
  issue = {6},
  pages = {064063},
  numpages = {30},
  year = {2025},
  month = {Sep},
  publisher = {American Physical Society},
  doi = {10.1103/yrc1-gmql},
  url = {https://link.aps.org/doi/10.1103/yrc1-gmql}
}

@article{Barbado_2016,
  author = {L. C. Barbado and C. Barceló and L. J. Garay and G. Jannes},
  title = {Hawking versus {U}nruh effects, or the difficulty of slowly crossing a black hole horizon},
  journal = {J. High Energy Phys.},
  volume = {10},
  number = {2016},
  pages = {161},
  year = {2016},
  doi = {10.1007/jhep10(2016)161},
}

@article{Banados_1992,
   title={Black hole in three-dimensional spacetime},
   volume={69},
   ISSN={0031-9007},
   url={http://dx.doi.org/10.1103/PhysRevLett.69.1849},
   DOI={10.1103/physrevlett.69.1849},
   number={13},
   journal={Phys. Rev. Lett.},
   publisher={American Physical Society (APS)},
   author={Bañados, Máximo and Teitelboim, Claudio and Zanelli, Jorge},
   year={1992},
   month=sep, pages={1849–1851} }

@article{Eyheralde_2020,
   title={Quantum fluctuating geometries and the information paradox II},
   volume={37},
   ISSN={1361-6382},
   url={http://dx.doi.org/10.1088/1361-6382/ab6e89},
   DOI={10.1088/1361-6382/ab6e89},
   number={6},
   journal={Class. Quantum Grav.},
   publisher={IOP Publishing},
   author={Eyheralde, Rodrigo and Gambini, Rodolfo and Pullin, Jorge},
   year={2020},
   month=feb, pages={065001} }

@article{Eyheralde_2017,
   title={Quantum fluctuating geometries and the information paradox},
   volume={34},
   ISSN={1361-6382},
   url={http://dx.doi.org/10.1088/1361-6382/aa8e30},
   DOI={10.1088/1361-6382/aa8e30},
   number={23},
   journal={Class. Quantum Grav.},
   publisher={IOP Publishing},
   author={Eyheralde, Rodrigo and Campiglia, Miguel and Gambini, Rodolfo and Pullin, Jorge},
   year={2017},
   month=nov, pages={235015} }

@inproceedings{Polchinski_2016,
   title={The Black Hole Information Problem},
   url={http://dx.doi.org/10.1142/9789813149441_0006},
   DOI={10.1142/9789813149441_0006},
   booktitle={New Frontiers in Fields and Strings},
   publisher={WORLD SCIENTIFIC},
   author={Polchinski, Joseph},
   year={2016},
   month=nov }

@Inbook{Kraus2008,
author="Kraus, P.",
title="Lectures on Black Holes and the {A}d{S}3/{CFT}2 Correspondence",
bookTitle="Supersymmetric Mechanics - Vol. 3: Attractors and Black Holes in Supersymmetric Gravity",
year="2008",
publisher="Springer Berlin Heidelberg",
address="Berlin, Heidelberg",
pages="1--55",
isbn="978-3-540-79523-0",
doi="10.1007/978-3-540-79523-0_4",
url="https://doi.org/10.1007/978-3-540-79523-0_4"
}

@misc{carlip2023quantumgravity21dimensions,
      title={Quantum Gravity in 2+1 Dimensions}, 
      author={S. Carlip},
      year={2023},
      eprint={2312.12596},
      archivePrefix={arXiv},
      primaryClass={gr-qc} 
}

@article{Carlip_1995,
   title={The (2 + 1)-dimensional black hole},
   volume={12},
   ISSN={1361-6382},
   url={http://dx.doi.org/10.1088/0264-9381/12/12/005},
   DOI={10.1088/0264-9381/12/12/005},
   number={12},
   journal={Class. Quantum Grav.},
   publisher={IOP Publishing},
   author={Carlip, S},
   year={1995},
   month=dec, pages={2853–2879} }

@article{giacomini2022second,
  title={Second-quantized {U}nruh-{D}e{W}itt detectors and their quantum reference frame transformations},
  author={Giacomini, Flaminia and Kempf, Achim},
  journal={Phys. Rev.  D},
  volume={105},
  number={12},
  pages={125001},
  year={2022},
  publisher={APS},
  doi={https://doi.org/10.1103/PhysRevD.105.125001}
}

@article{barbado2020unruh,
  title={Unruh effect for detectors in superposition of accelerations},
  author={Barbado, Luis C and Castro-Ruiz, Esteban and Apadula, Luca and Brukner, {\v{C}}aslav},
  journal={Phys. Rev.  D},
  volume={102},
  number={4},
  pages={045002},
  year={2020},
  publisher={APS}, 
  doi={https://doi.org/10.1103/PhysRevD.102.045002}
}

@article{kabel2024identification,
   title={Quantum coordinates, localisation of events, and the quantum hole argument},
   volume={8},
   ISSN={2399-3650},
   url={http://dx.doi.org/10.1038/s42005-025-02084-3},
   DOI={10.1038/s42005-025-02084-3},
   pages={185},
   journal={Commun. Phys.},
   publisher={Springer Science and Business Media LLC},
   author={Kabel, Viktoria and de la Hamette, Anne-Catherine and Apadula, Luca and Cepollaro, Carlo and Gomes, Henrique and Butterfield, Jeremy and Brukner, {\v{C}}aslav},
   year={2025},
   month=apr }

@article{Cepollaro_2024,
doi = {10.1088/1361-6382/ad6d26},
url = {https://dx.doi.org/10.1088/1361-6382/ad6d26},
year = {2024},
month = {aug},
publisher = {IOP Publishing},
volume = {41},
number = {18},
pages = {185009},
author = {Carlo Cepollaro and Flaminia Giacomini},
title = {Quantum generalisation of {E}instein’s equivalence principle can be verified with entangled clocks as quantum reference frames},
journal = {Class. Quantum Grav.}
}

@article{henderson2018harvesting,
  title={Harvesting entanglement from the black hole vacuum},
  author={Henderson, Laura J and Hennigar, Robie A and Mann, Robert B and Smith, Alexander RH and Zhang, Jialin},
  journal={Class. Quantum Grav.},
  volume={35},
  number={21},
  pages={21LT02},
  year={2018},
  publisher={IOP Publishing},
  doi={https://doi.org/10.1088/1361-6382/aae27e}
}

@article{wood2022quantized,
  title={Quantized mass-energy effects in an {U}nruh-{D}e{W}itt detector},
  author={Wood, Carolyn E and Zych, Magdalena},
  journal={Phys. Rev.  D},
  volume={106},
  number={2},
  pages={025012},
  year={2022},
  publisher={APS},
  doi={https://doi.org/10.1103/PhysRevD.106.025012}
}

@article{bose2017spin,
  title={Spin entanglement witness for quantum gravity},
  author={Bose, Sougato and Mazumdar, Anupam and Morley, Gavin W and Ulbricht, Hendrik and Toro{\v{s}}, Marko and Paternostro, Mauro and Geraci, Andrew A and Barker, Peter F and Kim, MS and Milburn, Gerard},
  journal={Phys. Rev. Lett.},
  volume={119},
  number={24},
  pages={240401},
  year={2017},
  publisher={APS}, 
  doi={https://doi.org/10.1103/PhysRevLett.119.240401}
}

@article{marletto2017gravitationally,
  title={Gravitationally induced entanglement between two massive particles is sufficient evidence of quantum effects in gravity},
  author={Marletto, Chiara and Vedral, Vlatko},
  journal={Phys. Rev. Lett.},
  volume={119},
  number={24},
  pages={240402},
  year={2017},
  publisher={APS},
  doi={https://doi.org/10.1103/PhysRevLett.119.240402}
}

@article{sudhir2021unruh,
  title={Unruh effect of detectors with quantized center of mass},
  author={Sudhir, Vivishek and Stritzelberger, Nadine and Kempf, Achim},
  journal={Phys. Rev.  D},
  volume={103},
  number={10},
  pages={105023},
  year={2021},
  publisher={APS}, 
  doi={https://doi.org/10.1103/PhysRevD.103.105023}
}

@article{Stritzelberger_2021,
   title = {Entanglement harvesting with coherently delocalized matter},
  author = {Stritzelberger, Nadine and Henderson, Laura J. and Baccetti, Valentina and Menicucci, Nicolas C. and Kempf, Achim},
  journal = {Phys. Rev. D},
  volume = {103},
  issue = {1},
  pages = {016007},
  numpages = {14},
  year = {2021},
  month = {Jan},
  publisher = {American Physical Society},
  doi = {10.1103/PhysRevD.103.016007},
  url = {https://link.aps.org/doi/10.1103/PhysRevD.103.016007}
}

@article{Hotta_2020,
    title = {Duality in the dynamics of {U}nruh-{D}e{W}itt detectors in conformally related spacetimes},
  author = {Hotta, Masahiro and Kempf, Achim and Mart\'{\i}n-Mart\'{\i}nez, Eduardo and Tomitsuka, Takeshi and Yamaguchi, Koji},
  journal = {Phys. Rev. D},
  volume = {101},
  issue = {8},
  pages = {085017},
  numpages = {10},
  year = {2020},
  month = {Apr},
  publisher = {American Physical Society},
  doi = {10.1103/PhysRevD.101.085017},
  url = {https://link.aps.org/doi/10.1103/PhysRevD.101.085017}
}

@article{Ju_rez_Aubry_2014,
   title={Onset and decay of the 1 + 1 {H}awking-{U}nruh effect: what the derivative-coupling detector saw},
   volume={31},
   ISSN={1361-6382},
   url={http://dx.doi.org/10.1088/0264-9381/31/24/245007},
   DOI={10.1088/0264-9381/31/24/245007},
   number={24},
   journal={Class. Quantum Grav.},
   publisher={IOP Publishing},
   author={Juárez-Aubry, Benito A and Louko, Jorma},
   year={2014},
   month=nov, pages={245007} }

@article{barrow1986three,
  title={Three-dimensional classical spacetimes},
  author={Barrow, John D and Burd, AB and Lancaster, David},
  journal={Class. Quantum Grav.},
  volume={3},
  number={4},
  pages={551},
  year={1986},
  doi = {10.1088/0264-9381/3/4/010}}

@article{Bacry:1968zf,
    author = "Bacry, H. and Levy-Leblond, J.",
    title = "{Possible kinematics}",
    doi = "10.1063/1.1664490",
    journal = "J. Math. Phys.",
    volume = "9",
    pages = "1605--1614",
    year = "1968"
}

@article{Bizo__2018,
  author = {Bizoń, Piotr and Evnin, Oleg and Ficek, Filip},
  title = {A nonrelativistic limit for {AdS} perturbations},
  journal = {J. High Energy Phys.},
  volume = {12},
  number = {2018},
  pages = {113},
  year = {2018},
  doi = {10.1007/jhep12(2018)113},
}

@article{DESER1984220,
title = {Three-dimensional {E}instein gravity: Dynamics of flat space},
journal = {Ann. Phys.},
volume = {152},
number = {1},
pages = {220-235},
year = {1984},
issn = {0003-4916},
doi = {https://doi.org/10.1016/0003-4916(84)90085-X},
url = {https://www.sciencedirect.com/science/article/pii/000349168490085X},
author = {S Deser and R Jackiw and G {'t Hooft}}
}

@article{Stritzelberger_2020,
   title = {Coherent delocalization in the light-matter interaction},
  author = {Stritzelberger, Nadine and Kempf, Achim},
  journal = {Phys. Rev. D},
  volume = {101},
  issue = {3},
  pages = {036007},
  numpages = {10},
  year = {2020},
  month = {Feb},
  publisher = {American Physical Society},
  doi = {10.1103/PhysRevD.101.036007},
  url = {https://link.aps.org/doi/10.1103/PhysRevD.101.036007}
}

@article{Ng_2014,
   title={{U}nruh-{D}e{W}itt detector response along static and circular-geodesic trajectories for {S}chwarzschild–anti-de {S}itter black holes},
  author = {Ng, Keith K. and Hodgkinson, Lee and Louko, Jorma and Mann, Robert B. and Mart\'{\i}n-Mart\'{\i}nez, Eduardo},
  journal = {Phys. Rev. D},
  volume = {90},
  issue = {6},
  pages = {064003},
  numpages = {13},
  year = {2014},
  month = {Sep},
  publisher = {American Physical Society},
  doi = {10.1103/PhysRevD.90.064003},
  url = {https://link.aps.org/doi/10.1103/PhysRevD.90.064003}
}

@article{bhattacharya2024probinghiddentopologyquantum,
   title={Probing hidden topology with quantum detectors},
   volume={111},
   ISSN={2470-0029},
   DOI={10.1103/physrevd.111.045005},
   pages={045005},
   journal={Phys. Rev. D},
   publisher={American Physical Society (APS)},
   author={Bhattacharya, Dyuman and Louko, Jorma and Mann, Robert B.},
   year={2025},
   month=feb }

@article{Lima_2019,
   title={Probing the {U}nruh effect with an accelerated extended system},
   volume={10},
   ISSN={2041-1723},
   url={http://dx.doi.org/10.1038/s41467-019-10962-y},
   DOI={10.1038/s41467-019-10962-y},
   number={1},
   journal={Nat. Commun.},
   publisher={Springer Science and Business Media LLC},
   author={Lima, Cesar A. Uliana and Brito, Frederico and Hoyos, José A. and Vanzella, Daniel A. Turolla},
   year={2019},
   pages = {3030},
number = {},
   month=jul }

@article{CAMPOS2021136198,
title = {The anti-Hawking effect on a {BTZ} black hole with Robin boundary conditions},
journal = {Phys. Lett. B},
volume = {816},
pages = {136198},
year = {2021},
issn = {0370-2693},
doi = {https://doi.org/10.1016/j.physletb.2021.136198},
url = {https://www.sciencedirect.com/science/article/pii/S0370269321001386},
author = {Lissa de Souza Campos and Claudio Dappiaggi}
}

@article{Henderson_2020,
   title={Anti-{H}awking phenomena},
   volume={809},
   ISSN={0370-2693},
   url={http://dx.doi.org/10.1016/j.physletb.2020.135732},
   DOI={10.1016/j.physletb.2020.135732},
   journal={Phys. Lett. B},
   publisher={Elsevier BV},
   author={Henderson, Laura J. and Hennigar, Robie A. and Mann, Robert B. and Smith, Alexander R.H. and Zhang, Jialin},
   year={2020},
   month=oct, pages={135732} }

@article{Polo_G_mez_2024,
   title={Nonperturbative method for particle detectors with continuous interactions},
   volume={109},
   ISSN={2470-0029},
   url={http://dx.doi.org/10.1103/PhysRevD.109.045014},
   DOI={10.1103/physrevd.109.045014},
   pages={045014},
   journal={Phys. Rev. D},
   publisher={American Physical Society (APS)},
   author={Polo-Gómez, José and Martín-Martínez, Eduardo},
   year={2024},
   month=feb }

@article{han2025reliablequantummasterequation,
  title = {Reliable quantum master equation of the Unruh-DeWitt detector},
  author = {Han, Si-Wei and Chen, Wenjing and Chen, Langxuan and Ouyang, Zhichun and Feng, Jun},
  journal = {Phys. Rev. D},
  volume = {112},
  issue = {8},
  pages = {085014},
  numpages = {15},
  year = {2025},
  month = {Oct},
  publisher = {American Physical Society},
  doi = {10.1103/c9wz-lm5m},
  url = {https://link.aps.org/doi/10.1103/c9wz-lm5m}
}

@article{Prabhu_2024,
  author = {Prabhu, Kartik and Satishchandran, Gautam},
  title = {Infrared finite scattering theory: scattering states and representations of the {BMS} group},
  journal = {J. High Energy Phys.},
  volume = {08},
  number = {2024},
  pages = {055},
  year = {2024},
  doi = {10.1007/jhep08(2024)055},
}

@article{prabhu_infrared_2024,
    title = {Infrared finite scattering theory: {Amplitudes} and soft theorems},
    volume = {110},
    issn = {2470-0010, 2470-0029},
    shorttitle = {Infrared finite scattering theory},
    url = {https://link.aps.org/doi/10.1103/PhysRevD.110.085022},
    doi = {10.1103/PhysRevD.110.085022},
    
    number = {8},
    journal = {Phys. Rev. D},
    author={Prabhu, K. and Satishchandran, G.},
    month = oct,
    year = {2024},
    pages = {085022},
}

@article{Thiemann_2001,
   title={Gauge field theory coherent states (GCS): I. General properties},
   volume={18},
   ISSN={1361-6382},
   url={http://dx.doi.org/10.1088/0264-9381/18/11/304},
   DOI={10.1088/0264-9381/18/11/304},
   number={11},
   journal={Class. Quantum Grav.},
   publisher={IOP Publishing},
   author={Thiemann, Thomas},
   year={2001},
   month=may, pages={2025–2064} }

@book{klauder1985coherent,
  title={Coherent states: applications in physics and mathematical physics},
  author={Klauder, John R and Skagerstam, Bo-Sture},
  year={1985},
  publisher={World scientific},
   doi={10.1142/0096}
}

@article{hartle1983wave,
  title = {Wave function of the Universe},
  author = {Hartle, J. B. and Hawking, S. W.},
  journal = {Phys. Rev. D},
  volume = {28},
  issue = {12},
  pages = {2960--2975},
  numpages = {0},
  year = {1983},
  month = {Dec},
  publisher = {American Physical Society},
  doi = {10.1103/PhysRevD.28.2960},
  url = {https://link.aps.org/doi/10.1103/PhysRevD.28.2960}
}

@book{Kiefer:2004xyv,
    author = "Kiefer, Claus",
    title = "{Quantum gravity}",
    isbn = "978-0-19-958520-5",
    publisher = "Clarendon",
    address = "Oxford",
    volume = "124",
    year = "2004"
}

@article{zhang1990coherent,
  title = {Coherent states: Theory and some applications},
  author = {Zhang, Wei-Min and Feng, Da Hsuan and Gilmore, Robert},
  journal = {Rev. Mod. Phys.},
  volume = {62},
  issue = {4},
  pages = {867--927},
  numpages = {0},
  year = {1990},
  month = {Oct},
  publisher = {American Physical Society},
  doi = {10.1103/RevModPhys.62.867},
  url = {https://link.aps.org/doi/10.1103/RevModPhys.62.867}
}

@article{ashtekar_null_2018,
    title = {Null infinity, the {BMS} group and infrared issues},
    volume = {50},
    issn = {0001-7701, 1572-9532},
    url = {http://arxiv.org/abs/1808.07093},
    doi = {10.1007/s10714-018-2464-3},
    
    number = {11},
    journal = {Gen. Relativ. Gravit.},
    author = {A. Ashtekar and M. Campiglia and A. Laddha},
    month = nov,
    year = {2018},
    note = {arXiv:1808.07093 [gr-qc]},
    pages = {140},
}

@misc{ashtekar2015geometryphysicsnullinfinity,
      title={Geometry and Physics of Null Infinity}, 
      author={Abhay Ashtekar},
      year={2015},
      eprint={1409.1800},
      archivePrefix={arXiv},
      primaryClass={gr-qc}}

@article{carney_dressed_2018,
    title = {Dressed infrared quantum information},
    volume = {97},
    issn = {2470-0010, 2470-0029},
    url = {https://link.aps.org/doi/10.1103/PhysRevD.97.025007},
    doi = {10.1103/PhysRevD.97.025007},
    
    number = {2},
    urldate = {2025-05-08},
    journal = {Phys. Rev. D},
    author = {D. Carney and L. Chaurette and D. Neuenfeld and G. W. Semenoff},
    month = jan,
    year = {2018},
    pages = {025007},
}

@article{Brown_2013,
   title={Detectors for probing relativistic quantum physics beyond perturbation theory},
   volume={87},
   ISSN={1550-2368},
   DOI={10.1103/physrevd.87.084062},
   pages={084062},
   journal={Phys. Rev. D},
   publisher={American Physical Society (APS)},
   author={Brown, Eric G. and Martín-Martínez, Eduardo and Menicucci, Nicolas C. and Mann, Robert B.},
   year={2013},
   month=apr }

@misc{bradler2018unitaryevolutionpairunruhdewitt,
      title={Unitary evolution of a pair of {Unruh}--{DeWitt} detectors calculated efficiently to an arbitrary perturbative order}, 
      author={Kamil Bradler},
      year={2018},
      eprint={1608.08274},
      archivePrefix={arXiv},
      primaryClass={hep-th}}

@article{Cong_2020,
  author = {Cong, Wan and Qian, Chen and Good, Michael R.R. and Mann, Robert B.},
  title = {Effects of horizons on entanglement harvesting},
  journal = {J. High Energy Phys.},
  volume = {10},
  number = {2020},
  pages = {067},
  year = {2020},
  doi = {10.1007/jhep10(2020)067},
}

@article{Ng_2018,
   title = {New techniques for entanglement harvesting in flat and curved spacetimes},
  author = {Ng, Keith K. and Mann, Robert B. and Mart\'{\i}n-Mart\'{\i}nez, Eduardo},
  journal = {Phys. Rev. D},
  volume = {97},
  issue = {12},
  pages = {125011},
  numpages = {8},
  year = {2018},
  month = {Jun},
  publisher = {American Physical Society},
  doi = {10.1103/PhysRevD.97.125011},
  url = {https://link.aps.org/doi/10.1103/PhysRevD.97.125011}
}

@article{jennings2010response,
  title={On the response of a particle detector in {A}nti-de {S}itter spacetime},
  author={Jennings, David},
  journal={Class. Quantum Grav.},
  volume={27},
  number={20},
  pages={205005},
  year={2010},
  publisher={IOP Publishing},
  doi={https://doi.org/10.1088/0264-9381/27/20/205005}
}

@article{vsoda2022acceleration,
  title={Acceleration-induced effects in stimulated light-matter interactions},
  author={{\v{S}}oda, Barbara and Sudhir, Vivishek and Kempf, Achim},
  journal={Phys. Rev. Lett.},
  volume={128},
  number={16},
  pages={163603},
  year={2022},
  publisher={APS},
  doi={https://doi.org/10.1103/PhysRevLett.128.163603}
}

@article{Bekenstein_1995,
   title={Spectroscopy of the quantum black hole},
   volume={360},
   ISSN={0370-2693},
   url={http://dx.doi.org/10.1016/0370-2693(95)01148-J},
   DOI={10.1016/0370-2693(95)01148-j},
   number={1–2},
   journal={Phys. Lett. B},
   publisher={Elsevier BV},
   author={Bekenstein, Jacob D. and Mukhanov, V.F.},
   year={1995},
   month=oct, pages={7–12} }

@inproceedings{bekenstein1998quantumblackholesatoms,
  author    = {J. D. Bekenstein},
  title     = {Quantum black holes as atoms},
  booktitle = {Proceedings of the 8th Marcel Grossmann Meeting on Recent Developments in Theoretical and Experimental General Relativity, Gravitation and Relativistic Field Theories (MG 8)},
  publisher = {World Scientific},
  address   = {Singapore},
  year      = {1997},
  pages     = {92--111},
  eprint    = {gr-qc/9710076},
  archivePrefix = {arXiv},
  primaryClass  = {gr-qc}
}

@article{hod1998bohr,
  title = {Bohr's Correspondence Principle and the Area Spectrum of Quantum Black Holes},
  author = {Hod, Shahar},
  journal = {Phys. Rev. Lett.},
  volume = {81},
  issue = {20},
  pages = {4293--4296},
  numpages = {0},
  year = {1998},
  month = {Nov},
  publisher = {American Physical Society},
  doi = {10.1103/PhysRevLett.81.4293},
  url = {https://link.aps.org/doi/10.1103/PhysRevLett.81.4293}
}

@incollection{bekenstein2020quantum,
  title={The quantum mass spectrum of the {Kerr} black hole},
  author={Bekenstein, Jacob D},
  booktitle={JACOB BEKENSTEIN: The Conservative Revolutionary},
  pages={331--334},
  year={2020},
  publisher={World Scientific},
  doi = {https://doi.org/10.1007/BF02762768}
}

@article{deser1997accelerated,
  title={Accelerated detectors and temperature in (anti-) de {S}itter spaces},
  author={Deser, Stanley and Levin, Orit},
  journal={Class. Quantum Grav.},
  volume={14},
  number={9},
  pages={L163},
  year={1997},
  publisher={IOP Publishing},
  doi={https://doi.org/10.1088/0264-9381/14/9/003}
}

@article{deser1998equivalence,
  title={Equivalence of {H}awking and {U}nruh temperatures and entropies through flat space embeddings},
  author={Deser, Stanley and Levin, Orit},
  journal={Class. Quantum Grav.},
  volume={15},
  number={12},
  pages={L85},
  year={1998},
  publisher={IOP Publishing},
  doi={https://doi.org/10.1088/0264-9381/15/12/002}
}

@article{deser1999mapping,
  title={Mapping {H}awking into {U}nruh thermal properties},
  author={Deser, Stanley and Levin, Orit},
  journal={Phys. Rev.  D},
  volume={59},
  number={6},
  pages={064004},
  year={1999},
  publisher={APS},
  doi={https://doi.org/10.1103/PhysRevD.59.064004}
}

@article{Paston_2014,
  author = {Paston, S. A.},
  title = {When does the {H}awking into {U}nruh mapping for global embeddings work?},
  journal = {J. High Energy Phys.},
  volume = {06},
  number = {2014},
  pages = {122},
  year = {2014},
  doi = {10.1007/jhep06(2014)122},
}

@article{sheykin2019global,
  title={Global embeddings of {BTZ} and {S}chwarzschild--{A}d{S} type black holes in a flat space},
  author={Sheykin, Anton and Solovyev, Dmitry and Paston, Sergey},
  journal={Symmetry},
  volume={11},
  number={7},
  pages={841},
  year={2019},
  publisher={MDPI},
  doi={https://doi.org/10.3390/sym11070841}
}

@article{sheykin2021global,
  title={Global embedding of {BTZ} spacetime using generalized method of symmetric embeddings construction},
    author = {Sheykin, A. A. and Markov, M. V. and Paston, S. A.},
    journal = {J. Math. Phys.},
    volume = {62},
    number = {10},
    pages = {102502},
    year = {2021},
    month = {10},
    issn = {0022-2488},
    doi = {10.1063/5.0062060},
    url = {https://doi.org/10.1063/5.0062060}
}

@article{pan2024gravity,
  title={Gravity-induced transparency},
  author={Pan, Yongjie and Zhang, Baocheng},
  journal={Phys. Rev.  D},
  volume={109},
  number={12},
  pages={125018},
  year={2024},
  publisher={APS}, 
  doi={https://doi.org/10.1103/PhysRevD.109.125018}
}

@article{russo2008accelerating,
  title={Accelerating branes and brane temperature},
  author={Russo, JG and Townsend, PK},
  journal={Class. Quantum Grav.},
  volume={25},
  number={17},
  pages={175017},
  year={2008},
  publisher={IOP Publishing}, 
  doi={https://doi.org/10.1088/0264-9381/25/17/175017}
}

@article{Henderson_2022,
    author = {Henderson, Laura J. and Ding, Su Yu and Mann, Robert B.},
    title = {Entanglement harvesting with a twist},
    journal = {AVS Quantum Science},
    volume = {4},
    number = {1},
    pages = {014402},
    year = {2022},
    month = {02},
    issn = {2639-0213},
    doi = {10.1116/5.0078314},
    url = {https://doi.org/10.1116/5.0078314}
}

@article{Gallock_Yoshimura_2021,
   title = {Harvesting entanglement with detectors freely falling into a black hole},
  author = {Gallock-Yoshimura, Kensuke and Tjoa, Erickson and Mann, Robert B.},
  journal = {Phys. Rev. D},
  volume = {104},
  issue = {2},
  pages = {025001},
  numpages = {19},
  year = {2021},
  month = {Jul},
  publisher = {American Physical Society},
  doi = {10.1103/PhysRevD.104.025001},
  url = {https://link.aps.org/doi/10.1103/PhysRevD.104.025001}
}

@article{Mendez_Avalos_2022,
   title={Entanglement harvesting of three {U}nruh-{D}e{W}itt detectors},
   volume={54},
    pages = {87},
   ISSN={1572-9532},
   url={http://dx.doi.org/10.1007/s10714-022-02956-x},
   DOI={10.1007/s10714-022-02956-x},
   number={8},
   journal={Gen. Relativ. Gravit.},
   publisher={Springer Science and Business Media LLC},
   author={Mendez-Avalos, Diana and Henderson, Laura J. and Gallock-Yoshimura, Kensuke and Mann, Robert B.},
   year={2022},
   month=aug }

@article{Smith_2014,
   title={Looking inside a black hole},
   volume={31},
   ISSN={1361-6382},
   url={http://dx.doi.org/10.1088/0264-9381/31/8/082001},
   DOI={10.1088/0264-9381/31/8/082001},
   number={8},
   journal={Class. Quantum Grav.},
   publisher={IOP Publishing},
   author={Smith, Alexander R H and Mann, Robert B},
   year={2014},
   month=apr, pages={082001} }

@article{Tjoa_2020,
  author = {Tjoa, Erickson and Mann, Robert B.},
  title = {Harvesting correlations in {S}chwarzschild and collapsing shell spacetimes},
  journal = {J. High Energy Phys.},
  volume = {08},
  number = {2020},
  pages = {155},
  year = {2020},
  doi = {10.1007/jhep08(2020)155},
}

@article{Hodgkinson_2012,
   title={Static, stationary, and inertial {U}nruh-{D}e{W}itt detectors on the {BTZ} black hole},
  author = {Hodgkinson, Lee and Louko, Jorma},
  journal = {Phys. Rev. D},
  volume = {86},
  issue = {6},
  pages = {064031},
  numpages = {16},
  year = {2012},
  month = {Sep},
  publisher = {American Physical Society},
  doi = {10.1103/PhysRevD.86.064031},
  url = {https://link.aps.org/doi/10.1103/PhysRevD.86.064031}
}

@misc{robbins2020entanglementamplificationrotatingblack,
      title={Entanglement Amplification from Rotating Black Holes}, 
      author={Matthew P. G. Robbins and Laura J. Henderson and Robert B. Mann},
      year={2020},
      eprint={2010.14517},
      archivePrefix={arXiv},
      primaryClass={hep-th}
}

@misc{niermann2024particle,
  title={Particle detectors in superposition in de {S}itter spacetime},
  author={Niermann, Laura and Barbado, Luis C},
  eprint={2403.02087},
archivePrefix={arXiv},
      primaryClass={gr-qc},
      year={2024}
}

@article{juarez2022quantum,
    author = {Juárez-Aubry, Benito A. and Louko, Jorma},
    title = {Quantum kicks near a {C}auchy horizon},
    journal = {AVS Quantum Science},
    volume = {4},
    number = {1},
    pages = {013201},
    year = {2022},
    month = {02},
    issn = {2639-0213},
    doi = {10.1116/5.0073373},
    url = {https://doi.org/10.1116/5.0073373}
}

@article{wang2024singular,
  title={Singular excitement beyond the horizon of a rotating black hole},
  author={Wang, Sijia and Preciado-Rivas, Mar{\'\i}a R and Spadafora, Massimiliano and Mann, Robert B},
  journal={Phys. Rev.  D},
  volume={110},
  number={6},
  pages={065013},
  year={2024},
  publisher={APS},
  doi={https://doi.org/10.1103/PhysRevD.110.065013}
}

@article{paczos2023hawking,
  title={Hawking radiation for detectors in superposition of locations outside a black hole},
  author={Paczos, Jerzy and Barbado, Luis C},
  journal={Phys. Rev.  D},
  volume={108},
  number={12},
  pages={125015},
  year={2023},
  publisher={APS}, 
  doi={https://doi.org/10.1103/PhysRevD.108.125015}
}

@article{Foo_2020,
   title={{U}nruh-{D}e{W}itt detectors in quantum superpositions of trajectories},
  author = {Foo, Joshua and Onoe, Sho and Zych, Magdalena},
  journal = {Phys. Rev. D},
  volume = {102},
  issue = {8},
  pages = {085013},
  numpages = {15},
  year = {2020},
  month = {Oct},
  publisher = {American Physical Society},
  doi = {10.1103/PhysRevD.102.085013},
  url = {https://link.aps.org/doi/10.1103/PhysRevD.102.085013}
}

@article{wang2025harvestinginformationhorizon,
      title={Harvesting Information Across the Horizon}, 
      author={S. Wang and M. R. Preciado-Rivas and R. B. Mann},
      year={2025},
      number={2025},
      journal={J. High Energy Phys.},
      volume = {07},
      pages={254},
      doi={10.1007/JHEP07(2025)254}}

@article{Membrere_2023,
   title={Tripartite Entanglement Extraction from the Black Hole Vacuum},
   volume={6},
   ISSN={2511-9044},
   url={http://dx.doi.org/10.1002/qute.202300125},
   DOI={10.1002/qute.202300125},
   pages={2300125},
   journal={Adv. Quantum Technol.},
   publisher={Wiley},
   author={Membrere, I. J. and Gallock‐Yoshimura, K. and Henderson, L. J. and Mann, R. B.},
   year={2023},
   month=jul }

@article{Barman_2023,
  author = {Barman, S. and Chakraborty, I. and Mukherjee, S.},
  title = {Entanglement harvesting for different gravitational wave burst profiles with and without memory},
  journal = {J. High Energy Phys.},
  volume = {09},
  number = {2023},
  pages = {180},
  year = {2023},
  doi = {10.1007/jhep09(2023)180},
}

@article{dubey2025harvesting,
  title = {Harvesting fermionic field entanglement in {S}chwarzschild spacetime},
  author = {Dubey, N. K. and Kolekar, S.},
  journal = {Phys. Rev. D},
  volume = {112},
  issue = {2},
  pages = {025019},
  numpages = {35},
  year = {2025},
  month = {Jul},
  publisher = {American Physical Society},
  doi = {10.1103/1mwx-7jmf},
  url = {https://link.aps.org/doi/10.1103/1mwx-7jmf}
}

@article{foo2021entanglement,
  title = {Entanglement amplification between superposed detectors in flat and curved spacetimes},
  author = {Foo, Joshua and Mann, Robert B. and Zych, Magdalena},
  journal = {Phys. Rev. D},
  volume = {103},
  issue = {6},
  pages = {065013},
  numpages = {19},
  year = {2021},
  month = {Mar},
  publisher = {American Physical Society},
  doi = {10.1103/PhysRevD.103.065013},
  url = {https://link.aps.org/doi/10.1103/PhysRevD.103.065013}
}

@article{Foo_2022,
   title={Schrödinger’s black hole cat},
   volume={31},
   ISSN={1793-6594},
   DOI={10.1142/s0218271822420160},
   number={14},
    pages={2242016},
   journal={Int. J. Modern Phys. D},
   publisher={World Scientific Pub Co Pte Ltd},
   author={Foo, Joshua and Mann, Robert B. and Zych, Magdalena},
   year={2022},
   month=oct }

@article{suryaatmadja2023signaturesrotatingblackholes,
  title = {Signatures of rotating black holes in quantum superposition},
  author = {Suryaatmadja, Cendikiawan and Arabaci, Cemile Senem and Robbins, Matthew P. G. and Foo, Joshua and Zych, Magdalena and Mann, Robert B.},
  journal = {Phys. Rev. D},
  volume = {110},
  issue = {6},
  pages = {066018},
  numpages = {16},
  year = {2024},
  month = {Sep},
  publisher = {American Physical Society},
  doi = {10.1103/PhysRevD.110.066018},
  url = {https://link.aps.org/doi/10.1103/PhysRevD.110.066018}
}

@article{bose2022mechanism,
  title={Mechanism for the quantum natured gravitons to entangle masses},
  author={Bose, Sougato and Mazumdar, Anupam and Schut, Martine and Toro{\v{s}}, Marko},
  journal={Phys. Rev.  D},
  volume={105},
  number={10},
  pages={106028},
  year={2022},
  publisher={APS}, 
  doi={https://doi.org/10.1103/PhysRevD.105.106028}
}

@article{tino2021testing,
  title={Testing gravity with cold atom interferometry: results and prospects},
  author={Tino, Guglielmo M},
  journal={Quantum Sci. Technol.},
  volume={6},
  number={2},
  pages={024014},
  year={2021},
  publisher={IOP Publishing},
  doi={https://doi.org/10.1088/2058-9565/abd83e}
}

@article{marshman2020locality,
  title={Locality and entanglement in table-top testing of the quantum nature of linearized gravity},
  author={Marshman, Ryan J and Mazumdar, Anupam and Bose, Sougato},
  journal={Phys. Rev.  A},
  volume={101},
  number={5},
  pages={052110},
  year={2020},
  publisher={APS},
  doi={https://doi.org/10.1103/PhysRevA.101.052110}
}

@article{tobar2024detecting,
  title={Detecting single gravitons with quantum sensing},
  author={Tobar, Germain and Manikandan, Sreenath K and Beitel, Thomas and Pikovski, Igor},
  journal={Nat. Commun.},
  volume={15},
  number={1},
  pages={7229},
  year={2024},
  publisher={Nature Publishing Group UK London},
  doi={https://doi.org/10.1038/s41467-024-51420-8}
}

@article{moore2021searching,
  title={Searching for new physics using optically levitated sensors},
  author={Moore, David C and Geraci, Andrew A},
  journal={Quantum Sci. Technol.},
  volume={6},
  number={1},
  pages={014008},
  year={2021},
  publisher={IOP Publishing},
  doi={https://doi.org/10.1088/2058-9565/abcf8a}
}

@article{margalit2021realization,
  title={Realization of a complete {S}tern-{G}erlach interferometer: {T}oward a test of quantum gravity},
  author={Margalit, Yair and Dobkowski, Or and Zhou, Zhifan and Amit, Omer and Japha, Yonathan and Moukouri, Samuel and Rohrlich, Daniel and Mazumdar, Anupam and Bose, Sougato and Henkel, Carsten and others},
  journal={Sci. Adv.},
  volume={7},
  number={22},
  pages={eabg2879},
  year={2021},
  publisher={American Association for the Advancement of Science},
  doi={https://doi.org/10.1126/sciadv.abg2879}
}

@book{carlip2003quantum,
  title={Quantum gravity in 2+ 1 dimensions},
  author={Carlip, Steven},
  volume={50},
  year={2003},
  publisher={Cambridge University Press},
collection={Cambridge Monographs on Mathematical Physics},
doi = {10.1017/CBO9780511564192}
}

@article{zych2018quantum,
  title={Quantum formulation of the {E}instein equivalence principle},
  author={Zych, Magdalena and Brukner, {\v{C}}aslav},
  journal={Nat. Phys.},
  volume={14},
  number={10},
  pages={1027--1031},
  year={2018},
  publisher={Nature Publishing Group UK London},
  doi ={https://doi.org/10.1038/s41567-018-0197-6}
}

@book{peskin2018introduction,
  title={An introduction to quantum field theory},
  author={Peskin, Michael E and Schroeder, Daniel V.},
  year={1995},
  publisher={CRC press}, 
  doi={https://doi.org/10.1201/9780429503559}
}

@misc{tong2009university,
  title={Quantum {F}ield {T}heory. {U}niversity of {C}ambridge {P}art {III} {M}athematical {T}ripos},
  author={Tong, David},
  year={2006},
  url ={https://www.damtp.cam.ac.uk/user/tong/qft.html}
}

@book{Weinberg_1995, place={Cambridge}, title={The Quantum Theory of Fields}, publisher={Cambridge University Press}, doi ={https://doi.org/10.1017/CBO9781139644167}, author={Weinberg, Steven}, year={1995}}

@article{Carlip_2005,
   title={Conformal field theory, (2 + 1)-dimensional gravity and the {BTZ} black hole},
   volume={22},
   ISSN={1361-6382},
   url={http://dx.doi.org/10.1088/0264-9381/22/12/R01},
   DOI={10.1088/0264-9381/22/12/r01},
   number={12},
   journal={Class. Quantum Grav.},
   publisher={IOP Publishing},
   author={Carlip, S},
   year={2005},
   month=jun, pages={R85–R123} }

@article{Shiraishi_1994,
   title={Quantum fluctuation of stress tensor and black holes in three dimensions},
   volume={49},
   ISSN={0556-2821},
   url={http://dx.doi.org/10.1103/PhysRevD.49.5286},
   DOI={10.1103/physrevd.49.5286},
   number={10},
   journal={Phys. Rev.  D},
   publisher={American Physical Society (APS)},
   author={Shiraishi, Kiyoshi and Maki, Takuya},
   year={1994},
   month=may, pages={5286–5294} }

@inproceedings{smith1990gravitational,
  title={Gravitational effects of an infinite straight cosmic string on classical and quantum fields: self-forces and vacuum fluctuations},
  author={Smith, AG},
  booktitle={Proceedings of the Symposium of the Formation and Evolution of Cosmic Strings},
  pages={263},
  year={1990},
  organization={Cambridge University Press}
}

@article{avis1978quantum,
  title={Quantum field theory in anti-de {S}itter space-time},
  author={Avis, SJ and Isham, CJ and Storey, D},
  journal={Phys. Rev.  D},
  volume={18},
  number={10},
  pages={3565},
  year={1978},
  publisher={APS}, 
  doi={https://doi.org/10.1103/PhysRevD.18.3565}
}

@article{Santiago_2019,
   title={Tolman temperature gradients in a gravitational field},
   volume={40},
   ISSN={1361-6404},
   url={http://dx.doi.org/10.1088/1361-6404/aaff1c},
   DOI={10.1088/1361-6404/aaff1c},
   number={2},
   journal={Eur. J. Phys.},
   publisher={IOP Publishing},
   author={Santiago, Jessica and Visser, Matt},
   year={2019},
   month=feb, pages={025604} }

@article{chen2024quantum,
  title = {Quantum Effects in Gravity Beyond the Newton Potential from a Delocalized Quantum Source},
  author = {Chen, Lin-Qing and Giacomini, Flaminia},
  journal = {Phys. Rev. X},
  volume = {15},
  issue = {3},
  pages = {031063},
  numpages = {19},
  year = {2025},
  month = {Sep},
  publisher = {American Physical Society},
  doi = {10.1103/hl1c-t8z9},
  url = {https://link.aps.org/doi/10.1103/hl1c-t8z9}
}

@article{chen2023quantum,
  title={Quantum states of fields for quantum split sources},
  author={Chen, Lin-Qing and Giacomini, Flaminia and Rovelli, Carlo},
  journal={Quantum},
  volume={7},
  pages={958},
  year={2023},
  publisher={Verein zur F{\"o}rderung des Open Access Publizierens in den Quantenwissenschaften},
  doi={https://doi.org/10.22331/q-2023-03-20-958}
}

@article{Carlip_2014,
   title={Black hole thermodynamics},
   volume={23},
   ISSN={1793-6594},
   url={http://dx.doi.org/10.1142/S0218271814300237},
   DOI={10.1142/s0218271814300237},
   number={11},
   journal={Int. J. Modern Phys. D},
   publisher={World Scientific Pub Co Pte Lt},
   author={Carlip, S.},
   year={2014},
   month=oct, pages={1430023} }

@article{emparan1998quantization,
  title={Quantization of ${A}d{S}_3$ Black Holes in External Fields: {S}emiclassical Results from Pure Gravity},
  author={Emparan, Roberto and Sachs, Ivo},
  journal={Phys. Rev. Lett.},
  volume={81},
  number={12},
  pages={2408},
  year={1998},
  publisher={APS},
  doi={https://doi.org/10.1103/PhysRevLett.81.2408}
}

@misc{walleghem2026wignersfriendsblackhole,
      title={Wigner's friend's black hole adventure: an argument for complementarity?}, 
      author={Laurens Walleghem},
      year={2026},
      eprint={2507.05369},
      archivePrefix={arXiv},
      primaryClass={gr-qc}}

@article{spadafora2024deepknottedblackhole,
   title={Deep in the knotted black hole},
   volume={111},
   ISSN={2470-0029},
   url={http://dx.doi.org/10.1103/PhysRevD.111.065013},
   DOI={10.1103/physrevd.111.065013},
   pages={065013},
   journal={Phys. Rev. D},
   publisher={American Physical Society (APS)},
   author={Spadafora, Massimiliano and Naeem, Manar and Preciado-Rivas, María R. and Mann, Robert B. and Louko, Jorma},
   year={2025},
   month=mar }

@article{chakraborty2024entanglementharvestingquantumsuperposed,
   title={Entanglement harvesting in quantum superposed spacetime},
   volume={111},
   ISSN={2470-0029},
   DOI={10.1103/physrevd.111.104052},
   pages={104052},
   journal={Phys. Rev. D},
   publisher={American Physical Society (APS)},
   author={Chakraborty, Anwesha and Hackl, Lucas and Zych, Magdalena},
   year={2025},
   month=may }

@article{Fewster_2013,
   title={The necessity of the {H}adamard condition},
   volume={30},
   ISSN={1361-6382},
   url={http://dx.doi.org/10.1088/0264-9381/30/23/235027},
   DOI={10.1088/0264-9381/30/23/235027},
   number={23},
   journal={Class. Quantum Grav.},
   publisher={IOP Publishing},
   author={Fewster, Christopher J and Verch, Rainer},
   year={2013},
   month=nov, pages={235027} }

@article{deppe2024echoesbeyonddetectinggravitational,
   title={Echoes from beyond: Detecting gravitational-wave quantum imprints with {LISA}},
   volume={111},
   ISSN={2470-0029},
   url={http://dx.doi.org/10.1103/k7jh-rhgw},
   DOI={10.1103/k7jh-rhgw},
   pages={124035},
   journal={Phys. Rev. D},
   publisher={American Physical Society (APS)},
   author={Deppe, Nils and Heisenberg, Lavinia and Inchauspé, Henri and Kidder, Lawrence E. and Maibach, David and Ma, Sizheng and Moxon, Jordan and Nelli, Kyle C. and Throwe, William and Vu, Nils L.},
   year={2025},
   month=jun }

@article{Mart_n_Mart_nez_2020,
   title={General relativistic quantum optics: Finite-size particle detector models in curved spacetimes},
  author = {Mart\'{\i}n-Mart\'{\i}nez, Eduardo and Perche, T. Rick and de S. L. Torres, Bruno},
  journal = {Phys. Rev. D},
  volume = {101},
  issue = {4},
  pages = {045017},
  numpages = {10},
  year = {2020},
  month = {Feb},
  publisher = {American Physical Society},
  doi = {10.1103/PhysRevD.101.045017},
  url = {https://link.aps.org/doi/10.1103/PhysRevD.101.045017}
}

@book{birrell1984quantum, place={Cambridge}, series={Cambridge Monographs on Mathematical Physics}, title={Quantum Fields in Curved Space}, publisher={Cambridge University Press}, author={Birrell, Nicholas David and Davies, Paul Charles William}, year={1982}, collection={Cambridge Monographs on Mathematical Physics}, doi = {https://doi.org/10.1017/CBO9780511622632}}

@article{hummer2016renorm,
   title={Renormalized {U}nruh-{D}e{W}itt particle detector models for boson and fermion fields},
   volume={93},
   url={http://dx.doi.org/10.1103/PhysRevD.93.024019},
   DOI={10.1103/physrevd.93.024019},
   number={2},
   pages={024019},
   journal={Phys. Rev.  D},
   publisher={American Physical Society (APS)},
   author={Hümmer, Daniel and Martín-Martínez, Eduardo and Kempf, Achim},
   year={2016},
   month=jan }

@article{Guedes_2024,
   title={A Study of Spin 1 {U}nruh–{D}e{W}itt Detectors},
   volume={10},
   ISSN={2218-1997},
   url={http://dx.doi.org/10.3390/universe10080307},
   DOI={10.3390/universe10080307},
   number={8},
   journal={Universe},
   publisher={MDPI AG},
   author={Guedes, F. M. and Guimaraes, M. S. and Roditi, I. and Sorella, S. P.},
   year={2024},
   month=jul, pages={307} }

@article{ali2021unruhdewittdetectorresponsescomplex,
      title={Unruh-{D}e{W}itt detector responses for complex scalar fields in de {S}itter spacetime}, 
      author={Md. Sabir Ali and Sourav Bhattacharya and Kinjalk Lochan},
      year={2021},
      number={2021},
      journal={J. High Energy Phys.},
      volume={03}, 
      pages={220},
      doi={10.1007/JHEP03(2021)220}}

@article{unruh1976notes,
  title = {Notes on black-hole evaporation},
  author = {Unruh, W. G.},
  journal = {Phys. Rev. D},
  volume = {14},
  issue = {4},
  pages = {870--892},
  numpages = {0},
  year = {1976},
  month = {Aug},
  publisher = {American Physical Society},
  doi = {10.1103/PhysRevD.14.870},
  url = {https://link.aps.org/doi/10.1103/PhysRevD.14.870}
}

@incollection{dewitt1979,
title = {},
  booktitle={General relativity: an {E}instein centenary survey},
 author = {B. Seligman DeWitt},
  editor={Hawking, Stephen and Israel, Werner},
  year={1979},
pages  = {680},
publisher ={Cambridge University Press}
}

@article{lima2023unruh,
   title={Unruh phenomena and thermalization for qudit detectors},
   volume={108},
   ISSN={2470-0029},
   url={http://dx.doi.org/10.1103/PhysRevD.108.105020},
   DOI={10.1103/physrevd.108.105020},
   number={10},
    pages={105020},
   journal={Phys. Rev.  D},
   publisher={American Physical Society (APS)},
   author={Lima, Caroline and Patterson, Everett and Tjoa, Erickson and Mann, Robert B.},
   year={2023},
   month=nov }

@article{brenna2016anti,
title = {Anti-{U}nruh phenomena},
journal = {Phys. Lett. B},
volume = {757},
pages = {307-311},
year = {2016},
issn = {0370-2693},
doi = {10.1016/j.physletb.2016.04.002},
url = {https://www.sciencedirect.com/science/article/pii/S0370269316300727},
author = {W.G. Brenna and R. B. Mann and E. Martín-Martínez}
}

@article{garay2016thermalization,
  title = {Thermalization of particle detectors: The {U}nruh effect and its reverse},
  author = {Garay, Luis J. and Mart\'{\i}n-Mart\'{\i}nez, Eduardo and de Ram\'on, Jos\'e},
  journal = {Phys. Rev. D},
  volume = {94},
  issue = {10},
  pages = {104048},
  numpages = {11},
  year = {2016},
  month = {Nov},
  publisher = {American Physical Society},
  doi = {10.1103/PhysRevD.94.104048},
  url = {https://link.aps.org/doi/10.1103/PhysRevD.94.104048}
}

@article{martin2013wavepacket,
  title={Wavepacket detection with the {U}nruh-{D}e{W}itt model},
  author={Mart{\'\i}n-Mart{\'\i}nez, Eduardo and Montero, Miguel and del Rey, Marco},
  journal={Phys. Rev. D},
  volume={87},
  number={6},
  pages={064038},
  year={2013},
  publisher={APS},
  doi={10.1103/PhysRevD.93.024019}
}

@article{raju2022failure,
  title={Failure of the split property in gravity and the information paradox},
  author={Raju, Suvrat},
  journal={Class. Quantum Grav.},
  volume={39},
  number={6},
  pages={064002},
  year={2022},
  publisher={IOP Publishing}, 
doi={10.1088/1361-6382/ac482b}
}

@article{Lifschytz_1994,
   title={Scalar field quantization on the (2+1)-dimensional black hole background},
   volume={49},
   ISSN={0556-2821},
   url={http://dx.doi.org/10.1103/PhysRevD.49.1929},
   DOI={10.1103/physrevd.49.1929},
   number={4},
   journal={Phys. Rev. D},
   publisher={American Physical Society (APS)},
   author={Lifschytz, Gilad and Ortiz, Miguel},
   year={1994},
   month=feb, pages={1929–1943} }

@misc{hodgkinson2013particledetectorscurvedspacetime,
      title={Particle detectors in curved spacetime quantum field theory}, 
      author={Lee Hodgkinson},
      year={2013},
      eprint={1309.7281},
      archivePrefix={arXiv},
      primaryClass={gr-qc}}

@article{braunstein2013better,
  title={Better late than never: information retrieval from black holes},
  author={Braunstein, Samuel L and Pirandola, Stefano and {\.Z}yczkowski, Karol},
  journal={Phys. Rev. Lett.},
  volume={110},
  number={10},
  pages={101301},
  year={2013},
  publisher={APS},
  doi={https://doi.org/10.1103/PhysRevLett.110.101301}
}

@article{braunstein2007quantum,
  title={Quantum information cannot be completely hidden in correlations: implications for the black-hole information paradox},
  author={Braunstein, Samuel L and Pati, Arun K},
  journal={Phys. Rev. Lett.},
  volume={98},
  number={8},
  pages={080502},
  year={2007},
  publisher={APS},
  doi={https://doi.org/10.1103/PhysRevLett.98.080502}
}

@article{hawking1975particle,
  title={Particle creation by black holes},
  author={Hawking, Stephen W},
  journal={Commun. Math. Phys.},
  volume={43},
  number={3},
  pages={199--220},
  year={1975},
  publisher={Springer},
 doi = {https://doi.org/10.1007/BF02345020}
}

@book{wald1994quantum,
  title={Quantum field theory in curved spacetime and black hole thermodynamics},
  author={Wald, Robert M},
  year={1994},
  publisher={University of Chicago press}
}

@article{calmet2022brief,
  title={A brief history of {H}awking’s information paradox},
  author={Calmet, Xavier and Hsu, Stephen DH},
  journal={EuroPhys. Lett.},
  year={2022},
  publisher={IOP Publishing}, 
  doi={10.1209/0295-5075/ac81e8},
  volume={139},
  pages={49001}
}

@article{braunstein2023analogue,
  title={Analogue simulations of quantum gravity with fluids},
  author={Braunstein, Samuel L and Faizal, Mir and Krauss, Lawrence M and Marino, Francesco and Shah, Naveed A},
  journal={Nat. Rev. Phys.},
  pages={612--622},
  volume = {5},
  year={2023},
  doi = {https://doi.org/10.1038/s42254-023-00630-y},
  publisher={Nature Publishing Group UK London}
}

@article{almheiri2019entropy,
  author = {Almheiri, Ahmed and Engelhardt, Netta and Marolf, Donald and Maxfield, Henry},
  title = {The entropy of bulk quantum fields and the entanglement wedge of an evaporating black hole},
  journal = {J. High Energy Phys.},
  volume = {12},
  number = {2019},
  pages = {063},
  year = {2019},
  doi = {10.1007/JHEP12(2019)063},
}

@article{wang2024entanglement,
  title={Entanglement transition and replica wormholes in the dissipative {Sachdev}--{Ye}--{Kitaev} model},
  author={Wang, Hanteng and Liu, Chang and Zhang, Pengfei and Garc{\'\i}a-Garc{\'\i}a, Antonio M},
  journal={Phys. Rev.  D},
  volume={109},
  number={4},
  pages={046005},
  year={2024},
  publisher={APS},
   doi={10.1103/PhysRevD.109.046005}
}

@article{de2024gravitational,
  title={Gravitational entropy is observer-dependent},
  author={De Vuyst, Julian and Eccles, Stefan and Hoehn, Philipp A and Kirklin, Josh},
  year={2025},
  number={2025},
  journal = {J. High Energy Phys.},
  pages = {146},
  volume = {07}, 
  doi={10.1007/JHEP07(2025)146}
}

@article{torres24particle,
  title = {Particle detector models from path integrals of localized quantum fields},
  author = {Torres, Bruno de S. L.},
  journal = {Phys. Rev. D},
  volume = {109},
  issue = {6},
  pages = {065004},
  numpages = {22},
  year = {2024},
  month = {Mar},
  publisher = {American Physical Society},
  doi = {10.1103/PhysRevD.109.065004},
  url = {https://link.aps.org/doi/10.1103/PhysRevD.109.065004}
}

@article{unruh1981experimental,
  title = {Experimental Black-Hole Evaporation?},
  author = {Unruh, W. G.},
  journal = {Phys. Rev. Lett.},
  volume = {46},
  issue = {21},
  pages = {1351--1353},
  numpages = {0},
  year = {1981},
  month = {May},
  publisher = {American Physical Society},
  doi = {10.1103/PhysRevLett.46.1351},
  url = {https://link.aps.org/doi/10.1103/PhysRevLett.46.1351}
}

@article{garay2000sonic,
  title = {Sonic Analog of Gravitational Black Holes in {B}ose--{E}instein Condensates},
  author = {Garay, L. J. and Anglin, J. R. and Cirac, J. I. and Zoller, P.},
  journal = {Phys. Rev. Lett.},
  volume = {85},
  issue = {22},
  pages = {4643--4647},
  numpages = {0},
  year = {2000},
  month = {Nov},
  publisher = {American Physical Society},
  doi = {10.1103/PhysRevLett.85.4643},
  url = {https://link.aps.org/doi/10.1103/PhysRevLett.85.4643}
}

@article{weinfurtner2011measurement,
  title = {Measurement of Stimulated {H}awking Emission in an Analogue System},
  author = {Weinfurtner, Silke and Tedford, Edmund W. and Penrice, Matthew C. J. and Unruh, William G. and Lawrence, Gregory A.},
  journal = {Phys. Rev. Lett.},
  volume = {106},
  issue = {2},
  pages = {021302},
  numpages = {4},
  year = {2011},
  month = {Jan},
  publisher = {American Physical Society},
  doi = {10.1103/PhysRevLett.106.021302},
  url = {https://link.aps.org/doi/10.1103/PhysRevLett.106.021302}
}

@article{barcelo2005analogue,
   title={Analogue Gravity},
   volume={8},
   ISSN={1433-8351},
   url={http://dx.doi.org/10.12942/lrr-2005-12},
   DOI={10.12942/lrr-2005-12},
   number={1},
    volume  = {8},
  pages = {12},
   journal={Living Rev. Relativity},
   publisher={Springer Science and Business Media LLC},
   author={Barceló, Carlos and Liberati, Stefano and Visser, Matt},
   year={2005},
   month=dec }

@article{Fewster_2016,
   title={Waiting for Unruh},
   volume={33},
   ISSN={1361-6382},
   url={http://dx.doi.org/10.1088/0264-9381/33/16/165003},
   DOI={10.1088/0264-9381/33/16/165003},
   number={16},
   journal={Class. Quantum Grav.},
   publisher={IOP Publishing},
   author={Fewster, Christopher J and Juárez-Aubry, Benito A and Louko, Jorma},
   year={2016},
   month=jul, pages={165003} }

@article{kabel2023quantum,
  title={Quantum reference frames at the boundary of spacetime},
  author={Kabel, Viktoria and Brukner, {\v{C}}aslav and Wieland, Wolfgang},
  journal={Phys. Rev.  D},
  volume={108},
  number={10},
  pages={106022},
  year={2023},
  publisher={APS},
  doi={10.1103/PhysRevD.108.106022}
}

@article{de_la_Hamette_2023,
   title={Quantum reference frames for an indefinite metric},
   volume={6},
pages = {231},
   ISSN={2399-3650},
   url={http://dx.doi.org/10.1038/s42005-023-01344-4},
   DOI={10.1038/s42005-023-01344-4},
   number={1},
   journal={Commun. Phys.},
   publisher={Springer Science and Business Media LLC},
   author={de la Hamette, Anne-Catherine and Kabel, Viktoria and Castro-Ruiz, Esteban and Brukner, {\v{C}}aslav},
   year={2023},
   month=aug }

@article{gale2023relativistic,
  title = {Relativistic {Unruh}--{DeWitt} detectors with quantized center of mass},
  author = {Gale, Evan P. G. and Zych, Magdalena},
  journal = {Phys. Rev. D},
  volume = {107},
  issue = {5},
  pages = {056023},
  numpages = {18},
  year = {2023},
  month = {Mar},
  publisher = {American Physical Society},
  doi = {10.1103/PhysRevD.107.056023},
  url = {https://link.aps.org/doi/10.1103/PhysRevD.107.056023}
}

@article{PhysRevResearch.3.043056,
  title = {Thermality, causality, and the quantum-controlled {Unruh}--{deWitt} detector},
  author = {Foo, Joshua and Onoe, Sho and Mann, Robert B. and Zych, Magdalena},
  journal = {Phys. Rev. Res.},
  volume = {3},
  issue = {4},
  pages = {043056},
  numpages = {9},
  year = {2021},
  month = {Oct},
  publisher = {American Physical Society},
  doi = {10.1103/PhysRevResearch.3.043056},
  url = {https://link.aps.org/doi/10.1103/PhysRevResearch.3.043056}
}

@misc{zych2018relativityquantumsuperpositions,
      title={Relativity of quantum superpositions}, 
      author={Magdalena Zych and Fabio Costa and Timothy C. Ralph},
      year={2018},
      eprint={1809.04999},
      archivePrefix={arXiv},
      primaryClass={quant-ph}}

@article{de_la_Hamette_2020,
   title={Quantum reference frames for general symmetry groups},
   volume={4},
   ISSN={2521-327X},
   url={http://dx.doi.org/10.22331/q-2020-11-30-367},
   DOI={10.22331/q-2020-11-30-367},
   journal={Quantum},
   publisher={Verein zur Forderung des Open Access Publizierens in den Quantenwissenschaften},
   author={de la Hamette, Anne-Catherine and Galley, Thomas D.},
   year={2020},
   month=nov, pages={367} }

@article{Vanrietvelde_2020,
   title={A change of perspective: switching quantum reference frames via a perspective-neutral framework},
   volume={4},
   ISSN={2521-327X},
   url={http://dx.doi.org/10.22331/q-2020-01-27-225},
   DOI={10.22331/q-2020-01-27-225},
   journal={Quantum},
   publisher={Verein zur Forderung des Open Access Publizierens in den Quantenwissenschaften},
   author={Vanrietvelde, Augustin and Hoehn, Philipp A. and Giacomini, Flaminia and Castro-Ruiz, Esteban},
   year={2020},
   month=jan, pages={225} }

@misc{delahamette2021perspectiveneutralapproachquantumframe,
      title={Perspective-neutral approach to quantum frame covariance for general symmetry groups}, 
      author={Anne-Catherine de la Hamette and Thomas D. Galley and Philipp A. Hoehn and Leon Loveridge and Markus P. Mueller},
      year={2021},
      eprint={2110.13824},
      archivePrefix={arXiv},
      primaryClass={quant-ph}
}

@article{Loveridge_2018,
   title={Symmetry, Reference Frames, and Relational Quantities in Quantum Mechanics},
   volume={48},
   ISSN={1572-9516},
   url={http://dx.doi.org/10.1007/s10701-018-0138-3},
   DOI={10.1007/s10701-018-0138-3},
   number={2},
   journal={Found. Phys.},
   publisher={Springer Science and Business Media LLC},
   author={Loveridge, Leon and Miyadera, Takayuki and Busch, Paul},
   year={2018},
   month=feb, pages={135–198} }

@article{carette2024operationalquantumreferenceframe,
   title={Operational Quantum Reference Frame Transformations},
   volume={9},
   ISSN={2521-327X},
   DOI={10.22331/q-2025-03-27-1680},
   journal={Quantum},
   publisher={Verein zur Forderung des Open Access Publizierens in den Quantenwissenschaften},
   author={Carette, Titouan and Glowacki, Jan and Loveridge, Leon},
   year={2025},
   month=mar, pages={1680} }

@misc{chen2026quantumreferencefieldstransformations,
      title={Quantum Reference Fields Transformations in Linearized Quantum Gravity}, 
      author={Lin-Qing Chen and Flaminia Giacomini},
      year={2026},
      eprint={2606.09344},
      archivePrefix={arXiv},
      primaryClass={gr-qc}}

@article{Bartlett_2006,
   title={Degradation of a quantum reference frame},
   volume={8},
   ISSN={1367-2630},
   url={http://dx.doi.org/10.1088/1367-2630/8/4/058},
   DOI={10.1088/1367-2630/8/4/058},
   number={4},
   journal={New J. Phys.},
   publisher={IOP Publishing},
   author={Bartlett, Stephen D and Rudolph, Terry and Spekkens, Robert W and Turner, Peter S},
   year={2006},
   month=Apr, pages={58} }

@article{Giacomini_2021,
   title={Spacetime Quantum Reference Frames and superpositions of proper times},
   volume={5},
   ISSN={2521-327X},
   url={http://dx.doi.org/10.22331/q-2021-07-22-508},
   DOI={10.22331/q-2021-07-22-508},
   journal={Quantum},
   publisher={Verein zur Forderung des Open Access Publizierens in den Quantenwissenschaften},
   author={Giacomini, Flaminia},
   year={2021},
   month=July, pages={508} }

@misc{fewster2025semilocalobservablesedgemodes,
      title={Semi-local observables, edge modes and quantum reference frames in quantum electromagnetism: an algebraic approach}, 
      author={Christopher J. Fewster and Daan W. Janssen and Kasia Rejzner},
      year={2025},
      eprint={2508.20939},
      archivePrefix={arXiv},
      primaryClass={math-ph}}

@article{castroruiz2023relativesubsystemsquantumreference,
      title={Relative subsystems and quantum reference frame transformations}, 
      author={Esteban Castro-Ruiz and Ognyan Oreshkov},
      year={2025},
      journal={Commun. Phys.},
      volume ={8}, 
pages={187},
doi={10.1038/s42005-025-02036-x}
}

@article{Bussola_2017,
     title = {Ground state for a massive scalar field in the BTZ spacetime with Robin boundary conditions},
  author = {Bussola, Francesco and Dappiaggi, Claudio and Ferreira, Hugo R. C. and Khavkine, Igor},
  journal = {Phys. Rev. D},
  volume = {96},
  issue = {10},
  pages = {105016},
  numpages = {18},
  year = {2017},
  month = {Nov},
  publisher = {American Physical Society},
  doi = {10.1103/PhysRevD.96.105016},
  url = {https://link.aps.org/doi/10.1103/PhysRevD.96.105016}
}

@article{Tambornino_2012,
   title={Relational Observables in Gravity: a Review},
   doi={10.3842/sigma.2012.017},
   year = {2012},
   volume = {8},
   pages = {017},
   journal={Symmetry, Integrability and Geometry: Methods and Applications},
   author={Tambornino, Johannes} }

@misc{langlois2005imprintsspacetimetopologyhawkingunruh,
      title={Imprints of Spacetime Topology in the {H}awking--{U}nruh Effect}, 
      author={Paul Langlois},
      year={2005},
      eprint={gr-qc/0510127},
      archivePrefix={arXiv},
      primaryClass={gr-qc}}

@article{hoehn2023matterrelativequantumhypersurfaces,
      title = {Matter relative to quantum hypersurfaces},
  author = {H\"ohn, Philipp A. and Russo, Andrea and Smith, Alexander R. H.},
  journal = {Phys. Rev. D},
  volume = {109},
  issue = {10},
  pages = {105011},
  numpages = {21},
  year = {2024},
  month = {May},
  publisher = {American Physical Society},
  doi = {10.1103/PhysRevD.109.105011},
  url = {https://link.aps.org/doi/10.1103/PhysRevD.109.105011}
}

@misc{thiemann2026quantumreferenceframesrelational,
      title={(Quantum) reference frames, relational observables, gauge reduction and physical interpretation}, 
      author={Thomas Thiemann},
      year={2026},
      eprint={2603.04072},
      archivePrefix={arXiv},
      primaryClass={quant-ph}}

@article{Doat_2026,
   title={What can we do in a symmetry-constrained perspective? {T}he importance of the total charge's status in quantum reference frame frameworks},
   volume={10},
   ISSN={2521-327X},
   url={http://dx.doi.org/10.22331/q-2026-06-08-2126},
   DOI={10.22331/q-2026-06-08-2126},
   journal={Quantum},
   publisher={Verein zur Forderung des Open Access Publizierens in den Quantenwissenschaften},
   author={Doat, Guilhem and Vanrietvelde, Augustin},
   year={2026},
   month=June, pages={2126} }

@article{castroruiz2025interpretingquantumreferenceframe,
      title = {Interpreting quantum-reference-frame transformations through a simple example},
  author = {Castro-Ruiz, Esteban and Galley, Thomas D. and Loveridge, Leon},
  journal = {Phys. Rev. A},
  volume = {114},
  issue = {3},
  pages = {032201},
  numpages = {14},
  year = {2026},
  month = {Sep},
  publisher = {American Physical Society},
  doi = {10.1103/czr8-dv1l},
  url = {https://link.aps.org/doi/10.1103/czr8-dv1l}
}

@article{Bartlett_2007,
   title={Reference frames, superselection rules, and quantum information},
   volume={79},
   ISSN={1539-0756},
   url={http://dx.doi.org/10.1103/RevModPhys.79.555},
   DOI={10.1103/revmodphys.79.555},
   number={2},
   journal={Rev. Mod. Phys.},
   publisher={American Physical Society (APS)},
   author={Bartlett, Stephen D. and Rudolph, Terry and Spekkens, Robert W.},
   year={2007},
   month=Apr, pages={555–609} }

@article{Ali_Ahmad_2022,
   title={Quantum Relativity of Subsystems},
   volume={128},
   ISSN={1079-7114},
   url={http://dx.doi.org/10.1103/PhysRevLett.128.170401},
   DOI={10.1103/physrevlett.128.170401},
   pages={170401},
   journal={Phys. Rev. Lett.},
   publisher={American Physical Society (APS)},
   author={Ali Ahmad, Shadi and Galley, Thomas D. and Höhn, Philipp A. and Lock, Maximilian P. E. and Smith, Alexander R. H.},
   year={2022},
   month=Apr }

@misc{aguilargutierrez2026relationalpathintegraleffective,
      title={Relational path integral, effective actions and quantum frame covariance in gravity}, 
      author={Sergio E. Aguilar-Gutierrez and Renata Ferrero and Philipp A. Hoehn and Luca Marchetti},
      year={2026},
      eprint={2607.21463},
      archivePrefix={arXiv},
      primaryClass={hep-th}}

@article{Loveridge:2016tnh,
    author = "Loveridge, Leon and Busch, Paul and Miyadera, Takayuki",
    title = "{Relativity of quantum states and observables}",
    eprint = "1604.02836",
    archivePrefix = "arXiv",
    primaryClass = "quant-ph",
    doi = "10.1209/0295-5075/117/40004",
    journal = "EPL",
    volume = "117",
    number = "4",
    pages = "40004",
    year = "2017"
}

@article{Busch:2016rui,
    author = "Busch, Paul and Loveridge, Leon and Miyadera, Takayuki",
    title = "{Approximating relational observables by absolute quantities: a quantum accuracy-size trade-off}",
    eprint = "1510.02063",
    archivePrefix = "arXiv",
    primaryClass = "quant-ph",
    doi = "10.1088/1751-8113/49/18/185301",
    journal = "J. Phys. A",
    volume = "49",
    number = "18",
    pages = "185301",
    year = "2016"
}

@article{Barbado_2011,
   title={{H}awking radiation as perceived by different observers},
   volume={28},
   ISSN={1361-6382},
   url={http://dx.doi.org/10.1088/0264-9381/28/12/125021},
   DOI={10.1088/0264-9381/28/12/125021},
   number={12},
   journal={Class. Quantum Grav.},
   publisher={IOP Publishing},
   author={Barbado, L C and Barceló, C and Garay, L J},
   year={2011},
   month=may, pages={125021} }

@article{preciado2024more,
     title = {More excitement across the horizon},
  author = {Preciado-Rivas, Mar\'{\i}a R. and Naeem, Manar and Mann, Robert B. and Louko, Jorma},
  journal = {Phys. Rev. D},
  volume = {110},
  issue = {2},
  pages = {025002},
  numpages = {15},
  year = {2024},
  month = {Jul},
  publisher = {American Physical Society},
  doi = {10.1103/PhysRevD.110.025002},
  url = {https://link.aps.org/doi/10.1103/PhysRevD.110.025002}
}

@article{foo2023quantum,
  title={Quantum superpositions of {M}inkowski spacetime},
  author={Foo, Joshua and Arabaci, Cemile Senem and Zych, Magdalena and Mann, Robert B},
  journal={Phys. Rev.  D},
  volume={107},
  number={4},
  pages={045014},
  year={2023},
  publisher={APS},
  doi={https://doi.org/10.1103/PhysRevD.107.045014}
}

@article{goel2024accelerateddetectorsuperposedspacetime,
     title={Accelerated detector in a superposed spacetime},
   volume={111},
   ISSN={2470-0029},
   url={http://dx.doi.org/10.1103/PhysRevD.111.025015},
   DOI={10.1103/physrevd.111.025015},
   pages={025015},
   journal={Phys. Rev. D},
   publisher={American Physical Society (APS)},
   author={Goel, Lakshay and Patterson, Everett A. and Preciado-Rivas, María Rosa and Torabian, Mahdi and Mann, Robert B. and Afshordi, Niayesh},
   year={2025},
   month=jan }

@article{almheiri2013black,
  author = {Almheiri, Ahmed and Marolf, Donald and Polchinski, Joseph and Sully, James},
  title = {Black holes: complementarity or firewalls?},
  journal = {J. High Energy Phys.},
  volume = {02},
  number = {2013},
  pages = {062},
  year = {2013},
  doi = {10.1007/JHEP02(2013)062},
}

@article{Ng_2022,
   title={A little excitement across the horizon},
   volume={24},
   ISSN={1367-2630},
   url={http://dx.doi.org/10.1088/1367-2630/ac9547},
   DOI={10.1088/1367-2630/ac9547},
   number={10},
   journal={New J. Phys.},
   publisher={IOP Publishing},
   author={Ng, Keith K and Zhang, Chen and Louko, Jorma and Mann, Robert B},
   year={2022},
   month=oct, pages={103018} }

@article{hawking1976breakdown,
  title={Breakdown of predictability in gravitational collapse},
  author={Hawking, Stephen W},
  journal={Phys. Rev.  D},
  volume={14},
  number={10},
  pages={2460},
  year={1976},
  publisher={APS}, 
  doi = {10.1103/PhysRevD.14.2460},
}

@article{giacomini2019quantum,
  title={Quantum mechanics and the covariance of physical laws in quantum reference frames},
  author={Giacomini, Flaminia and Castro-Ruiz, Esteban and Brukner, {\v{C}}aslav},
  journal={Nat. Commun.},
  volume={10},
  number={1},
  pages={494},
  year={2019},
  publisher={Nature Publishing Group UK London},
  doi={https://doi.org/10.1038/s41467-018-08155-0}
}

@article{hayden2007black,
  author = {Hayden, Patrick and Preskill, John},
  title = {Black holes as mirrors: quantum information in random subsystems},
  journal = {J. High Energy Phys.},
  volume = {09},
  number = {2007},
  pages = {120},
  year = {2007},
  doi = {10.1088/1126-6708/2007/09/120},
}

@article{harlow2016jerusalem,
  title={Jerusalem lectures on black holes and quantum information},
  author={Harlow, Daniel},
  journal={Rev. Mod. Phys.},
  volume={88},
  number={1},
  pages={015002},
  year={2016},
  publisher={APS},
  doi={https://doi.org/10.1103/RevModPhys.88.015002}
}

@article{geng2022inconsistency,
  author = {Geng, Hao and Karch, Andreas and Perez-Pardavila, Carlos and Raju, Suvrat and Randall, Lisa and Riojas, Marcos and Shashi, Sanjit},
  title = {Inconsistency of islands in theories with long-range gravity},
  journal = {J. High Energy Phys.},
  volume = {01},
  number = {2022},
  pages = {182},
  year = {2022},
  doi = {10.1007/JHEP01(2022)182},
}

@article{penington2020entanglement,
  author = {Penington, Geoffrey},
  title = {Entanglement wedge reconstruction and the information paradox},
  journal = {J. High Energy Phys.},
  volume = {09},
  number = {2020},
  pages = {002},
  year = {2020},
  doi = {10.1007/JHEP09(2020)002},
}

@article{geng2021information,
  title={Information transfer with a gravitating bath},
  author={Geng, Hao and Karch, Andreas and Perez-Pardavila, Carlos and Raju, Suvrat and Randall, Lisa and Riojas, Marcos and Shashi, Sanjit},
  journal={SciPost Phys.},
  volume={10},
  number={5},
  pages={103},
  year={2021},
  doi={10.21468/SciPostPhys.10.5.103}
}

@article{maldacena1999large,
  title={The large-{N} limit of superconformal field theories and supergravity},
  author={Maldacena, Juan},
  journal={Int. J. Theor. Phys.},
  volume={38},
  number={4},
  pages={1113--1133},
  year={1999},
  publisher={Springer},
  doi={https://doi.org/10.1023/A%3A1026654312961}
}

@article{mertens2023solvable,
  title={Solvable models of quantum black holes: a review on {Jackiw}--{Teitelboim} gravity},
  author={Mertens, Thomas G and Turiaci, Gustavo J},
  journal={Living Rev. Relativity},
  volume={26},
  number={1},
  pages={4},
  year={2023},
  publisher={Springer},
  doi={https://doi.org/10.1007/s41114-023-00046-1}
}

@article{almheiri2020page,
  author = {Almheiri, Ahmed and Mahajan, Raghu and Maldacena, Juan and Zhao, Ying},
  title = {The {P}age curve of {H}awking radiation from semiclassical geometry},
  journal = {J. High Energy Phys.},
  volume = {03},
  number = {2020},
  pages = {149},
  year = {2020},
  doi = {10.1007/JHEP03(2020)149},
}

@article{page1993information,
  title={Information in black hole radiation},
  author={Page, Don N},
  journal={Phys. Rev. Lett.},
  volume={71},
  number={23},
  pages={3743},
  year={1993},
  publisher={APS},
  doi={https://doi.org/10.1103/PhysRevLett.71.3743}
}

@article{page1993average,
  title={Average entropy of a subsystem},
  author={Page, Don N},
  journal={Phys. Rev. Lett.},
  volume={71},
  number={9},
  pages={1291},
  year={1993},
  publisher={APS}, 
  doi={https://doi.org/10.1103/PhysRevLett.71.1291}
}

@article{raju2022lessons,
  title={Lessons from the information paradox},
  author={Raju, Suvrat},
  journal={Phys. Rep.},
  volume={943},
  pages={1--80},
  year={2022},
  publisher={Elsevier},
   doi={10.1016/j.physrep.2021.10.001}
}

@article{hawking2016soft,
  title={Soft hair on black holes},
  author={Hawking, Stephen W and Perry, Malcolm J and Strominger, Andrew},
  journal={Phys. Rev. Lett.},
  volume={116},
  number={23},
  pages={231301},
  year={2016},
  publisher={APS}
}

@article{Crispino_2008,
   title={The {U}nruh effect and its applications},
   volume={80},
   ISSN={1539-0756},
   url={http://dx.doi.org/10.1103/RevModPhys.80.787},
   DOI={10.1103/revmodphys.80.787},
   number={3},
   journal={Rev. Mod. Phys.},
   publisher={American Physical Society (APS)},
   author={Crispino, Luís C. B. and Higuchi, Atsushi and Matsas, George E. A.},
   year={2008},
   month=jul, pages={787–838} }

@article{cepollaro2024entanglement,
   title={Sum of Entanglement and Subsystem Coherence Is Invariant under Quantum Reference Frame Transformations},
   volume={135},
   ISSN={1079-7114},
   DOI={10.1103/h6b3-y4vt},
   pages={010201},
   journal={Phys. Rev. Lett.},
   publisher={American Physical Society (APS)},
   author={Cepollaro, Carlo and Akil, Ali and Cieśliński, Paweł and de la Hamette, Anne-Catherine and Brukner,  {\v{C}}aslav},
   year={2025},
   month=jun }

@article{Bengyat_2024,
   title={Gravity-mediated entanglement between oscillators as quantum superposition of geometries},
   volume={110},
   ISSN={2470-0029},
   DOI={10.1103/physrevd.110.056046},
   pages={056046},
   journal={Phys. Rev.  D},
   publisher={American Physical Society (APS)},
   author={Bengyat, Ofek and Di Biagio, Andrea and Aspelmeyer, Markus and Christodoulou, Marios},
   year={2024},
   month=sep }

@article{foo_schrodingers_2021,
    title = {Schrödinger’s cat for de {S}itter spacetime},
    volume = {38},
    issn = {0264-9381, 1361-6382},
    url = {https://iopscience.iop.org/article/10.1088/1361-6382/abf1c4},
    doi = {10.1088/1361-6382/abf1c4},
    number = {11},
    urldate = {2024-10-14},
    journal = {Class. Quantum Grav.},
    author = {Foo, Joshua and Mann, Robert B and Zych, Magdalena},
    month = jun,
    year = {2021},
    pages = {115010},
}

@article{hohn2020switch,
  title={How to switch between relational quantum clocks},
  author={H{\"o}hn, Philipp A and Vanrietvelde, Augustin},
  journal={New J. Phys.},
  volume={22},
  number={12},
  pages={123048},
  year={2020},
  publisher={IOP Publishing},
  doi={10.1088/1367-2630/abd1ac}
}

@article{louko_hamiltonian_1998,
    title = {Hamiltonian spacetime dynamics with a spherical null-dust shell},
    volume = {57},
    copyright = {http://link.aps.org/licenses/aps-default-license},
    issn = {0556-2821, 1089-4918},
    url = {https://link.aps.org/doi/10.1103/PhysRevD.57.2279},
    doi = {10.1103/PhysRevD.57.2279},
    number = {4},
    urldate = {2025-04-21},
    journal = {Phys. Rev. D},
    author = {Louko, Jorma and Whiting, Bernard F. and Friedman, John L.},
    month = feb,
    year = {1998},
    pages = {2279--2298}
}

@article{susskind1993stretched,
  title = {The stretched horizon and black hole complementarity},
  author = {L. Susskind and L. Thorlacius and J. Uglum},
  journal = {Phys. Rev. D},
  volume = {48},
  issue = {8},
  pages = {3743--3761},
  numpages = {0},
  year = {1993},
  month = {Oct},
  publisher = {American Physical Society},
  doi = {10.1103/PhysRevD.48.3743},
  url = {https://link.aps.org/doi/10.1103/PhysRevD.48.3743}
}

@misc{goeller2022diffeomorphisminvariantobservablesdynamicalframes,
      title={Diffeomorphism-invariant observables and dynamical frames in gravity: reconciling bulk locality with general covariance}, 
      author = {C. Goeller and P. A. Höhn and J. Kirklin},
      year={2022},
      eprint={2206.01193},
      archivePrefix={arXiv},
      primaryClass={hep-th}}

@misc{strominger2018lecturesinfraredstructuregravity,
      title={Lectures on the Infrared Structure of Gravity and Gauge Theory}, 
      author = {A. Strominger},
      year={2018},
      eprint={1703.05448},
      archivePrefix={arXiv},
      primaryClass={hep-th}}

@article{Araujo_Regado_2025,
  author = {Araujo-Regado, G. and Höhn, P. A. and Sartini, F. and Tomova, B.},
  title = {Soft edges: the many links between soft and edge modes},
  journal = {J. High Energy Phys.},
  volume = {07},
  number = {2025},
  pages = {180},
  year = {2025},
  doi = {10.1007/jhep07(2025)180},
}

@article{carney_infrared_2017,
    title = {Infrared quantum information},
    volume = {119},
    issn = {0031-9007, 1079-7114},
    url = {http://arxiv.org/abs/1706.03782},
    doi = {10.1103/PhysRevLett.119.180502},   
    number = {18},
    journal = {Phys. Rev. Lett.},
    author = {D. Carney and L. Chaurette and D. Neuenfeld and G. W. Semenoff},
    month = oct,
    year = {2017},
    note = {arXiv:1706.03782 [hep-th]},
    pages = {180502},
}

@misc{strominger2017blackholeinformationrevisited,
      title={Black Hole Information Revisited}, 
      author = {A. Strominger},
      year={2017},
      eprint={1706.07143},
      archivePrefix={arXiv},
      primaryClass={hep-th}}

@article{bondi1962gravitational,
  title={Gravitational waves in general relativity, VII. Waves from axi-symmetric isolated system},
  author = {H. Bondi and M. G. J. Van der Burg and A. Metzner},
  journal={Proc. R. Soc. London, Ser. A},
  volume={269},
  number={1336},
  pages={21--52},
  year={1962},
  publisher={The Royal Society London},
doi = {10.1098/rspa.1962.0161}
}

@article{sachs1962gravitational,
  title={Gravitational waves in general relativity VIII. Waves in asymptotically flat space-time},
  author = {R. K. Sachs},
  journal={Proc. R. Soc. London, Ser. A},
  volume={270},
  number={1340},
  pages={103--126},
  year={1962},
  publisher={The Royal Society London}, 
  doi = {10.1098/rspa.1962.0206}
}

@article{compere_poincare_2020,
  author = {G. Compère and R. Oliveri and A. Seraj},
  title = {The {Poincaré} and {BMS} flux-balance laws with application to binary systems},
  journal = {J. High Energy Phys.},
  volume = {10},
  number = {2020},
  pages = {116},
  year = {2020},
  doi = {10.1007/JHEP10(2020)116},
}

@misc{compere2019advancedlecturesgeneralrelativity,
      title={Advanced Lectures on General Relativity}, 
      author = {G. Compère and A. Fiorucci},
      year={2019},
      eprint={1801.07064},
      archivePrefix={arXiv},
      primaryClass={hep-th}}

@article{Zeldovich:1974gvh,
    author = "Zel'dovich, Y. B. and Polnarev, A. G.",
    title = "{Radiation of gravitational waves by a cluster of superdense stars}",
    journal = "Sov. Astron.",
    volume = "18",
    pages = "17",
    year = "1974"
}

@article{Bieri_2024,
   title={Gravitational wave displacement and velocity memory effects},
   volume={41},
   ISSN={1361-6382},
   url={http://dx.doi.org/10.1088/1361-6382/ad4dfe},
   DOI={10.1088/1361-6382/ad4dfe},
   number={13},
   journal={Class. Quantum Grav.},
   publisher={IOP Publishing},
   author={Bieri, L. and Polnarev, A.},
   year={2024},
   month=jun, pages={135012} }

@article{christodoulou1991nonlinear,
  title={Nonlinear nature of gravitation and gravitational-wave experiments},
  author={Christodoulou, D.},
  journal={Phys. Rev. Lett.},
  volume={67},
  number={12},
  pages={1486},
  year={1991},
  publisher={APS}, 
  doi = {10.1103/PhysRevLett.67.1486}
}

@misc{maibach2026balancefluxlawsgeneral,
      title={Balance flux laws beyond general relativity}, 
      author={David Maibach and Jann Zosso},
      year={2026},
      eprint={2601.07091},
      archivePrefix={arXiv},
      primaryClass={gr-qc}}

@article{thorne1992gravitational,
  title = {Gravitational-wave bursts with memory: The Christodoulou effect},
  author = {Thorne, Kip S.},
  journal = {Phys. Rev. D},
  volume = {45},
  issue = {2},
  pages = {520--524},
  numpages = {0},
  year = {1992},
  month = {Jan},
  publisher = {American Physical Society},
  doi = {10.1103/PhysRevD.45.520},
  url = {https://link.aps.org/doi/10.1103/PhysRevD.45.520}
}

@article{Freidel_2020,
  author = {Freidel, Laurent and Geiller, Marc and Pranzetti, Daniele},
  title = {Edge modes of gravity. {Part I}. {Corner} potentials and charges},
  journal = {J. High Energy Phys.},
  volume = {11},
  number = {2020},
  pages = {026},
  year = {2020},
  doi = {10.1007/JHEP11(2020)026},
}

@article{Donnelly_2016,
  author = {W. Donnelly and L. Freidel},
  title = {Local subsystems in gauge theory and gravity},
  journal = {J. High Energy Phys.},
  volume = {09},
  number = {2016},
  pages = {102},
  year = {2016},
  doi = {10.1007/jhep09(2016)102},
}

@article{Donnay_2016,
  author = {L. Donnay and G. Giribet and H. A. González and M. Pino},
  title = {Extended symmetries at the black hole horizon},
  journal = {J. High Energy Phys.},
  volume = {09},
  number = {2016},
  pages = {100},
  year = {2016},
  doi = {10.1007/jhep09(2016)100},
}

@article{Carrozza_2022,
  author = {Carrozza, Sylvain and Höhn, Philipp A.},
  title = {Edge modes as reference frames and boundary actions from post-selection},
  journal = {J. High Energy Phys.},
  volume = {02},
  number = {2022},
  pages = {172},
  year = {2022},
  doi = {10.1007/jhep02(2022)172},
}

@incollection{gomes2025boundariesframesissuephysical,
    author = {Gomes, Henrique and Langenscheidt, Simon and Oriti, Daniele},
    editor = {Cuffaro, Michael E and Hartmann, Stephan},
    isbn = {9780198929246},
    title = {Boundaries, Frames, and the Issue of Physical Covariance},
    booktitle = {Open Systems: Physics, Metaphysics, and Methodology},
    publisher = {Oxford University Press},
    year = {2026},
    month = {05},
    doi = {10.1093/9780198929277.003.0009},
    url = {https://doi.org/10.1093/9780198929277.003.0009},
}

@article{Donnelly_2015,
   title={Entanglement Entropy of Electromagnetic Edge Modes},
   volume={114},
   ISSN={1079-7114},
   DOI={10.1103/physrevlett.114.111603},
   pages={111603},
   journal={Phys. Rev. Lett.},
   publisher={American Physical Society (APS)},
   author={Donnelly, William and Wall, Aron C.},
   year={2015},
   month=mar }

@article{Speranza_2018,
  author = {Speranza, Antony J.},
  title = {Local phase space and edge modes for diffeomorphism-invariant theories},
  journal = {J. High Energy Phys.},
  volume = {02},
  number = {2018},
  pages = {021},
  year = {2018},
  doi = {10.1007/jhep02(2018)021},
}

@article{Pal_2025,
   title={On the magnetic 2+1- D space-time and its non-relativistic counterpart},
   volume={3152},
   ISSN={1742-6596},
   url={http://dx.doi.org/10.1088/1742-6596/3152/1/012028},
   DOI={10.1088/1742-6596/3152/1/012028},
   number={1},
   journal={J. Phys. Conf. Ser.},
   publisher={IOP Publishing},
   author={Pal, Sayan Kumar},
   year={2025},
   month=dec, pages={012028} }

@article{Gibbons_2003,
   title={Newton–Hooke spacetimes, Hpp-waves and the cosmological constant},
   volume={20},
   ISSN={1361-6382},
   url={http://dx.doi.org/10.1088/0264-9381/20/23/016},
   DOI={10.1088/0264-9381/20/23/016},
   number={23},
   journal={Class. Quantum Grav.},
   publisher={IOP Publishing},
   author={Gibbons, G W and Patricot, C E},
   year={2003},
   month=oct, pages={5225–5239} }

@article{cornish1991gravitation,
  title = {Gravitation in 2+1 dimensions},
  author = {Cornish, N. J. and Frankel, N. E.},
  journal = {Phys. Rev. D},
  volume = {43},
  issue = {8},
  pages = {2555--2565},
  numpages = {0},
  year = {1991},
  month = {Apr},
  publisher = {American Physical Society},
  doi = {10.1103/PhysRevD.43.2555},
  url = {https://link.aps.org/doi/10.1103/PhysRevD.43.2555}
}

@article{susskind1994gedanken,
  title = {Gedanken experiments involving black holes},
  author = {L. Susskind and L. Thorlacius},
  journal = {Phys. Rev. D},
  volume = {49},
  issue = {2},
  pages = {966--974},
  numpages = {0},
  year = {1994},
  month = {Jan},
  publisher = {American Physical Society},
  doi = {10.1103/PhysRevD.49.966},
  url = {https://link.aps.org/doi/10.1103/PhysRevD.49.966}
}

@article{page2013time,
  title={Time dependence of {H}awking radiation entropy},
  author = {D. N. Page},
  journal={J. Cosmol. Astropart. Phys.},
  volume={2013},
  number={09},
  pages={028},
  year={2013},
  publisher={IOP Publishing},
   DOI={10.1088/1475-7516/2013/09/028}
}

@article{Uhlemann2022information,
  author = {Uhlemann, C. F.},
  title = {Information transfer with a twist},
  journal = {J. High Energy Phys.},
  volume = {01},
  number = {2022},
  pages = {126},
  year = {2022},
  doi = {10.1007/jhep01(2022)126},
}

@article{Uhlemann_2021,
  author = {Uhlemann, Christoph F.},
  title = {Islands and Page curves in 4d from Type {IIB}},
  journal = {J. High Energy Phys.},
  volume = {08},
  number = {2021},
  pages = {104},
  year = {2021},
  doi = {10.1007/jhep08(2021)104},
}

@article{Hartman_2020,
  author = {T. Hartman and E. Shaghoulian and A. Strominger},
  title = {Islands in asymptotically flat 2D gravity},
  journal = {J. High Energy Phys.},
  volume = {07},
  number = {2020},
  pages = {022},
  year = {2020},
  doi = {10.1007/jhep07(2020)022},
}

@article{geng2026mechanisminformationencodingislands,
      title={The Mechanism behind the Information Encoding for Islands}, 
      author = {H. Geng},
      year={2026},
      number={2026},
      journal={J. High Energy Phys.},
      volume={03},
      pages={37},
      doi={10.1007/JHEP03(2026)037}}

@article{Krishnan_2021,
  author = {C. Krishnan},
  title = {Critical islands},
  journal = {J. High Energy Phys.},
  volume = {01},
  number = {2021},
  pages = {179},
  year = {2021},
  doi = {10.1007/jhep01(2021)179},
}

@article{Van_Raamsdonk_2021,
  author = {M. Van Raamsdonk},
  title = {Comments on wormholes, ensembles, and cosmology},
  journal = {J. High Energy Phys.},
  volume = {12},
  number = {2021},
  pages = {156},
  year = {2021},
  doi = {10.1007/jhep12(2021)156},
}

@article{flanagan_infrared_2021,
  author = {E. E. Flanagan},
  title = {Infrared {Effects} in the {Late} {Stages} of {Black} {Hole} {Evaporation}},
  journal = {J. High Energy Phys.},
  volume = {07},
  number = {2021},
  pages = {137},
  year = {2021},
  doi = {10.1007/JHEP07(2021)137},
}

@article{Bao_2018,
   title={Branches of the black hole wave function need not contain firewalls},
   volume={97},
   ISSN={2470-0029},
   url={http://dx.doi.org/10.1103/PhysRevD.97.126014},
   DOI={10.1103/physrevd.97.126014},
   pages={126014},
   journal={Phys. Rev. D},
   publisher={American Physical Society (APS)},
   author = {N. Bao and S. M. Carroll and A. Chatwin-Davies and J. Pollack and G. N. Remmen},
   year={2018},
   month=jun }

@misc{akil2025quantumsuperpositionblackhole,
      title={A Quantum Superposition of Black Hole Evaporation Histories: Recovering Unitarity}, 
      author = {A. Akil and L. Giannelli and L. Modesto and O. Dahlsten and G. Chiribella and {\v{C}}. Brukner},
      year={2025},
      eprint={2507.17031},
      archivePrefix={arXiv},
      primaryClass={gr-qc}}

@article{Hutchinson_2016,
   title={Icezones instead of firewalls: extended entanglement beyond the event horizon and unitary evaporation of a black hole},
   volume={33},
   ISSN={1361-6382},
   url={http://dx.doi.org/10.1088/0264-9381/33/13/135006},
   DOI={10.1088/0264-9381/33/13/135006},
   pages={135006},
   journal={Class. Quantum Grav.},
   publisher={IOP Publishing},
   author={Hutchinson, J. and Stojkovic, D.},
   year={2016},
   month=may, pages={135006} }

@article{geng2025makingcasemassiveislands,
      title={Making the Case for Massive Islands}, 
      author = {H. Geng},
      year={2026},
      number={2026},
      journal={J. High Energy Phys.},
      volume = {08},
      pages={214},
      doi={10.1007/JHEP08(2026)214}}

@article{page1980is,
  title = {Is Black-Hole Evaporation Predictable?},
  author = {D. N. Page},
  journal = {Phys. Rev. Lett.},
  volume = {44},
  issue = {5},
  pages = {301--304},
  numpages = {0},
  year = {1980},
  month = {Feb},
  publisher = {American Physical Society},
  doi = {10.1103/PhysRevLett.44.301},
  url = {https://link.aps.org/doi/10.1103/PhysRevLett.44.301}
}

@article{geng2021informationparadoxandits,
  title = {Information paradox and its resolution in de {S}itter holography},
  author = {H. Geng and Y. Nomura and H. Sun},
  journal = {Phys. Rev. D},
  volume = {103},
  issue = {12},
  pages = {126004},
  numpages = {8},
  year = {2021},
  month = {Jun},
  publisher = {American Physical Society},
  doi = {10.1103/PhysRevD.103.126004},
  url = {https://link.aps.org/doi/10.1103/PhysRevD.103.126004}
}

@article{pasterski_hps_2021,
  author = {Pasterski, S. and Verlinde, H.},
  title = {{HPS} meets {AMPS}: {How} {Soft} {Hair} {Dissolves} the {Firewall}},
  journal = {J. High Energy Phys.},
  volume = {09},
  number = {2021},
  pages = {099},
  year = {2021},
  doi = {10.1007/JHEP09(2021)099},
}

@misc{geng2026seeingpagecurvesislands,
      title={Seeing {P}age Curves and Islands with Blinders On}, 
      author={Hao Geng and Andreas Karch and Carlos Perez-Pardavila and Suvrat Raju and Lisa Randall and Marcos Riojas},
      year={2026},
      eprint={2602.06543},
      archivePrefix={arXiv},
      primaryClass={hep-th}}

@article{wang2021page,
  title = {{P}age curves for a family of exactly solvable evaporating black holes},
  author = {X. Wang and R. Li and J. Wang},
  journal = {Phys. Rev. D},
  volume = {103},
  issue = {12},
  pages = {126026},
  numpages = {17},
  year = {2021},
  month = {Jun},
  publisher = {American Physical Society},
  doi = {10.1103/PhysRevD.103.126026},
  url = {https://link.aps.org/doi/10.1103/PhysRevD.103.126026}
}

@article{Maldacena_2013,
   title={Cool horizons for entangled black holes},
   volume={61},
   ISSN={1521-3978},
   url={http://dx.doi.org/10.1002/prop.201300020},
   DOI={10.1002/prop.201300020},
   number={9},
   journal={Fortschr. Phys.},
   publisher={Wiley},
   author={Maldacena, J. and Susskind, L.},
   year={2013},
   month=aug, pages={781–811} }

@article{Geng_2020,
  author = {H. Geng and A. Karch},
  title = {Massive islands},
  journal = {J. High Energy Phys.},
  volume = {09},
  number = {2020},
  pages = {121},
  year = {2020},
  doi = {10.1007/jhep09(2020)121},
}

@article{Balasubramanian_2021,
  author = {V. Balasubramanian and A. Kar and T. Ugajin},
  title = {Islands in de {S}itter space},
  journal = {J. High Energy Phys.},
  volume = {02},
  number = {2021},
  pages = {072},
  year = {2021},
  doi = {10.1007/jhep02(2021)072},
}

@article{kibe2022holographic,
  title={Holographic spacetime, black holes and quantum error correcting codes: a review},
  author = {T. Kibe and P. Mandayam and A. Mukhopadhyay},
  journal={Eur. Phys. J. C},
  volume={82},
  number={5},
  pages={463},
   DOI={10.1140/epjc/s10052-022-10382-1},
  year={2022},
  publisher={Springer}
}

@article{chen2020quantumextremalislandseasyI,
      title={Quantum Extremal Islands Made Easy, Part {I}: Entanglement on the Brane}, 
      author = {H. Z. Chen and R. C. Myers and D. Neuenfeld and I. A. Reyes and J. Sandor},
      year={2020},
       journal={J. High Energy Phys.},
       volume = {10},
       pages={166},
       number={2020},
       doi={10.1007/JHEP10(2020)166}}

@misc{almheiri2023islandsoutsidehorizon,
      title={Islands outside the horizon}, 
      author = {A. Almheiri and R. Mahajan and J. Maldacena},
      year={2019},
      eprint={1910.11077},
      archivePrefix={arXiv},
      primaryClass={hep-th}}

@article{chen2020quantumextremalislandseasyII,
      title={Quantum Extremal Islands Made Easy, Part {II}: Black Holes on the Brane}, 
      author = {H. Z. Chen and R. C. Myers and D. Neuenfeld and I. A. Reyes and J. Sandor},
     journal={J. High Energy Phys.},
     year={2020},
     number={2020},
     volume = {12},
     pages={025},
     doi={10.1007/JHEP12(2020)025}}

@Article{almheiri2020entanglementislandsinhigher,
	title={Entanglement islands in higher dimensions},
	author = {A. Almheiri and R. Mahajan and J. E. Santos},
	journal={SciPost Phys.},
	volume={9},
	pages={001},
	year={2020},
	publisher={SciPost},
	doi={10.21468/SciPostPhys.9.1.001},
	url={https://scipost.org/10.21468/SciPostPhys.9.1.001},
}

@article{Hashimoto_2020,
  author = {K. Hashimoto and N. Iizuka and Y. Matsuo},
  title = {Islands in {S}chwarzschild black holes},
  journal = {J. High Energy Phys.},
  volume = {06},
  number = {2020},
  pages = {085},
  year = {2020},
  doi = {10.1007/jhep06(2020)085},
}

@article{geng2025revisitingrecentprogresskarchrandall,
      title={Revisiting Recent Progress in the {K}arch--{R}andall Braneworld}, 
      author = {H. Geng},
      year={2026},
      journal={Class. Quantum Grav.},
      volume={43},
pages={095014},
DOI={10.1088/1361-6382/ae62ec}
}

@article{bousso_islands_2023,
    title = {Islands far outside the horizon},
year={2024},
number={2024},
     journal={J. High Energy Phys.},
     volume = {11},
pages={164},
doi={10.1007/JHEP11(2024)164},
    author = {R. Bousso and G. Penington},
    }

@misc{susskind2014ereprghzconsistencyquantum,
      title={{ER}={EPR}, {GHZ}, and the Consistency of Quantum Measurements}, 
      author={Leonard Susskind},
      year={2014},
      eprint={1412.8483},
      archivePrefix={arXiv},
      primaryClass={hep-th}}

@article{Susskind_2016,
   title={Copenhagen vs {E}verett, Teleportation, and {ER}={EPR}},
   volume={64},
   ISSN={0015-8208},
   url={http://dx.doi.org/10.1002/prop.201600036},
   DOI={10.1002/prop.201600036},
   pages={551-564},
   journal={Fortschr. Phys.},
   publisher={Wiley},
   author={Susskind, Leonard},
   year={2016},
   month=jun, pages={551–564} }

@article{Jacobson_2019,
   title={Diffeomorphism invariance and the black hole information paradox},
   volume={100},
   ISSN={2470-0029},
   url={http://dx.doi.org/10.1103/PhysRevD.100.046002},
   DOI={10.1103/physrevd.100.046002},
   pages={046002},
   journal={Phys. Rev. D},
   publisher={American Physical Society (APS)},
   author = {T. Jacobson and P. Nguyen},
   year={2019},
   month=aug }

@article{JACOBSON_2013,
   title={Boundary unitarity and the black hole information paradox},
   volume={22},
   ISSN={1793-6594},
   url={http://dx.doi.org/10.1142/S0218271813420029},
   DOI={10.1142/s0218271813420029},
   number={12},
   journal={Int. J. Modern Phys. D},
   publisher={World Scientific Pub Co Pte Lt},
   author = {T. Jacobson},
   year={2013},
   month=oct, pages={1342002} }

@article{fewster2024quantumreferenceframesmeasurement,
   title={Quantum Reference Frames, Measurement Schemes and the Type of Local Algebras in Quantum Field Theory},
   volume={406},
   ISSN={1432-0916},
   url={http://dx.doi.org/10.1007/s00220-024-05180-7},
   DOI={10.1007/s00220-024-05180-7},
   pages={19},
   journal={Commun. Math. Phys.},
   publisher={Springer Science and Business Media LLC},
   author = {C. J. Fewster and D. W. Janssen and L. D. Loveridge and K. Rejzner and J. Waldron},
   year={2024},
   month=dec }

@article{Brown_1995,
   title={Dust as a standard of space and time in canonical quantum gravity},
   volume={51},
   ISSN={0556-2821},
   url={http://dx.doi.org/10.1103/PhysRevD.51.5600},
   DOI={10.1103/physrevd.51.5600},
   pages={5600},
   journal={Phys. Rev. D},
   publisher={American Physical Society (APS)},
   author={Brown, J. D. and Kucha{\v{r}}, K. V.},
   year={1995},
   month=may, pages={5600–5629} }

@article{Rovelli:1990ph,
    author = "Rovelli, Carlo",
    title = "{What Is Observable in Classical and Quantum Gravity?}",
    reportNumber = "PITT-90-10",
    doi = "10.1088/0264-9381/8/2/011",
    journal = "Class. Quant. Grav.",
    volume = "8",
    pages = "297--316",
    year = "1991"
}

@article{Dittrich_2007,
   title={Partial and complete observables for Hamiltonian constrained systems},
   volume={39},
   ISSN={1572-9532},
   url={http://dx.doi.org/10.1007/s10714-007-0495-2},
   DOI={10.1007/s10714-007-0495-2},
   number={11},
   journal={Gen. Relativ. Gravit.},
   publisher={Springer Science and Business Media LLC},
   author = {B. Dittrich},
   year={2007},
   month=aug, pages={1891–1927} }

@article{Giddings_2006,
   title={Observables in effective gravity},
   volume={74},
   ISSN={1550-2368},
   url={http://dx.doi.org/10.1103/PhysRevD.74.064018},
   DOI={10.1103/physrevd.74.064018},
   pages={064018},
   journal={Phys. Rev. D},
   publisher={American Physical Society (APS)},
   author = {S. B. Giddings and D. Marolf and J. B. Hartle},
   year={2006},
   month=sep }

@article{Giddings:2022hba,
    author = "Giddings, S. B. and Perkins, J.",
    title = "{Perturbative quantum evolution of the gravitational state and dressing in general backgrounds}",
    eprint = "2209.06836",
    archivePrefix = "arXiv",
    primaryClass = "hep-th",
    doi = "10.1103/PhysRevD.110.026012",
    journal = "Phys. Rev. D",
    volume = "110",
    number = "2",
    pages = "026012",
    year = "2024"
}

@article{Giddings:2019hjc,
    author = "Giddings, S. B.",
    title = "{Gravitational dressing, soft charges, and perturbative gravitational splitting}",
    eprint = "1903.06160",
    archivePrefix = "arXiv",
    primaryClass = "hep-th",
    reportNumber = "CERN-TH-2019-074",
    doi = "10.1103/PhysRevD.100.126001",
    journal = "Phys. Rev. D",
    volume = "100",
    number = "12",
    pages = "126001",
    year = "2019"
}

@article{Donnelly:2016rvo,
    author = "Donnelly, W. and Giddings, S. B.",
    title = "{Observables, gravitational dressing, and obstructions to locality and subsystems}",
    doi = "10.1103/PhysRevD.94.104038",
    journal = "Phys. Rev. D",
    volume = "94",
    number = "10",
    pages = "104038",
    year = "2016"
}

@article{Donnelly:2018nbv,
    author = "Donnelly, W. and Giddings, S. B.",
    title = "{Gravitational splitting at first order: Quantum information localization in gravity}",
    doi = "10.1103/PhysRevD.98.086006",
    journal = "Phys. Rev. D",
    volume = "98",
    number = "8",
    pages = "086006",
    year = "2018"
}

@misc{Giddings:2025bkp,
    author = "Giddings, S. B.",
    title = "{Gravitational dressing: from the crossed product to more general algebraic and mathematical structure}",
    eprint = "2510.24833",
    archivePrefix = "arXiv",
    primaryClass = "hep-th",
    month = "10",
    year = "2025"
}

@article{Giddings:2025xym,
    author = "Giddings, S. B.",
    title = "{Quantum gravity observables: observation, algebras, and mathematical structure}",
    doi = "10.1088/1751-8121/ae0b12",
    journal = "J. Phys. A",
    volume = "58",
    number = "41",
    pages = "415401",
    year = "2025"
}

@misc{deboer2022frontiersquantumgravityshared,
      title={Frontiers of Quantum Gravity: shared challenges, converging directions}, 
      author = {J. d. Boer and B. Dittrich and A. Eichhorn and S. B. Giddings and S. Gielen and S. Liberati and E. R. Livine and D. Oriti and K. Papadodimas and A. D. Pereira and M. Sakellariadou and S. Surya and H. Verlinde},
      year={2022},
      eprint={2207.10618},
      archivePrefix={arXiv},
      primaryClass={hep-th}}

@misc{Giddings:2024qcf,
    author = "Giddings, S. B.",
    title = "{The unitarity crisis, nonviolent unitarization, and implications for quantum spacetime}",
    eprint = "2412.18650",
    archivePrefix = "arXiv",
    primaryClass = "hep-th",
    month = "12",
    year = "2024"
}

@article{fronsdal1959completion,
  title = {Completion and Embedding of the {Schwarzschild} Solution},
  author = {Fronsdal, C.},
  journal = {Phys. Rev.},
  volume = {116},
  issue = {3},
  pages = {778--781},
  numpages = {0},
  year = {1959},
  month = {Nov},
  publisher = {American Physical Society},
  doi = {10.1103/PhysRev.116.778},
  url = {https://link.aps.org/doi/10.1103/PhysRev.116.778}
}

@article{Radzikowski:1996pa,
    author = "Radzikowski, M. J.",
    title = "{Micro-local approach to the Hadamard condition in quantum field theory on curved space-time}",
    doi = "10.1007/BF02100096",
    journal = "Commun. Math. Phys.",
    volume = "179",
    pages = "529--553",
    year = "1996"
}

@article{Kay:1988mu,
    author = "Kay, Bernard S. and Wald, Robert M.",
    title = "{Theorems on the Uniqueness and Thermal Properties of Stationary, Nonsingular, Quasifree States on Space-Times with a Bifurcate Killing Horizon}",
    reportNumber = "PRINT-88-0840 (CHICAGO)",
    doi = "10.1016/0370-1573(91)90015-E",
    journal = "Phys. Rept.",
    volume = "207",
    pages = "49--136",
    year = "1991"
}

@article{akil2022semiclassical,
  title = {Semiclassical spacetimes at super-Planckian scales from delocalized sources},
  author = {Akil, A. and Cadoni, M. and Modesto, L. and Oi, M. and Sanna, A. P.},
  journal = {Phys. Rev. D},
  volume = {108},
  issue = {4},
  pages = {044051},
  numpages = {23},
  year = {2023},
  month = {Aug},
  publisher = {American Physical Society},
  doi = {10.1103/PhysRevD.108.044051},
  url = {https://link.aps.org/doi/10.1103/PhysRevD.108.044051}
}

@misc{howl2022quantum,
      title={Quantum gravity as a communication resource}, 
      author={Richard Howl and Ali Akil and Hlér Kristjánsson and Xiaobin Zhao and Giulio Chiribella},
      year={2024},
      eprint={2203.05861},
      archivePrefix={arXiv},
      primaryClass={quant-ph}}

\newpage

\onecolumngrid

\makeatletter
\long\def\@makefntext#1{%
    \noindent
    \setlength{\parindent}{1em} 
    \setlength{\leftskip}{0em} 
    \setlength{\rightskip}{0em} 
    \setlength{\parfillskip}{0pt} 
    \setlength{\parskip}{0pt} 
    \raggedright 
    \raggedbottom
    \small 
    #1%
}
\makeatother
\appendix

\section*{Appendix}
\vspace{0.5cm}

\section{BTZ correlation functions} \label{sec:app_W_BTZ}
The correlation functions in the BTZ black hole can be obtained from the correlation functions of $AdS_3$ using the method of images~\cite{smith1990gravitational,Carlip_1995}, as the BTZ solution is a quotient space thereof. We use the results of Refs.~\cite{Carlip_1995,Hodgkinson_2012,foo2022quantum}.
The correlation functions in the scalar vacuum of $AdS_3$ are given by  \begin{equation}
    \begin{split}
        W_{\mathrm{AdS}}\left(x, x^{\prime}\right)=\frac{1}{4 \pi l \sqrt{2}}\left[\frac{1}{\sqrt{\rho\left(x, x^{\prime}\right)}}-\frac{\zeta}{\sqrt{\rho\left(x, x^{\prime}\right)+2}}\right],
    \end{split}
\end{equation} with $\rho$ the squared geodesic distance in the 2+2 embedding space of $AdS_3$:
\begin{equation}
\begin{split}
    \rho\left(x, x^{\prime}\right)&=\frac{1}{2 l^2}\left[\left(X_1-X_1^{\prime}\right)^2-\left(T_1-T_1^{\prime}\right)^2+\left(X_2-X_2^{\prime}\right)^2-\left(T_2-T_2^{\prime}\right)^2\right].  
\end{split}
\end{equation} Furthermore, $\zeta \in\{-1,0,1\}$ specifies the boundary conditions at asymptotic infinity: Neumann, transparent or Dirichlet~\cite{carlip2003quantum,Lifschytz_1994,Shiraishi_1994}. We take transparent boundary conditions $\zeta=0$ for now, calculations for other boundary conditions are analogous. 
The BTZ correlation function is then given by \begin{equation}
\begin{split} \label{eq:BTZ_corr_1}
    W_{\mathrm{BTZ}}\left(x, x^{\prime}\right)&=\frac{1}{\sum_n \Upsilon^{2 n}} \sum_n \sum_m \Upsilon^n \Upsilon^m W_{\mathrm{AdS}}\left(\Gamma_M^n x, \Gamma_M^m x^{\prime}\right) \\ &= \sum_m \Upsilon^m W_{\mathrm{AdS}}\left(x, \Gamma_M^m x^{\prime}\right) ,    
\end{split}
\end{equation} with $\Gamma_M^n$ the identification action $\phi \rightarrow \phi+2 \pi n  \sqrt{M}$ and $\Upsilon=+1,-1$ for untwisted and twisted fields; we will take $\Upsilon=1$.
The Wightman functions are actually distributions, and are given by the taking the limit $\epsilon \rightarrow 0+$ of \cref{eq:BTZ_corr_1} with $\rho$ replaced by~\cite{foo2022quantum,preciado2024more}  
\begin{equation} \label{eq:BTZ_corr_ieps}
    \begin{aligned}
\rho_\epsilon\left({x}, \Gamma^n {x}^{\prime}\right) & =\frac{r r^{\prime}}{r_h^2} \cosh \left[\frac{r_h}{l}(\Delta \varphi-2 \pi n)\right]-1 
 -\frac{\sqrt{\left(r^2-r_h^2\right)\left(r^{\prime 2}-r_h^2\right)}}{r_h^2} \cosh \left(\frac{r_h}{l^2} \Delta t-\mathrm{i} \epsilon\right)
\end{aligned}
\end{equation} with $r_h=l\sqrt{M}$, with $\Delta \varphi, \Delta t$ the differences in angular and time coordinates of $x,x'$ whose radial components are denoted by $r,r'$ (with coordinates giving the metric in \cref{eq:BTZ_metric}). 

Using this expression for $\rho$ in the BTZ correlation function \cref{eq:BTZ_corr_1}, we obtain \begin{equation} \label{eq:BTZ_corr_app}
\begin{split}
    W_{BTZ}(x_1,x_2) &=  \frac{1}{4\pi l  \sqrt{2}}\sum_m \frac{1}{\sqrt{ \frac{R_1 R_2}{r_h^2} \cosh \left[(\theta - 2 \pi m)\frac{r_h}{ l }\right]-1 -\sqrt{\left(\frac{R_1^2}{r_h^2}-1\right)\left(\frac{R_2^2}{r_h^2}-1\right)} \cosh \left(\frac{r_h}{ l^2} \Delta t-\mathrm{i} \epsilon\right) }} \\
    &=   \frac{1}{4\pi l \sqrt{2}} \frac{1}{\sqrt[\leftroot{-2}\uproot{2}4]{\left(\frac{R_1^2}{r_h^2}-1\right)\left(\frac{R_2^2}{r_h^2}-1\right)}} \sum_m \frac{1}{\sqrt{ \alpha_m^{(12)} - \cosh \left(\frac{r_h}{ l^2} \Delta t-\mathrm{i} \epsilon\right) }}
    \\ & =: W^{12}_{BTZ}(t-t'),
\end{split}
\end{equation} with \begin{equation} \label{eq:alpha_m}
    \alpha_m^{(12)} = \frac{ \frac{R_1 R_2}{r_h^2} \cosh \left[(\theta - 2 \pi m)\frac{r_h}{ l}\right]-1}{\sqrt{\left(\frac{R_1^2}{r_h^2}-1\right)\left(\frac{R_2^2}{r_h^2}-1\right)}}.
\end{equation} Recall that, unless stated otherwise, we are working under transparent boundary conditions, see \cref{eq:BTZ_correlation}; for Neumann and Dirichlet boundary conditions we will have an additional term, giving   \begin{equation} \label{eq:W_BTZ_app_NeuDiri}
\begin{split}
    W_{BTZ}(x_1,x_2) = &  \frac{1}{4\pi l \sqrt{2}} \frac{1}{\sqrt[\leftroot{-2}\uproot{2}4]{\left(\frac{R_1^2}{r_h^2}-1\right)\left(\frac{R_2^2}{r_h^2}-1\right)}} \sum_m \bigg( \frac{1}{\sqrt{ \alpha_m^{(12)} - \cosh \left(\frac{r_h}{ l^2} \Delta t-\mathrm{i} \epsilon\right) }} - \frac{\zeta}{\sqrt{ \alpha_m^{'(12)} - \cosh \left(\frac{r_h}{ l^2} \Delta t-\mathrm{i} \epsilon\right) }}\bigg) 
\end{split}
\end{equation} with \begin{equation} \label{eq:alpha'_m}
    \alpha_m^{'(12)} = \frac{ \frac{R_1 R_2}{r_h^2} \cosh \left[(\theta - 2 \pi m)\frac{r_h}{ l}\right]+1}{\sqrt{\left(\frac{R_1^2}{r_h^2}-1\right)\left(\frac{R_2^2}{r_h^2}-1\right)}}.
\end{equation}

\section{Interaction picture calculation} \label{sec:app_interaction}
In this section we briefly sketch the calculation for the detector's transition amplitudes \cref{eq:single_detector_amplitude,eq:transition_amplitude_superp} in the interaction picture. The interaction picture (see for example \cite{birrell1984quantum,peskin2018introduction,tong2009university,Weinberg_1995}) is an intermediate picture between the Schrödinger representation, where states carry all time dependence but fields are time independent, and the Heisenberg representation, where operators carry all time dependence but states are time independent. 

More specifically, let the total Hamiltonian of the system be \begin{equation}
    \hat H = \hat H_{0,S}+ \hat H_\text{int}
\end{equation} where $\hat H_\text{int}$ is the interaction Hamiltonian between the detector and quantum fields in the Schrödinger picture; $\hat H_{0,S}$ is the free Hamiltonian in the Schrödinger picture.
In the interaction picture, operators evolve according to the free time evolution (without interaction, given by Hamiltonian $\hat H_{0,S}$) and states evolve according to the freely-time-evolved interaction Hamiltonian $\leftindex_I {\hat H}_\text{int}(t) = e^{i \hat H_{0,S} t} \hat H_\text{int} e^{-i\hat H_{0,S} t}$. Here the pre-subscript $I$ denotes the subscript for `interaction picture'. For simplicity of notation we will denote $\leftindex_I {\hat H}_\text{int}(t)$ simply by $\hat H_\text{int}(t)$ from now on as we will perform our calculations only in the interaction picture. 
States in the Schrödinger and interaction pictures are related by $\ket{\psi(t)}_I = e^{i\hat H_{0,S} t} \ket{\psi(t)}_S$.
Thus, in the interaction picture operators solve the equations of motion with respect to $\hat H_0$, and states evolve according to \begin{equation}
    i \frac{d}{dt} |\psi(t) \rangle_I =  {\hat H}_\text{int}(t) \ket{\psi(t)}_I
\end{equation} for time slicing $t$ defining the states $\ket{\psi(t)}_I$. Defining a unitary operator $U(t,t_0)$ such that \begin{equation} \label{eq:interaction_U}
    |\psi(t) \rangle_I = \hat U(t,t_0) \ket{\psi(t_0)}_I,
\end{equation} this unitary must satisfy 
\begin{equation}
     {\hat H}_\text{int}(t) \hat U(t,t_0) = i \frac{d}{dt} \hat U(t,t_0)
\end{equation} with initial condition $\hat U(t_0,t_0) = 1$. Simply exponentiating does not work because ${\hat H}_\text{int}$ is an operator, with ${\hat H}_\text{int}(t)$ not necessarily commuting at different times, and thus there is an ordering ambiguity. A formal solution of this equation can be written as the Dyson series \begin{equation}
    U(t,t_0) = T e^{-i  \int_{t_0}^t \hat H_\text{int}(t') dt'} 
\end{equation} with $T$ time ordering, which to second order in a perturbation expansion in the interaction is given by \begin{equation}
    \begin{split}
        {\hat U}(t_f,t_i)= \mathbb{1}-i  \int_{t_i}^{t_f} \mathrm{~d} t \hat H_{\text{int}}(t)- \int_{t_i}^{t_f} \mathrm{~d} t \int_{t_i}^t \mathrm{d} t^\prime \hat H_{\text{int}}(t) \hat H_{\text{int}}\left(t^{\prime}\right)+\mathcal{O}\left(\lambda^3\right)
    \end{split}
\end{equation} 
One can also think of the Dyson series as iteratively using the integral equation \begin{equation}
    \hat U\left(t_f, t_i\right)=\mathbb{1}-i \int_{t_i}^{t_f} d t' \hat H_\text{int}(t') \hat U\left(t', t_i\right).
\end{equation}

In our case, the free Hamiltonian governs the evolution of the (free) field $\phi$ and detector monopole moment $\hat \mu$, and the interaction Hamiltonian (with respect to proper time $\tau$) consisting of a coupling between the detector's monopole moment and the field along the detector's path, given by  \begin{equation}
    \hat H^\tau_{int} = \lambda \hat \mu \otimes \hat \phi(\mathbf{x})
\end{equation} in the Schrödinger picture, with $\mathbf {x}$ the position of the detector,
so in the interaction picture we have \begin{equation}
    \hat H^\tau_{int}(t) = \lambda \hat \mu(\tau) \otimes \hat \phi(x(\tau)).
\end{equation}
Let the total initial state be $\ket{\psi(\tau_i)}_S =|0\rangle_\phi |E_0\rangle_\mu$ of the scalar field vacuum and detector ground state (the detector's internal degree of freedom), and denote the final total state by $\ket{\psi(\tau_f)}_S$ in the Schrödinger picture. We take the detector to be held at a fixed spatial location. These states carry full time evolution dependence, being states in the Schrödinger picture. They are related to the states in the interaction picture by 
\begin{equation} \label{eq:interaction_pic_states1}
    \ket{\psi(\tau_i)}_I = e^{i \hat H^\tau_{0,S} \tau_i} \ket{\psi(\tau_i)}_S,
\end{equation} so that using \cref{eq:interaction_U} we have \begin{equation} \label{eq:interaction_pic_states2}
    \ket{\psi(\tau_f)}_S = e^{-i\hat H^\tau_{0,S} \tau_f} \hat U(t_f,t_i) e^{i\hat H^\tau_{0,S} \tau_i} \ket{\psi(\tau_i)}_S.
\end{equation} To first order in $\hat H_{\text{int}}$ we then have
\begin{equation} \label{eq:interaction_pic_states3}
    \begin{split} 
    \left|\psi\left(\tau_f\right)\right\rangle_S=e^{-i \hat H^\tau_{0,S} \tau_f} e^{i \hat H^\tau_{0,S} \tau_i}\left|\psi\left(\tau_i\right)\right\rangle_S-e^{-i \hat H^\tau_{0, S} \tau_f} i \int_{\tau_i}^{\tau_f} \mathrm{~d} \tau \hat H_{\mathrm{int}}(\tau) e^{i \hat H^\tau_{0,S} \tau_i}\left|\psi\left(\tau_i\right)\right\rangle_S.
    \end{split}
\end{equation} 
Then, writing out $H_\text{int}$ and post-selecting on the detector making a transition to energy level $E$ we obtain at time $\tau_f$ 
\begin{equation}
    \begin{split}
       - i \lambda e^{-i \Phi}\langle E | \int_{\tau_i}^{\tau_f} \mathrm{d}  \tau \mu(\tau) \hat \phi[x(\tau)] \left|0, E_0\right\rangle,
    \end{split}
\end{equation} with $\Phi$ a potential phase factor upon which we comment below, and coordinate times $t_i,t_f$ related to the detector's proper time by $t_i = \tau_i/\gamma_J, t_f = \tau_f/\gamma_J$ with redshift factor $\gamma_J$ given in \cref{eq:tau_to_t}. As time evolution for $\hat \mu(\tau)$ is given by $$\hat \mu(\tau) = e^{i\hat H_{0,\mu} \tau} \hat \mu e^{-i\hat H_{0,\mu}\tau}$$ with $\hat H_{0,\mu}|E_i\rangle = E_i \ket{E_i}$ we obtain for the transition amplitude at time $t_f$ \begin{equation}
    \begin{split}
       - i \lambda  e^{-i \Phi}  \langle E| \mu(0)  \left|E_0\right\rangle \int_{\tau_i}^{\tau_f} \mathrm{d} \tau \mathrm{e}^{\mathrm{i}\left(E-E_0\right) \tau}\langle\psi| \hat \phi(x)\left|0\right\rangle .
    \end{split}
\end{equation}
If we have a single path or geometry, this overall phase $\Phi$ is not important, but when we have a superposition of amplitudes a relative phase $\Delta \Phi$ might contribute; in our limited treatment we take the phase to be zero, or alternatively, we could absorb any potential phase in the definition of the states.

\vspace{0.5cm}

Next we consider a detector in a superposition of classical geometries with black hole at positions $\mathbf{x}_1,\mathbf{x}_2$ with total initial state in the detector frame given by $\ket{\psi(\tau_i)}_S =  \ket{\mathbf{0}}_D \left(\left|\mathbf{x}_1\right\rangle_B +\left|\mathbf{x}_2\right\rangle_B  \right) \ket{0}_\phi  |E_0\rangle_\mu / \sqrt{2}$ with $\ket{0}_\phi$ denoting the Hartle--Hawking scalar field vacuum.
In the case of a superposition of geometries (or detector location under a QRF transformation) one needs to be careful regarding coordinates, especially the choice of time parametrization used. 
In the detector QRF the proper time $\tau$ is single-valued and well-defined, but the black hole is in a superposition of locations so that the relation between proper time and coordinate times differ, whereas in the QRF with the black hole at the origin the coordinate time $t$ is well-defined but the proper time of the detector is indefinite. The relation between coordinate time $t$ and proper time $\tau$ is given by the time dilation of \cref{eq:tau_to_t}, $\tau = t \gamma(R)$ with $R$ the radial coordinate distance between the detector and black hole.

In the detector frame (i.e. putting the detector at the origin), the black hole is in a superposition of locations. Taking the interaction picture representation of the QRF-transformed UDW interaction \cref{eq:H_int_QRF_transformed} and post-selecting on the detector making a transition to energy level $E$ we obtain at time $\tau_f$ 
\begin{equation}  \label{eq:app_superp_detector_amplitude}
    \begin{split}
     \frac{1}{\sqrt{2}}\bigg( |\mathbf{x}_{1} \rangle_B  \Theta_1  + e^{i \Delta \Phi} |\mathbf{x}_{2}\rangle_B  \Theta_2  \bigg),
    \end{split}
\end{equation} where \begin{equation}
    \Theta_J =   -i \lambda \langle E |\hat \mu |E_0\rangle\int_{\tau_{i}}^{\tau_{f}} \mathrm{d}\tau  \eta(\tau) e^{i(E-E_0)\tau} \langle \vartheta| \hat \phi(x_J(\tau)) |0\rangle 
\end{equation} with $\phi$ the scalar field coupled to the metric $g$ with black hole at position $\mathbf{x}_J$ and $\ket{\vartheta}$ the final state of the scalar field which we will sum over. Note that if another clock were used to turn the detectors on and off, the proper time intervals for the UDW interaction could differ between the two branches.
We have omitted the detector's position state here as we are in the frame where this position is factorised as being the origin.

Within our simplified treatment, we do not model the dynamics of either the black hole or the external control degrees of freedom associated with the apparatus that maintains the detector on its prescribed trajectory. Consequently, no branch-dependent phase associated with these degrees of freedom appears explicitly.

In a more complete treatment, the relative phase $\Delta\Phi$ could receive contributions from several sources. One possible source is the evolution of the black hole itself. If its Hilbert space and state were included explicitly in the initial-state description, an approximately static BTZ black hole could be modeled as an approximate eigenstate of the gravitational Hamiltonian generating evolution with respect to the coordinate time $t$, with an eigenvalue determined by its ADM mass $M$~\cite{foo2022quantum}. Within this approximation, $ H_0^t\ket{M}_{BH}=M\ket{M}_{BH}$. The gravitational Hamiltonian would then contribute to $\hat H^\tau_{0,S}$ in \cref{eq:interaction_pic_states1,eq:interaction_pic_states2,eq:interaction_pic_states3}, with $ \hat H_{\mathrm{grav}}^\tau = \frac{dt}{d\tau}\hat H_{\mathrm{grav}}^t$. Because the relation between coordinate time and the detector's proper time differs between the two branches, this evolution could generate a relative phase when expressed in terms of $\tau$. Such a contribution would be particularly relevant for a black hole in a superposition of different masses, since the two branches would then acquire phases associated with different energy eigenvalues. In our setup, by contrast, the black hole has the same mass in both branches, which differ only in its position. We therefore introduce an explicit Hilbert space for the black hole's position but not for its mass.

A second possible contribution to $\Delta\Phi$ could arise from the external control degrees of freedom and from the apparatus that maintains the detector on its prescribed static trajectory. These dynamics are likewise neglected here. Any phases arising from the omitted degrees of freedom can be incorporated into the definition of the branch states, or equivalently into the phase convention adopted for the measurement basis below. Whether such an additional phase affects the experimental statistics ultimately depends on the precise implementation of the interferometric measurement of the black hole's position.

If we now consider a joint measurement of the detector transitioning to the excited state $\ket{E}_\mu$ at time $\tau_f$ and an interferometric measurement on the black hole position, i.e. measuring in the basis $(\ket{\mathbf{x}_{1}}\pm \ket{\mathbf{x}_{2}})_B/\sqrt{2}$, we obtain the post-selected detector response given in \cref{eq:probabilities}.

\section{Calculating the integrals for the measurement probabilities} \label{sec:app_integrals}

\subsection{Calculating the cross-term $\mathcal{F}_{12}(\Omega)$} 

We now insert the BTZ correlation function \cref{eq:BTZ_corr_app} (for transparent boundary conditions, calculations for other boundary conditions are analogous) into the cross-term for the probability \cref{eq:interference_term}, defining a response function $\mathcal{F}_{12} = P_{12}/\mathcal{N'}$ with $\mathcal{N'}=\lambda^2 |\langle E |\mu |E_0\rangle|^2$, obtaining\footnote{The calculations here resemble those of the Supplemental Material of Ref.~\cite{foo2022quantum} closely, but are slightly different as we consider a position-superposed BTZ black hole, as opposed to the mass-superposed case of Ref.~\cite{foo2022quantum}.} \begin{equation} \label{eq:cross_term_app_1}
    \begin{split}
         \mathcal{F}_{12}(\Omega) &= \int_{\tau_i}^{\tau_f} d\tau \int_{\tau_i}^{\tau_f} d\tau'  \eta(\tau) \eta(\tau') e^{-i\Omega(\tau-\tau')} W_{BTZ}(x_1(\tau),x_2(\tau')) \\
    \end{split}
\end{equation} where the paths $x_1(\tau)=(\tau/\gamma_1,R_1,0),x_2(\tau')=(\tau'/\gamma_2,R_2,\theta)$ with $\gamma_i=\gamma(R_i)$ and $\Omega = \Delta E= E-E_0$.
Now we move to coordinate time, obtaining
\begin{equation} \label{eq:cross_term_app_2}
    \begin{split}
         \mathcal{F}_{12}(\Omega) &= \gamma_1 \gamma_2 \int_{\tau_i/\gamma_1}^{\tau_f/\gamma_1} dt \int_{\tau_i/\gamma_2}^{\tau_f/\gamma_2} dt'  \eta(\gamma_1 t) \eta(\gamma_2 t') e^{-i\Omega(\gamma_1 t-\gamma_2 t')} W_{BTZ}(x_1(\tau),x_2(\tau')) \\
         &= \gamma_1 \gamma_2 \int_{-\tau_f/\gamma_1}^{\tau_f/\gamma_1} du \int_{-\tau_f/\gamma_2+u}^{\tau_f/\gamma_2+u} ds  \eta(\gamma_1 u) \eta(\gamma_2( u - s)) e^{-i\Omega \big(\gamma_1 u - \gamma_2(u- s)\big)} W_{BTZ}^{12}(s)
    \end{split}
\end{equation} where we have chosen $\tau_i=-\tau_f$ for simplicity, and changed variables $u=t$, $s = t-t'$ and $W^{12}_{BTZ}(t-t')$ is defined in \cref{eq:BTZ_corr_app}.
Next, we make the same approximation as in Appendix D (S31) of \cite{foo2022quantum}:
\begin{equation} \label{eq:approx_foo}
    \int_{-t_{f1}}^{t_{f1}} \mathrm{~d} u \int_{u-t_{f2}}^{u+t_{f2}} \mathrm{~d} s \eta\left(\gamma_D u\right) \eta\left(\gamma_D(u-s)\right) \ldots \simeq \int_{-t_{f1}}^{t_{f1}} \mathrm{~d} u \int_{-t_{f2}}^{t_{f2}} \mathrm{~d} s \eta\left(\gamma_D u\right) \eta\left(\gamma_D(u-s)\right) \ldots,
\end{equation} with $t_{fJ}=\tau_f/\gamma_J$. 
This approximation would be exact if the integration limits were $-\infty$ to $\infty$; recall that we required that $\Delta \tau \gg \sigma$ such that the switching functions have small support outside $[-\tau_f/\gamma_J,\tau_f/\gamma_J]$.
Using \cref{eq:approx_foo} in \cref{eq:cross_term_app_2}, we find \begin{equation} \label{eq:cross_term_app_3}
    \begin{split}
        \mathcal{F}_{12}(\Omega) &\approx \gamma_1 \gamma_2 \int_{-\tau_f/\gamma_1}^{\tau_f/\gamma_1} du \int_{-\tau_f/\gamma_2}^{\tau_f/\gamma_2} ds  \eta(\gamma_1 u) \eta(\gamma_2( u - s)) e^{-i\Omega \big(\gamma_1 u - \gamma_2(u- s)\big)} W_{BTZ}^{12}(s) \\
        &= \gamma_1 \gamma_2  \int_{-\tau_f/\gamma_2}^{\tau_f/\gamma_2} ds I(s) e^{-\frac{\gamma_2^2 s^2}{2\sigma^2}}  e^{-i\Omega \gamma_2 s} W_{BTZ}^{12}(s)
    \end{split}
\end{equation} where $I(s)$ is an integral over $u$: \begin{equation} \label{eq:u_integration}
    \begin{split}
        I(s) &= \int_{-\tau_f/\gamma_1}^{\tau_f/\gamma_1} du  e^{-\frac{\gamma_1^2 u^2}{2\sigma^2}} e^{-\frac{\gamma_2^2 u^2 - 2 \gamma_2^2 u s}{2\sigma^2}} e^{-i\Omega(\gamma_1-\gamma_2)u}  \\
        &=  \int_{-\tau_f/\gamma_1}^{\tau_f/\gamma_1} du e^{-\frac{\big( \sqrt{\gamma_1^2 + \gamma_2^2} u - \frac{\gamma_2^2}{\sqrt{\gamma_1^2+\gamma_2^2}}s + i \Omega \frac{(\gamma_1-\gamma_2)}{\sqrt{\gamma_1^2+\gamma_2^2}} \sigma^2  \big)^2}{2\sigma^2}} e^{\frac{\gamma_2^4 s^2}{2\sigma^2(\gamma_1^2+\gamma_2^2)}} 
        e^{-\Omega^2 \sigma^2 \frac{(\gamma_1 - \gamma_2)^2}{2(\gamma_1^2+\gamma_2^2)}} e^{i\Omega \frac{\gamma_2(\gamma_1-\gamma_2)s}{\gamma_1^2+\gamma_2^2}}      \\
        &= \frac{\sqrt{2}\sigma}{\sqrt{\gamma_1^2+\gamma_2^2}} e^{\frac{\gamma_2^4 s^2}{2\sigma^2(\gamma_1^2+\gamma_2^2)}} e^{-\Omega^2 \sigma^2 \frac{(\gamma_1 - \gamma_2)^2}{2(\gamma_1^2+\gamma_2^2)}} e^{-i\Omega \frac{\gamma_2^2(\gamma_1-\gamma_2)s}{\gamma_1^2+\gamma_2^2}}   \int^{\frac{\sqrt{\gamma_1^2+\gamma_2^2}\tau_f}{\sqrt{2}\sigma \gamma_1}-\frac{\gamma_2^2 s}{\sqrt{\gamma_1^2+\gamma_2^2}\sqrt{2}\sigma}+i\Omega\sigma \frac{\gamma_1-\gamma_2}{\sqrt{2}\sqrt{\gamma_1^2+\gamma_2^2}}}_{-\frac{\sqrt{\gamma_1^2+\gamma_2^2}\tau_f}{\sqrt{2}\sigma \gamma_1}-\frac{\gamma_2^2 s}{\sqrt{\gamma_1^2+\gamma_2^2}\sqrt{2}\sigma}+i\Omega\sigma \frac{\gamma_1-\gamma_2}{\sqrt{2}\sqrt{\gamma_1^2+\gamma_2^2}}} dz e^{-z^2}           \\
        &=: \frac{\sqrt{2}\sigma}{\sqrt{\gamma_1^2+\gamma_2^2}} e^{\frac{\gamma_2^4 s^2}{2\sigma^2(\gamma_1^2+\gamma_2^2)}} e^{-\Omega\sigma^2 \frac{(\gamma_1 - \gamma_2)^2}{2(\gamma_1^2+\gamma_2^2)}} e^{-i\Omega \frac{\gamma_2^2(\gamma_1-\gamma_2)s}{\gamma_1^2+\gamma_2^2}}  \frac{\sqrt{\pi}}{2} H(s).
    \end{split}
\end{equation} Here we changed variables $z= u\frac{\sqrt{\gamma_1^2+\gamma_2^2}}{\sqrt{2}\sigma}+...$ and $H(s)$ is given by \begin{equation} \label{eq:app_H(s)}
    \begin{split}
        H(s) &= \text{erf}\bigg( \frac{(\gamma_1^2+\gamma_2^2)\tau_f - \gamma_1 \gamma_2^2 s + i\Omega \sigma^2 \gamma_1 (\gamma_1-\gamma_2) }{\sqrt{2}\sigma  \gamma_1 \sqrt{\gamma_1^2+\gamma_2^2}}\bigg) +  \text{erf}\bigg( \frac{(\gamma_1^2+\gamma_2^2)\tau_f + \gamma_1 \gamma_2^2 s - i\Omega \sigma^2 \gamma_1 (\gamma_1-\gamma_2) }{\sqrt{2}\sigma  \gamma_1 \sqrt{\gamma_1^2+\gamma_2^2}}\bigg)
    \end{split}
\end{equation} with $\text{erf}$ the Gaussian error function.
Using \cref{eq:u_integration} in \cref{eq:cross_term_app_3} we find \begin{equation} \label{eq:cross_term_app_3.2}
    \begin{split}
         \mathcal{F}_{12}(\Omega) &\approx \frac{1}{4\pi l \sqrt{2}} \frac{\sqrt{M}}{\sqrt{\gamma_1 \gamma_2}}  \frac{\gamma_1 \gamma_2 \sigma}{\sqrt{\gamma_1^2+\gamma_2^2}} \sqrt{\frac{\pi}{2}} e^{-\Omega^2 \sigma^2\frac{(\gamma_1-\gamma_2)^2}{2(\gamma_1^2+\gamma_2^2)}} \sum_n \int^{\tau_f/\gamma_2}_{-\tau_f/\gamma_2} ds \frac{ e^{-\frac{\gamma_1^2\gamma_2^2}{2\sigma^2(\gamma_1^2+\gamma_2^2)} s^2} e^{-iEs \frac{\gamma_1 \gamma_2 (\gamma_1+\gamma_2)}{\gamma_1^2+\gamma_2^2}}  H(s)}{\sqrt{\alpha_n^{(12)} - \cosh(\frac{r_h}{l^2} s - i\epsilon)}}
         \\ 
         &= \sigma \frac{ \sqrt{M}}{2l} Y_0 \sum_n \int^{\tau_f/\gamma_2}_{-\tau_f/\gamma_2} ds \frac{Z_0(s)H(s)}{\sqrt{\alpha_n^{(12)}-\cosh(\frac{r_h}{l^2} s - i\epsilon)}}
          \\ 
        &= \sigma {Y_0} \sum_n \text{ Re}  \int_{0}^{\frac{\tau_f \sqrt{M}}{ l\gamma_2}} dz \frac{Z_0( l z/\sqrt{M})H( l z/\sqrt{M})}{\sqrt{\alpha_n^{(12)}-\cosh( z - i\epsilon)}} 
    \end{split}
\end{equation} with $H(s)$ given in \cref{eq:app_H(s)}, $\alpha^{12}_n$ given in \cref{eq:alpha_m} and $Y_0$ and $Z_0(s)$ given by \begin{align} 
 Y_0 &=     \frac{\sqrt{\gamma_1 \gamma_2} \sqrt{\pi}}{4\pi \sqrt{\gamma_1^2+\gamma_2^2}} e^{-\Omega^2 \sigma^2\frac{(\gamma_1-\gamma_2)^2}{2(\gamma_1^2+\gamma_2^2)}}, \label{eq:Y_0_app} \\
 Z_0(s) &=  e^{-\frac{\gamma_1^2\gamma_2^2}{2\sigma^2(\gamma_1^2+\gamma_2^2)} s^2} e^{-i\Omega s \frac{\gamma_1 \gamma_2 (\gamma_1+\gamma_2)}{\gamma_1^2+\gamma_2^2}}. \label{eq:Z_0_app} 
\end{align} 
The cross-term will typically also contain singularities, but integrable ones.

For Neumann and Dirichlet boundary conditions, we have an extra term, namely \begin{equation} \label{eq:cross-term_NeuDiri}
    \begin{split}
          \mathcal{F}_{12}(\Omega) &= \sigma Y_0 \sum_n \text{ Re}  \int^{\frac{\tau_f \sqrt{M}}{ l\gamma_2}}_0 dz Z_0( l z/\sqrt{M})H( l z/\sqrt{M}) \bigg( \frac{1}{\sqrt{\alpha_n^{(12)}-\cosh( z - i\epsilon)}}\pm \frac{1}{\sqrt{\alpha'^{12}_n-\cosh( z - i\epsilon)}} \bigg)
    \end{split}
\end{equation} with $+,-$ for Neumann and Dirichlet boundary conditions, respectively.

\vspace{0.5cm}

\subsection{Calculating $\mathcal{F}_J(E)$} 

The integrals for the diagonal terms $\mathcal{F}_J(E)$ are exactly those of \cite{foo2022quantum}, we briefly repeat them here as slightly different normalizations were used; they are analogous to calculation of $\mathcal{F}_{12}(\Omega)$ just presented. We omit the label $J=1,2$ here as we are now in the case of a single non-superposed black hole and localized detector.

The response function is given by \begin{equation} \label{eq:appendix_response_single}
     \mathcal{F}(\Omega) = \int_{\tau_i}^{\tau_f} d\tau \int_{\tau_i}^{\tau_f} d\tau' \eta(\tau) \eta(\tau') e^{-i\Omega(\tau-\tau')}  W_{BTZ}(x(\tau), x(\tau'))  . 
\end{equation}
We first move to coordinate time, obtaining 
\begin{equation}
    \begin{split}
        \mathcal{F}(\Omega) =& \gamma^2 \int_{\tau_i/\gamma}^{\tau_f/\gamma} dt \int_{\tau_i/\gamma}^{\tau_f/\gamma}  dt' \eta(\gamma t) \eta(\gamma t') e^{-i\Omega\gamma(t-t')} W_{BTZ}(x(\gamma t), x(\gamma t'))  \\ &=  \gamma^2 \int_{\tau_i/\gamma}^{\tau_f/\gamma} du \int_{\tau_i/\gamma+u}^{\tau_f/\gamma+u}  ds \eta(\gamma u) \eta(\gamma (u-s)) e^{-i\Omega\gamma s} W_{BTZ}(s)
    \end{split}
\end{equation} with $\gamma = \gamma(R) = \sqrt{-g_{00}}$, we chose $\tau_i = -\tau_f$, and we have performed the change of variables $u=t, s= t-t'$.
Next, we require $\Delta \tau >> \sigma$ such that we can use the approximation of \Cref{eq:approx_foo}, taking $\tau_i=-\tau_f$ giving 
\begin{equation}
\begin{split}
        \mathcal{F}(\Omega) &\approx  \gamma^2 \int_{-\tau_f/\gamma}^{\tau_f/\gamma} du \int_{-\tau_f/\gamma}^{\tau_f/\gamma}  ds \eta(\gamma u) \eta(\gamma (u-s)) e^{-i\Omega\gamma s} W_{BTZ}(s) \\ &= \gamma^2 \int_{-\tau_f/\gamma}^{\tau_f/\gamma}  ds J(s) e^{-\gamma^2 \frac{s^2}{4\sigma^2}} e^{-i\Omega\gamma s} W_{BTZ}(s) 
\end{split}
\end{equation}  with  \begin{equation}
    \begin{split}
        J(s) &= \int_{-\tau_f/\gamma}^{\tau_f/\gamma} du e^{-\gamma^2\frac{(u-1/2 s)^2}{\sigma^2}} \\ &= \frac{\sigma \sqrt{\pi}}{ 2 \gamma } \bigg( \erf{\frac{\tau_f-\gamma  s /2 }{\sigma}} + \erf{\frac{ \tau_f+\gamma s /2}{\sigma}} \bigg) \\ &= \frac{\sigma \sqrt{\pi}}{ 2 \gamma } H_0(s). 
    \end{split}
\end{equation} Here we defined \begin{equation}
    \begin{split}
        H_0(s) &= \erf{\frac{\tau_f-\gamma s/2 }{\sigma}} + \erf{\frac{ \tau_f+\gamma s /2}{\sigma}}.
    \end{split}
\end{equation} If we take the integration limits $\tau_f$ to infinity we can approximate this integral $J(s)$ simply by $\sigma \sqrt{\pi}/\gamma$.

We thus find that 
\begin{equation}
    \begin{split} \label{eq:F_E_app_us}
        \mathcal{F}(\Omega) &\approx  \frac{\gamma \sigma \sqrt{\pi}}{2} \int_{-\tau_f/\gamma}^{\tau_f/\gamma}  ds H_0(s) e^{-\gamma^2 \frac{s^2}{4\sigma^2}} e^{-i\Omega\gamma s} W_{BTZ}(s) \\
        &=  \frac{\gamma \sigma \sqrt{\pi}}{2} \frac{\sqrt{M}}{4\pi \gamma l \sqrt{2}} \sum_m \int_{-\tau_f/\gamma}^{\tau_f/\gamma}  ds \frac{ H_0(s) e^{-\gamma^2 \frac{s^2}{4\sigma^2}} e^{-i\Omega\gamma s}}{\sqrt{\alpha_m-\cosh(\frac{r_h}{l^2}s - i\epsilon) }}
        \\ &= \frac{\sigma \sqrt{M}}{8 l \sqrt{2\pi}} \sum_m \int_{-\tau_f/\gamma}^{\tau_f/\gamma}  ds \frac{ H_0(s) e^{-\gamma^2 \frac{s^2}{4\sigma^2}} e^{-i\Omega\gamma s}}{\sqrt{\alpha_m-\cosh(\frac{r_h}{l^2}s - i\epsilon) }} 
    \end{split}
\end{equation} where we have inserted the BTZ correlation function of \Cref{eq:BTZ_corr_app} and $\alpha_m$ is given by 
\begin{equation}
    \alpha_m =  \frac{R^2}{\gamma^2 l^2} \cosh(2\pi m \sqrt{M}) -\frac{M}{\gamma^2} .
\end{equation}
Note that $Y_0,Z_0( l z)$ are the same as for the cross term for the superposed black hole mass case of Ref.~\cite{foo2022quantum}, but $H( l z)$ and $\alpha^{12}_n$ are different.

\subsection{Relating to the coordinates of Ref.~\cite{foo2022quantum} and final expressions} \label{sec:foo_time_rescaling} 
The calculations in Ref.~\cite{foo2022quantum} were done in a rescaled AdS time coordinate.
Namely, Ref.~\cite{foo2022quantum} work with the time coordinates $\overline{t}$ of $AdS_3$, of which the BTZ black hole is obtained by a periodic identification.
This means that our coordinates $t,R$ are related to the coordinate $\overline{t},\overline{R}$ of 3d AdS--Rindler space of Ref.~\cite{foo2022quantum} by 
\begin{equation}
    \begin{split}
        \sqrt{M} t = \overline{t}, \\
        R/\sqrt{M} = \overline{R}.
    \end{split}
\end{equation} 
Our coordinates $t,R$ are referred to  as $\tilde{s},\tilde{R}$ in the Supplemental Materials of Ref.~\cite{foo2022quantum}, but the tildes are omitted there in the Supplemental Sections D and E.
Ref.~\cite{foo2022quantum} also defines a rescaled redshift factor which we refer to as $\widetilde{\gamma}$, defined by \begin{equation}
    \widetilde{\gamma}(R) = \sqrt{\frac{R^2}{ M l^2}-1} = \sqrt{\frac{\bar{R}^2}{ l^2}-1}, 
\end{equation} and thus related to our BTZ redshift factor $\gamma$ of \cref{eq:tau_to_t} by
\begin{equation}
     \gamma(R) = \widetilde{\gamma}(R) \sqrt{M}.
\end{equation}
Note that Ref.~\cite{foo2022quantum}  still uses the BTZ radial coordinate. Furthermore, because Ref.~\cite{foo2022quantum} rescaled the times by $\sqrt{M}$, the time span in $\tilde{t_f}$ is equal to the time span in coordinate $t$ times $\sqrt{M}$, so that our definition of the phase being $M \cdot t$ is dimensionally consistent with their definition of the phase being $\sqrt{M} \cdot \overline{t}$. 

We can put our results for $\mathcal{F}_{12}(\Omega)$ and $\mathcal{F}(\Omega)$ in terms of the coordinates and conventions of Ref.~\cite{foo2022quantum} as follows. In the remaining part of this section, $t,\gamma$ should be understood as $\overline{t},\widetilde{\gamma}$, finding 
\begin{equation} \label{eq:F12_us_app_rescaled}
    \begin{split}
        \frac{\mathcal{F}_{12}(\Omega)}{\sigma} \approx Y_0 \sum_n \text{ Re}  \int^{\frac{\tau_f}{ l\widetilde{\gamma}_2}}_{0} dz \frac{Z_0( l z )H( l z)}{\sqrt{\alpha_n-\cosh( z - i\epsilon)}}
    \end{split}
\end{equation} with 
\begin{align} 
\alpha^{12}_n &= \frac{R_1 R_2}{\widetilde{\gamma}_1 \widetilde{\gamma}_2 M l^2} \cosh (2 \pi n \sqrt{M})-\frac{1}{\widetilde{\gamma}_1 \widetilde{\gamma}_2}  \\
 Y_0 &=     \frac{\sqrt{\widetilde{\gamma}_1 \widetilde{\gamma}_2} \sqrt{\pi}}{4\pi \sqrt{\widetilde{\gamma}_1^2+\widetilde{\gamma}_2^2}} e^{-\Omega^2 \sigma^2\frac{(\widetilde{\gamma}_1-\widetilde{\gamma}_2)^2}{2(\widetilde{\gamma}_1^2+\widetilde{\gamma}_2^2)}}, \label{eq:Y_0_app2} \\
 Z_0(s) &=  e^{-\frac{\widetilde{\gamma}_1^2\widetilde{\gamma}_2^2}{2\sigma^2(\widetilde{\gamma}_1^2+\widetilde{\gamma}_2^2)} s^2} e^{-i \Omega s \frac{\widetilde{\gamma}_1 \widetilde{\gamma}_2 (\widetilde{\gamma}_1+\widetilde{\gamma}_2)}{\widetilde{\gamma}_1^2+\widetilde{\gamma}_2^2}}, \label{eq:Z_0_app2} \\
        H(s) &= \text{erf}\bigg( \frac{(\widetilde{\gamma}_1^2+\widetilde{\gamma}_2^2)\tau_f - \widetilde{\gamma}_1 \widetilde{\gamma}_2^2 s + i\Omega \sigma^2 \widetilde{\gamma}_1 (\widetilde{\gamma}_1-\widetilde{\gamma}_2) }{\sqrt{2}\sigma  \widetilde{\gamma}_1 \sqrt{\widetilde{\gamma}_1^2+\widetilde{\gamma}_2^2}}\bigg) +  \text{erf}\bigg( \frac{(\widetilde{\gamma}_1^2+\widetilde{\gamma}_2^2)\tau_f + \widetilde{\gamma}_1 \widetilde{\gamma}_2^2 s - i\Omega \sigma^2 \widetilde{\gamma}_1 (\widetilde{\gamma}_1-\widetilde{\gamma}_2) }{\sqrt{2}\sigma  \widetilde{\gamma}_1 \sqrt{\widetilde{\gamma}_1^2+\widetilde{\gamma}_2^2}}\bigg)
\end{align}  and $\gamma_i(R) = \sqrt{R_i^2/(M l^2) -1}$ for $i=1,2$.
If $\arccosh{\alpha_n} < \tau_f/(l \widetilde{\gamma}_2)$, we need to make sure to take the correct branch of the square root, i.e. 
\begin{equation} \label{eq:F12_us_app_rescaled_branch_squareroot}
    \begin{split}
        \frac{\mathcal{F}_{12}(\Omega)}{\sigma} \approx Y_0 \sum_n \text{ Re}  \int^{\arccosh(\alpha_n)}_{0} dz \frac{Z_0( l z )H( l z)}{\sqrt{\alpha_n-\cosh( z - i\epsilon)}} +  Y_0 \sum_n \text{ Re}  \int_{\arccosh(\alpha_n)}^{\tau_f/(l \widetilde{\gamma}_2)} dz \frac{Z_0( l z )H( l z)}{i \sqrt{\cosh( z - i\epsilon)-\alpha_n}}
    \end{split}
\end{equation}

For $\mathcal{F}(\Omega)$ from \Cref{eq:F_E_app_us}, we take a single branch with detector at radial coordinate distance $R$ from the black hole (so $R=R_1,R_2$ for the two branches of our position-superposition), and $\widetilde{\gamma}$ is the related (rescaled) redshift, so $\widetilde{\gamma} = \widetilde{\gamma}_1,\widetilde{\gamma}_2$ for the two branches. Using the rescaled coordinates of Ref.~\cite{foo2022quantum} as explained above, we obtain
\begin{equation} \label{eq:F_E_Foo}
    \begin{split}
        \mathcal{F}(\Omega) \approx  \frac{\sigma}{8 l \sqrt{2\pi}} \sum_m \int_{-\tau_f /\widetilde{\gamma}}^{\tau_f  /\widetilde{\gamma}}  ds \frac{ H_0(s) e^{-\widetilde{\gamma}^2 \frac{s^2}{4\sigma^2}} e^{-i\Omega\widetilde{\gamma} s}}{\sqrt{\alpha_m-\cosh(s/l - i\epsilon) }},
    \end{split}
\end{equation} which, as shown in Ref.~\cite{foo2022quantum} leads to the following 
\begin{equation} \label{eq:foo_FE_single}
    \frac{\mathcal{F}(\Omega)}{\sigma} \approx \frac{\sqrt{\pi} H_0(0)}{8}
    -\frac{i}{8 \sqrt{\pi}} \mathrm{PV} \int_{-\tau_f / (2 l \widetilde{\gamma})}^{\tau_f / (2 l \widetilde{\gamma})} \mathrm{~d} z \frac{X_0(2 l z) H_0(2 l z)}{\sinh (z)}
    +\frac{1}{4 \sqrt{2 \pi}} \sum_{n \neq 0} \operatorname{Re} \int_{0}^{\tau_f / (l\widetilde{\gamma})} \mathrm{~d} z \frac{X_0(l z) H_0(l z)}{\sqrt{\alpha_{n}-\cosh (z)}}.
\end{equation} with 
\begin{align}
    X_0(s)& = e^{-\frac{\widetilde{\gamma}^2 s^2}{4 \sigma^2}} e^{-i \Omega \widetilde{\gamma} s}, \\
    \alpha_n &= \frac{R^2}{M \widetilde{\gamma}^2 l^2} \cosh (2 \pi n \sqrt{M})-\frac{1}{\widetilde{\gamma}^2}, \label{eq:alpha_n_foo_corrected} \\
    H_0(s) &= \operatorname{erf}\left[\frac{\widetilde{\gamma} \left(s+2  \tau_f/\widetilde{\gamma} \right)}{2 \sigma}\right]-\operatorname{erf}\left[\frac{\widetilde{\gamma} \left(s-2 \tau_f / \widetilde{\gamma} \right)}{2 \sigma}\right].
\end{align} The first two terms of \Cref{eq:foo_FE_single} arise from a pole at $n=0$ using the Sokhotski formula 
\begin{equation}
    \frac{1}{\sinh (z-i \epsilon)}=i \pi \delta(z)+\mathrm{PV} \frac{1}{\sinh (z)}.
\end{equation}

Here we have corrected a small inconsistency in the Supplemental Material of Ref.~\cite{foo2022quantum}: in Eq. (S51) of Ref.~\cite{foo2022quantum}, the first term in $\beta_{nm}$ should be divided by $M$, i.e. the expression of $\beta_{nm}$ should be \begin{equation}
    \beta_{nm} = \frac{R_D^2}{{M}\widetilde{\gamma}_D^2 l^2 } \cosh(2\pi(n-m)\sqrt{M}) - \frac{1}{\widetilde{\gamma}_D^2}
\end{equation} where we have inserted the missing term $M$. The expression $\beta_{n0}$ corresponds to our $\alpha_n$ in \cref{eq:alpha_n_foo_corrected}.

Similarly to \cref{eq:F12_us_app_rescaled_branch_squareroot}, if the term inside the square root in the denominator of the third term in \cref{eq:F_E_Foo} becomes negative, taking care of the square root leads to 
\begin{equation} \label{eq:third_term_F_E_Foo_squareroot}
    \begin{split}
        \frac{1}{4 \sqrt{2 \pi}} \sum_{n \neq 0} \operatorname{Re} \int_{0}^{\arccosh \alpha_n} \mathrm{~d} z \frac{X_0(l z) H_0(l z)}{\sqrt{\alpha_{n}-\cosh (z)}} + \frac{1}{4 \sqrt{2 \pi}} \sum_{n \neq 0} \operatorname{Re} \int_{\arccosh \alpha_n}^{\tau_f / (l\widetilde{\gamma})} \mathrm{~d} z \frac{X_0(l z) H_0(l z)}{i \sqrt{\cosh (z)-\alpha_n}}.
    \end{split}
\end{equation}

\section{A treatment of the position-superposition using reference fields} \label{app:new_reference_fields}

This section gives an alternative description of a quantum reference frame change using physical (coordinate) reference fields. 

In \Cref{sec:setup} we described the set-up of a detector-black hole system where the relative coordinate distance between the detector and black hole is in a superposition. 
To give an operational meaning to such a position superposition, we introduce coordinate reference fields~\cite{kabel2024identification,chen2026quantumreferencefieldstransformations,hoehn2023matterrelativequantumhypersurfaces,thiemann2026quantumreferenceframesrelational}. 
This can be seen as a more precise use of quantum reference frames as outlined in the previous section, and clarifies the meaning of the coordinate operator $\mathbf{\hat{x}}$ in the Unruh--DeWitt interaction and its QRF transform. 
While we restrict to our specific set-up of two quantum reference fields restricted to specific sets of states, a quantum-controlled detector worldline, a fixed metric and conformally coupled scalar field and no backreaction, seeking a QRF transformation that works in this limited setting, a more complete quantum treatment of QRFs and their transformations is an active topic of research in literature; we comment on what a more complete treatment may involve, and refer the reader to Refs.~\cite{chen2026quantumreferencefieldstransformations,aguilargutierrez2026relationalpathintegraleffective,thiemann2026quantumreferenceframesrelational} for work in this direction. We comment upon this in the paragraph \textit{More complete treatment} below.

\vspace{0.5cm}

\textit{QRFs of reference fields.} In 2+1 dimensions, a classical frame of reference fields can be given by a locally invertible map $X^A: U \subset M \rightarrow X(U) \subset \mathbb{R}^3$ of three reference fields $X^0,X^1,X^2$ that we take to be three scalar fields. 
In a complete quantum treatment of reference fields, a quantum coordinate reference frame consists of quantum scalar fields,
restricted to a subspace formally spanned by field eigenstates with eigenfunctions that define locally valid coordinates. 
In our limited treatment, we will restrict the fields to states $\ket{\Psi_X}$ sharply peaked around a classical field configuration $X^A$ with negligible fluctuation $\delta \hat{X}^A = \hat{X}^A - X^A$. We will nevertheless allow for these quantum reference fields to be in states sharply peaked around a superposition of classical field configurations, i.e. $\ket{\Psi_{X,1}}+\ket{\Psi_{X,2}}$.
We will neglect the backreaction of the reference fields on all other systems including the black hole spacetime, and will assume factorization of the different systems and fields involved.\footnote{For a treatment in linearized gravity that does not neglect backreaction see Ref.~\cite{chen2026quantumreferencefieldstransformations}.}

A reference frame can be used to dress fields and observables into gauge-invariant operators (i.e. invariant under diffeomorphisms on $M$)~\cite{goeller2022diffeomorphisminvariantobservablesdynamicalframes,chen2026quantumreferencefieldstransformations,Loveridge:2016tnh,Dittrich_2007,Tambornino_2012}. 
For instance, given a spacetime scalar $F$, its dressing by $X$ has value $F_X(\xi^A)$ at the physical coordinate reading $X^A = \xi^A$ given by $F_X(\xi^A) = F(X^{-1}(\xi^A))$. 
Note that $F_X$ is indeed invariant under diffeomorphisms on $M$, as both $F$ and $X$ transform.
In the case of a quantum reference frame $\hat{X}^A$, such dressing can be defined formally through its eigenfunctions $\ket{X}_X$, $F_{\hat{X}}(\xi) \ket{X}_X = F(\hat{X}^{-1}(\xi)) \ket{X}_X = F({X}^{-1}(\xi)) \ket{X}$. One can define such eigenfunctions and eigenvectors using an equal-time hypersurface in an auxiliary foliation~\cite{chen2026quantumreferencefieldstransformations}. In the case of quantum gravity with superposed metrics, for instance, such a foliation and timelike-spacelike split could be defined branch per branch. The inverse $\hat{X}^{-1}$ is thus defined branchwise, applying the inverse of the classical field configuration, $X^{-1}$, in each spectral branch.
A similar construction holds for spacetime tensors, with the $X$-dressing obtained by pulling back the tensor along $X^{-1}$; for instance, $(g_X)_\xi(u,v) = (X^{-1})^* g_\xi(u,v)$. 
In the following we will consider a single metric $g$ on $M$, with two different (quantum) dressings.

\textit{The detector worldline.} The detector's path $z: I \subset \mathbb{R} \rightarrow M: \tau \rightarrow z(\tau)$ will be presented in a specific reference frame by specifying $z^{(X)} := X \circ z: I \subset \mathbb{R} \rightarrow \mathbb{R}^3$.
We will allow $z^{(X)}$ to be a quantum operator $\hat{z}^{(X)}$ with, formally, eigenfunctions denoting detector worldlines $z^{(X)}$. In particular, we will take the radial coordinate position of the detector $z_r$ to be an operator, and consider detector paths that remain at fixed values of the spatial parts $X^1,X^2$ of the reference frame.
We will take the detector path to be externally fixed (but potentially in a quantum superposed semiclassical state), while ignoring all dynamics related to this external force and the detector worldline $\hat{z}_D^{(X)}$ (see Ref.~\cite{gale2023relativistic,wood2022quantized,sudhir2021unruh} for treatments taking into account the dynamics of a detector with quantized center of mass). Therefore, it can be seen as an example of a quantum-controlled detector path.

\vspace{0.5cm}

We will consider reference frames $X_B$, associated with the black hole, and $Y_{D,i}$, associated with the detector in branch $i$, and construct a quantum reference frame transformation between them, which can be seen as the equivalent of \cref{eq:QRF_transformation_field}.

\vspace{0.5cm}

\textit{QRF associated with the black hole.} The first reference frame $X_B^0,X_B^1,X_B^2$ that we consider is one associated with the BTZ black hole exterior, and in the coordinate system $t,r,\varphi$ of \cref{eq:BTZ_metric}, we simply take to be \begin{equation}
    X_B^0(x) = t(x), \quad X_B^1(x) = r(x), \quad X_B^2(x) = \varphi(x).
\end{equation} This can be seen as an eigenfunction in the formal eigenstate $\ket{X}$ of the operator $\hat{X}$.
In this treatment of reference fields, we do not require our reference fields to satisfy a prescribed equation of motion (and thus they can be considered `idealized' or `test' reference fields). If, instead, for instance, we wanted our reference fields to be massless scalar fields on the BTZ spacetime satisfying $\Box X_B^A = 0$, we would have to change $X_B^1$, for example taking $X_B^1(x) = C_1 \ln |1- \frac{r_h^2}{r(x)^2}| + C_2$, for some constants $C_1,C_2$.

Our choice of reference frame $X$ means the metric in $X$-coordinates (the $X$-dressed metric $g_X$) is given by \begin{equation}
     \mathrm{d} s^2=-f(r) \, \mathrm{d} t^2+f(r)^{-1} \mathrm{~d} r^2+r^2 \mathrm{~d} \varphi^2,
\end{equation} with $f(r)=r^2/l^2-M$.

\textit{Specifying the detector in the $X$-frame.} 
To specify the detector location, we start in the $X$-reference frame, and establish the detector location through the reference frame $X$, considering $z^{(X)}(\tau) = X \circ z (\tau)$.
We will allow our detector location to be in a superposition of radial locations $X^1 = r_{D,1}$ and $X^1 = r_{D,2}$. Therefore, the radial component of $z^{(X)}$ is promoted to an operator denoted $\hat{z}_{r}(\tau)$. We will write its value in branch $i$ as $r_{D,i}$.
Formally, $\hat{z}_r$ is an operator acting on some control system that controls the path of the detector (keeps it fixed at some quantum radial location), and so acts on the Hilbert space of a particle on the coordinate line $r$, i.e. $\mathcal{H} = L_2(\mathbb{R},dr)$ (or more precisely, taking an open part $I \subset \mathbb{R}$, $L_2(I,dr)$).

The relational radial coordinate position between the detector (held fixed at certain radial and angular coordinates) and black hole can, for instance, be given meaning through the redshift factor on the detector worldline $z_i(\tau)$, \begin{equation}
    \frac{d\tau}{d X^0(z_i(\tau))} = \gamma(R_i),
\end{equation} which is a function of the radial coordinate on that worldline. 
Other ways to give it an operational meaning include using signals sent from a clock at the boundary.

\textit{Witnessing branch coherence.} To certify that the detector is in a superposition of worldlines, or under a QRF transformation, the black hole in a superposition of radial locations as seen from a frame associated with the detector, one possible way to witness this coherence is to recombine the branches and perform a measurement that erases which-branch information, similarly to the $\ket{\pm}$ measurement of \Cref{sec:new_results}, or, more generally, perform a measurement sensitive to the off-diagonal coherence. 
In fact, the interference of \Cref{sec:new_results} can be viewed as constituting such superposition witness as it is absent in the case of the corresponding incoherent classical mixture.
Operationally, the measurement and initial preparation themselves may require additional control systems. 
If these require additional quantum reference systems but the corresponding operations employ idealized absolute quantities, the localizability of the relevant reference quantities (represented in general by POVMs) and the localization of the reference system's state determine how accurately the idealized measurement and state preparation can be implemented~\cite{Loveridge_2018,Loveridge:2016tnh,Busch:2016rui}. 
Additionally, repeated use of the same finite-resource reference system may degrade its state and consequently its suitability as a (good) reference~\cite{Bartlett_2006,Bartlett_2007}. 
We will not go deeper into these issues here.

\vspace{0.5cm}

\textit{QRF associated with the detector.} The second reference frame $Y_D^0,Y_D^1,Y_D^2$ that we consider is a reference frame associated with the detector, meaning it is well-localized with respect to the detector, taking in branch $i$ \begin{equation}
    Y_{D,i}^0(x) = t(x), \quad Y_{D,i}^1(x) = r(x)-r_{D,i}, \quad Y_{D,i}^2(x) = \varphi(x),
\end{equation} with $t,r,\phi$ the BTZ coordinates of \cref{eq:BTZ_metric}, with detector at $Y_{D,i}^1(x) = 0, \varphi(x) = 0$. Note that the condition $Y_{D,i}^1(x) = 0$ means that $r(x) = r_{D,i}$ with $r_{D,i}$ the (static) radial location of the detector in the $t,r,\phi$-coordinates in branch $i$ (recall that the detector is held fixed at some fixed radial distance in each branch). The coordinates $Y_{D,i}$ can be thought of as eigenfunctions of an operator $\hat Y$, restricted to the `semiclassical' subspace generated by superpositions of eigenvectors corresponding to $Y_{D,i}$.

Note that we could as well have adopted the choice to instead consider the first reference field $Y_{D,i}^0$ (timelike in the black hole exterior that we are probing) to be a reference field that reduces to the detector's proper time at the detector's location (and is appropriately extended away from it to the whole open region considered), i.e. $Y_{D,i}^0(x) = \tau(x)$ with $\tau(x)|_{\text{detector location}}=\tau$. Instead, we have opted to implement the additional redshift factor from relating the time coordinate and proper time of the Hamiltonians in each frame as an additional step, as presented in \Cref{sec:QRFS}. Had we chosen this detector's proper time coordinate extension as the reference field $Y_{D,i}^0$, this branch-dependent redshift factor would result from the QRF transformation between these two frames of reference fields $X_B$ and $Y_{D,i}$ in each branch outlined below.

Note that the two branches are not related by a diffeomorphism because the relational distance $r_{D,i}=X^1_B( z_i)$ in the two branches is different. 

\vspace{0.5cm}

\textit{Transforming between QRFs of reference fields.} The classical reference frame transformation between these two frames in branch $i$ is determined by $X_B \circ Y_{D,i}^{-1}$, loosely stated, determined by a change of reference coordinate fields given by $r \rightarrow r+r_{D,i}$ in each branch. Now let us make this more precise.

Classically, a transformation from reference frame fields $X^A$ to $Y^A$ is determined by \begin{equation}
    Q_{X \rightarrow Y}: \mathbb{R}^3 \rightarrow \mathbb{R}^3: \xi^A \rightarrow X \circ Y^{-1} (\xi),
\end{equation} acting on dressed configuration space operators by its pullback.
For instance, the spacetime scalar $F$ with $F_X: \mathbb{R}^3 \rightarrow \mathbb{R}$ defined by $F_X(\xi) = F(X^{-1}(\xi))$ does transform correctly under this pullback: \begin{equation}
\begin{split}
        F_Y(\xi) &= F(Y^{-1}(\xi)) = F( X^{-1} \circ X \circ Y^{-1} (\xi)) \\ &= F\big( X^{-1} ( X \circ Y^{-1})(\xi) \big) = F_X (X \circ Y^{-1}(\xi)) \\ &= Q_{X \rightarrow Y}^* F_X (\xi) 
\end{split}
\end{equation} 
If, in our superposed BTZ case, we now apply this transformation coherently in each branch $i$, we get in branch $i$ the map \begin{equation}
    Q_{X_B \rightarrow {Y_D,i}} = X_B \circ Y_{D,i}^{-1}: \xi^A \rightarrow \xi^A + \delta^A_r \, r_{D,i}.
\end{equation}  We can thus make this work in each branch by considering the operator \begin{equation}
     \hat{Q}_{X \rightarrow \hat{Y}}: \xi^A \rightarrow \xi^A + \delta^A_r \,\hat{r}_D^{(X)},
\end{equation} with $\hat{r}_D^{(X)}$ the operator for the radial coordinate position between the detector and black hole.

In branch $i$ in the reference frame associated with the detector, the $Y_{D,i}$-dressed metric $g_{Y_{D,i}}$ can be written as \begin{equation} \label{eq:BTZ_metric_2}
    \mathrm{d} s^2=-f_i(\tilde r) \, \mathrm{d} t^2+f_i(\tilde r)^{-1} \mathrm{~d} \tilde r^2+(\tilde r+r_{D,i})^2 \mathrm{~d} \varphi^2,
\end{equation} with $f_i(\tilde r)=(\tilde r+r_{D,i})^2/l^2-M$, with $t,\tilde r,\varphi$ now denoting $Y_{D,i}^0, Y_{D,i}^1, Y_{D,i}^2$. 
Our set-up can thus be interpreted as a black hole in a position-superposition as seen from the $Y$-frame.
The equations of motion are covariant under this change of frame.

Concluding, we thus see that this map $\hat{Q}_{X \rightarrow Y}$ and its pullback reproduce the branchwise coordinate relabelling underlying the QRF transformation of \cref{eq:QRF_transformation_field}: \begin{equation}
    \hat S^{(B)\to(D)} = \int d\mathbf{x} \,\ket{-\mathbf{x}}_{B}\bra{\mathbf{x}}_D \otimes \mathbb{1}_\phi,
\end{equation} where the generalized position eigenbasis is normalized using the measure $d\mathbf{x}$.
The identity on $\phi$ arises as the QRF transformation acts as the identity on the abstract field-state factor (the undressed field $\phi$). Had we considered the dressed field $\phi_X$, it would change accordingly, as explained above.

The QRF map $\hat{Q}_{X \rightarrow \hat{Y}}$ we found is a reduction of the full quantum reference frame transformation, only intended to work correctly in the limited set-up we consider. 
It arises from the operator nature of $\hat{Y}$, and more specifically, $\hat{Y}^1$ through the definition of the detector worldline $\hat{z}^{X^1}$ in the $X$-reference frame. 
Note that this treatment clarifies the meaning of the coordinate operator $\mathbf{x}$ used in \cref{eq:H_int_time_changes,eq:H_int_QRF_transformed} and the choice of assigning a Hilbert space $\mathcal{H}_B = L^2(\mathbb{R},dr)$ to the black hole-detector system that captures this relational degree of freedom in the radial coordinate; it is the relational parameter describing the transformation between the reference fields, which happens to be a translation in the radial $r$-coordinate, $r \rightarrow r+r_{D,i}$, which is why in the main text we refer to this as a `translation'. (It is thus not intended to refer to a translation symmetry as there is none in the BTZ spacetime.)

A more complete quantum reference frame transformation, not restricted to a specific set of reference field states, might consider (pullbacks of) $\hat{Q}_{\hat{X} \rightarrow \hat{Y}} =  \hat{X} \circ \hat{Y}^{-1}$, which can be defined (formally) through its eigenfunctions.
Moreover, such QRF transformation may be defined through a perspective-neutral approach that takes into account the full dynamics and constraints~\cite{chen2026quantumreferencefieldstransformations,delahamette2021perspectiveneutralapproachquantumframe,Vanrietvelde_2020,aguilargutierrez2026relationalpathintegraleffective}, see also the paragraph \textit{More complete QRF treatment} below.
The frames we consider are idealized, sharply localized frames within the restricted configuration subspace considered here, resembling the ideal-frame limit of operational QRF constructions~\cite{Loveridge:2016tnh,Loveridge_2018,de_la_Hamette_2020,delahamette2021perspectiveneutralapproachquantumframe,carette2024operationalquantumreferenceframe}, although we do not construct a covariant PVM or invariant measure for the full diffeomorphism group. 

\vspace{0.5cm}

\textit{Connecting to literature.} We can connect this treatment also to that of Ref.~\cite{kabel2024identification}. Namely, being in the reference frame of $X_B$, we write the initial state as $\ket{X_B}  (\ket{x_1}_D \ket{Y_{D,1}^A} + \ket{x_2}_D \ket{Y_{D,2}^A} )/\sqrt{2}$. Going to the reference frame of $Y_D$, we perform a branch-dependent diffeomorphism $d^{(i)}$ on the manifold $M$ (or the open region we care about), given by $r \rightarrow r+r_{D,i}$ in branch $i$, or more precisely, \begin{equation} 
\label{eq:define_diffeo_phi_i}
\begin{split}
    d^{(i)}: U \subset M \rightarrow d^{(i)}(U) \subset M: p& = X_B^{-1}(t_p,r_p,\varphi_p) \\ &\mapsto d^{(i)}(p) = X_B^{-1}(t_p,r_p+r_{D,i},\varphi_p).
\end{split}
\end{equation} 
This indeed sets $\left(d^{(1)}\right)^*\left(Y_{D,1}\right)=\left(d^{(2)}\right)^*\left(Y_{D,2}\right)=\tilde{Y_D}_*$, so that we end up in a state where $\hat{Y}_D$ is single-valued (note that there are more diffeomorphisms that achieve this).
The branch-dependent diffeomorphisms are related to the above QRF transformation through $X_B \circ d^{(i)} \circ X_B^{-1} = X_B \circ Y_{D,i}^{-1}$.

\vspace{0.5cm}

\textit{More complete QRF treatment.} We have taken the reference fields to have zero backreaction, non-interacting with other fields and ignored their dynamics. In a more complete treatment, we should specify their dynamics and include their backreaction. 
Different formalisms then provide different counterparts of the reduced frame transformation we considered here.

In particular, the frame change map $\hat{X} \circ \hat{Y}^{-1}$ and the change of operator-dressing for the configuration variables may not determine the complete transformation; it may be necessary to further obtain its action on conjugate momenta, constraints, physical
Hamiltonians, or the physical inner product.
For instance, the perspective-neutral approach~\cite{Vanrietvelde_2020,hohn2020switch,Giacomini_2021,delahamette2021perspectiveneutralapproachquantumframe} considers the physical space encoding all different QRF perspectives, from which one reduces to a QRF $X$'s perspective through a reduction map $R_X$, with the QRF transformation given by $R_Y R_X^\dagger$, a unitary if the reduction maps are isometric isomorphisms onto the relevant reduced physical spaces.
This idea can be implemented canonically~\cite{chen2026quantumreferencefieldstransformations,Giacomini_2021,hohn2020switch,de2024gravitational,Ali_Ahmad_2022,Vanrietvelde_2020}, through the path integral~\cite{aguilargutierrez2026relationalpathintegraleffective} and quantization of classical covariant phase-space methods~\cite{goeller2022diffeomorphisminvariantobservablesdynamicalframes,giesel2025linkingedgemodesgeometrical}. 
Other, related approaches are the operational or algebraic approach~\cite{carette2024operationalquantumreferenceframe,Loveridge_2018,fewster2024quantumreferenceframesmeasurement,fewster2025semilocalobservablesedgemodes} and the extra-particle approach~\cite{castroruiz2023relativesubsystemsquantumreference}, whose relation to the perspective-neutral approach remains an active topic of investigation~\cite{castroruiz2025interpretingquantumreferenceframe,Doat_2026}. 

Moreover, we have neglected the action of the reference frame transformation on boundary data; in the presence of (asymptotic) boundaries, edge modes might appear if a boundary is held fixed~\cite{Carrozza_2022,kabel2023quantum,giesel2025linkingedgemodesgeometrical}.

\section{Derivation of the interaction Hamiltonian for a superposed black hole} \label{app:H_int_superp_BH}
In this section we derive the form of the Unruh--DeWitt detector Hamiltonian in the laboratory frame, i.e. in the presence of a superposition of black holes. 
The usual UDW interaction in the black hole frame (but according to the detector's time) is, in the Schrödinger picture, given by
\begin{equation}
    H_\text{int}^{(B)} = \lambda  \, \gamma(\hat{\mathbf{x}})  \, \mu \otimes \phi(\hat{\mathbf{x})} = \lambda \, \mu \otimes \int d\nu(\mathbf{x})\,\ket{\mathbf{x}}  \bra{\mathbf{x}}_D \,  \gamma(\mathbf{x}) \otimes \phi(\mathbf{x})
\end{equation}
acting on the Hilbert space $\mathcal{H}_{D}\otimes \mathcal{H}_{\mu}\otimes \mathcal{H}_\phi$ corresponding to the external degrees of freedom of the detector (its position), the internal degrees of freedom of the detector (the $n$-level energy system) and the scalar field sector, respectively. 
Recall that $\gamma(\mathbf{x})$ is the redshift factor.
A QRF transformation maps the perspective of the black hole to the perspective of the detector (i.e. the laboratory), with the transformation 
\begin{equation}
    S^{(B\to D)} = \int d\nu(\mathbf{x}) \, \ket{-\mathbf{x}}_B\bra{\mathbf{x}}_{D} \otimes \mathbb{I}_\phi \otimes \mathbb{I}_{\mu}
\end{equation}
After the QRF transformation, the UDW interaction Hamiltonian in the detector frame becomes
\begin{equation}
    H^{(D)}_{\text{int}} = S^{(B\to D)}\gamma^{-1}(\hat{\mathbf{x}}_D) H^{(B)}_\text{int} S^{\dagger(B\to D)} = \lambda \int d\nu(\mathbf{x}) \, \ket{\mathbf{x}} \bra{\mathbf{x}}_B \,  \otimes \, \mu \otimes  \phi(-\mathbf{x}) = \lambda \int d\nu(\mathbf{x}) \ket{\mathbf{x}} \bra{\mathbf{x}}_B \otimes H_{int,\mathbf{x}_B},
\end{equation} with $H_{int,\mathbf{x}_B}$ the Unruh--DeWitt interaction for a detector with a scalar field in a background with black hole at position $\mathbf{x}$ (and with detector now at the origin $\mathbf{0}$).

\section{Derivation of the analytical expressions of \Cref{sec:analytical}}
\label{sec:analytical_appendix}
\subsection{Summary of expressions} \label{sec:analytical_summary_app}
In this section we present the analytical expressions for the spectrum probed by the detector from \Cref{sec:analytical}. Before presenting detailed derivations, we provide a summary of the relevant expressions here. We consider a static detector near a (potentially superposed) BTZ black hole. 

\vspace{0.5cm}

\textit{Single-spacetime detector response.} In a non-superposed spacetime, the detector response from \cref{eq:transition_prob} can be expressed as \begin{equation} \label{eq:Fewster_eq_chi_hatW_app_expressions}
    \mathcal{F}(E)=\frac{1}{2 \pi} \int_{-\infty}^{\infty} d \omega|\widehat{\eta}({\scriptstyle\frac{\omega}{\gamma}})|^2 \widehat{W}_{\scriptstyle BTZ}(E \gamma +\omega),
\end{equation} 
The explicit form of the BTZ correlation function along the detector's path is given by taking $R_1=R_2=R$ to be the detector's radial coordinate and $\theta=0$ in \cref{eq:W_BTZ_app_NeuDiri}: 
\begin{equation}
     \begin{split}
       {W}_{\scriptstyle BTZ}(s) = \frac{1}{4 \pi l \sqrt{2} \widetilde{\gamma}} & \bigg[ \frac{1}{\sqrt{\alpha_m - \cosh(s \sqrt{M}/l - i \epsilon )}}  -  \frac{\zeta}{\sqrt{\alpha'_m - \cosh(s \sqrt{M}/l - i \epsilon)}} \bigg] 
    \end{split}
\end{equation} with limit $\epsilon \rightarrow 0^+$, $s= t-t', \widetilde{\gamma}=\gamma(R)/\sqrt{M}$ and
\begin{align}
     \alpha_m &= \frac{ \frac{R^2}{M l^2} \cosh [ 2 \pi m\sqrt{M}]-1}{\frac{R^2}{M l^2}-1 } \\
     \alpha'_m &= \frac{ \frac{R^2}{M l^2} \cosh [ 2 \pi m\sqrt{M}]+1}{\frac{R^2}{M l^2}-1 } .
\end{align}

Following Ref.~\cite{Lifschytz_1994}, an analytical expression for $\widehat{W}$ can be found, giving\footnote{Compared to Ref.~\cite{Lifschytz_1994} we have added the first term arising from the pole singularity using the Sokhotski formula.}
\begin{equation}
    \begin{split}
        \widehat{W}_{BTZ}(K) =  \frac{1}{4  {\gamma}} +  \frac{1}{2 {\gamma}} \frac{1}{e^{ K / T_H}+1} &\sum_n \bigg[ P_{\frac{i K }{2\pi T_H}-\frac{1}{2}}(\alpha_n)   -\zeta  P_{\frac{i K }{2 \pi T_H}-\frac{1}{2}}(\alpha'_n) \bigg]
    \end{split}
\end{equation} with Hawking temperature $T_H = \sqrt{M}/(2\pi l)$ and $P_\nu$ the Legendre function of the first kind.

\vspace{0.5cm}

\textit{Interference term in detector response.} The interference term $\mathcal{F}_{12}(E)$, both for mass- and position-superposed black holes now, in a similar form as \cref{eq:Fewster_eq_chi_hatW_app_expressions}:
\begin{equation}
\begin{split}
    \mathcal{F}_{12}(E) = \frac{1}{2\pi} \int_{-\infty}^{+ \infty}  \mathrm{d}\omega  \hspace{0.05cm} \widehat{\eta}({\scriptstyle\frac{\omega}{\widetilde{\gamma_1}}})^* \hspace{0.05cm} \widehat{\eta}({\scriptstyle\frac{\omega}{\widetilde{\gamma_2}} + E [\frac{\widetilde{\gamma_1}}{\widetilde{\gamma_2}} -1] } )  \hspace{0.05cm} \widehat{W}^{(12)}_{BTZ}(\omega+\widetilde{\gamma_1} E). 
\end{split}
\end{equation}
To calculate the Fourier transform of $\widehat{W}^{(12)}$, we will first perform a change of coordinates in each branch of the superposition. 
Namely, we want to be able to treat also the mass-superposed case, in which case we need coordinates to compare spacetimes for different black hole masses. 
Following Ref.~\cite{foo2022quantum}, we do so by considering AdS time coordinates $\overline{t}$ related to the BTZ time coordinates $t$ by $\overline{t}=t \sqrt{M_i}$ in branch $i$. In the mass-superposed black hole, we now need the additional assumption that the Unruh--DeWitt interaction is a quantum-controlled version of the single-spacetime interaction, with control on the black hole properties, its position and mass, as in Ref.~\cite{foo2022quantum}. 
We have made this assumption more plausible by deriving this interaction form in the case of the black hole position in \cref{eq:H_int_QRF_transformed}.
We denote the position and mass by $R_1,M_1$ and $R_2,M_2$ in the two branches of the superposition. 
We also introduce $\widetilde{\gamma_i}(R_i) = \sqrt{R_i^2/(M_i l^2) -1}$.

The scalar field correlation function between two BTZ spacetimes is then given by~\cite{foo2022quantum}:  
\begin{equation} \label{eq:W_BTZ_mass_and_position_superp}
    \begin{split}
       \widehat{W}^{(12)}_{BTZ}(\overline{s}) =  \frac{1}{4 \pi l \sqrt{2 \widetilde{\gamma_1} \widetilde{\gamma_2}} } & \frac{1}{\sum_k \Upsilon^{2k}} \sum_{m,n} \bigg[ \frac{1}{\sqrt{\beta_{mn}^{(12)} - \cosh(\overline{s} /l)}}  -  \frac{\zeta }{\sqrt{\beta_{mn}^{'(12)} - \cosh(\overline{s} /l)}} \bigg], 
    \end{split}
\end{equation} with $\overline{s}=\overline{t}-\overline{t}' - i \epsilon$ and
\begin{align}
     \beta_{mn}^{(12)} &= \frac{ \frac{R_1 R_2}{\sqrt{M_1 M_2} l^2} \cosh [ 2 \pi (m\sqrt{M_1}- n \sqrt{M_2})]-1}{\sqrt{\frac{R_1^2}{M_1 l^2}-1}\sqrt{\frac{R_2^2}{M_2 l^2}-1} }, \\
     \beta_{mn}^{'(12)} &=  \frac{ \frac{R_1 R_2}{\sqrt{M_1 M_2} l^2} \cosh [ 2 \pi (m\sqrt{M_1}- n \sqrt{M_2})]+1}{\sqrt{\frac{R_1^2}{M_1 l^2}-1}\sqrt{\frac{R_2^2}{M_2 l^2}-1} }.
\end{align}

Using these coordinates $\overline{t}$, we can obtain an analytical expression of $\widehat{W}^{(12)}_{BTZ}$, with the additional subtlety that for $\sqrt{M_1/M_2} \in \mathbb{Q}$ and $R_1/\sqrt{M_1} = R_2/\sqrt{M_2}$ an additional pole arises, giving a singular term $\textsf{sing.}$:
\begin{equation}
    \begin{split}
        \widehat{W}^{(12)}_{BTZ}(K) = \textsf{sing.} + \frac{1}{2 \sqrt{\widetilde{\gamma_1} \widetilde{\gamma_2}}} \frac{1}{\sum_k \Upsilon^{2k}} \hspace{0.07cm} \frac{1}{e^{K/T_H^{AdS}}+1} \\ \sum_{m,n} \bigg[ P_{\frac{i K}{2 \pi T_H^{AdS}}-\frac{1}{2}}\big(\beta_{mn}^{(12)} \big) - \zeta  P_{\frac{i K}{2 \pi T_H^{AdS}}-\frac{1}{2}}\big(\beta_{mn}^{'(12)}\big) \bigg],
    \end{split}
\end{equation} 
with $T_H^{AdS}$ the Hawking temperature of the AdS–Rindler horizon with respect to AdS–Rindler time, $T_H^{AdS} = 1/(2 \pi l)$.
If $\sqrt{M_1/M_2} \notin \mathbb{Q} \backslash \{ 1\}$, then we have $\textsf{sing.} = 0$, else it gives the following non-zero contribution if also $R_1 / \sqrt{M_1} = R_2 / \sqrt{M_2}$:
\begin{equation} \label{eq:sing.}
    \textsf{sing.} = \frac{1}{4  \sqrt{\widetilde{\gamma_1}\widetilde{\gamma_2} p q}}  \quad  \text{ if } \sqrt{M_1/M_2} = p/q \in \mathbb{Q} \backslash \{ 1\} \text{ and } R_1 / \sqrt{M_1} = R_2 / \sqrt{M_2}
\end{equation} with $\text{gcd}(p,q)=1$.

\subsection{Analytic expressions for single term}

Consider the detector response function 
\begin{equation}
\begin{split}
        \mathcal{F}(E) &= \int_{-\infty}^{\infty}   \int_{-\infty}^{\infty}  \mathrm{d}\tau \mathrm{d}\tau' \eta(\tau) \eta(\tau') e^{-iE (\tau-\tau')} W(\tau,\tau') 
    \\ 
    &=  \gamma^2 \int_{-\infty}^{\infty}   \int_{-\infty}^{\infty}  \mathrm{d}t \mathrm{d}t' \eta(\gamma t) \eta( \gamma t') e^{-iE \gamma (t-t')} W(t-t') 
    \end{split}
\end{equation} assuming $W(t,t') = W(t-t')$, with $\gamma = \gamma(R)$ with $R$ the fixed radial coordinate of the static detector. Here and everywhere below $W$ should be understood as $W_{BTZ}$.
By applying the inverse Fourier transform to $W$, and Fourier transforming the switching functions $\eta,\eta$, we obtain \begin{equation} \label{eq:Fewster_eq_chi_hatW_app}
    \mathcal{F}(E)=\frac{1}{2 \pi} \int_{-\infty}^{\infty} d \omega|\widehat{\eta}({\scriptstyle\frac{\omega}{\gamma}})|^2 \widehat{W}(E \gamma +\omega),
\end{equation} with the Fourier transform defined as \begin{equation}
    \widehat{f}(\omega)=\int_{-\infty}^{\infty} d s f(s) \mathrm{e}^{-\mathrm{i} \omega s} 
\end{equation} and inverse by \begin{equation}
    f(s) = \int_{-\infty}^{+\infty} \frac{d\omega }{2\pi} e^{i\omega s} \hat{f}(\omega).
\end{equation}  
See Ref.~\cite{Fewster_2016} for a more precise treatment.
For our Gaussian switching function, we have, for our detector at fixed location $r=R$,
\begin{equation}
    |\widehat{\eta}({\scriptstyle\frac{\omega}{\gamma}})|^2 =  2 \pi \sigma^2 e^{- \frac{\omega^2 \sigma^2}{ \gamma^2} } =   2 \pi \sigma^2 e^{- \frac{\omega^2 \sigma^2}{ \widetilde{\gamma}^2 M} }.
\end{equation} 
If we choose $\sigma$ to be very large, i.e. adiabatic switching, $\widehat{\eta}(\omega)$ becomes approximately a delta function,  and we are basically probing 
$\widehat{\mathcal{W}}(E \gamma)$ for frequencies $\omega/\gamma$. 
This $1/\gamma$ factor arising from the redshift is what leads to the local Tolman temperature $T(R)$ for a static detector at $R$ being related to the Hawking temperature $T_H$ by $T = T_H/\gamma(R)$ with $\gamma(R) = \sqrt{R^2/l^2 - M}$.

\vspace{0.5cm}

Now, to calculate $\widehat{\mathcal{W}}(K)$, we have, \begin{equation} \label{eq:w_K_1}
    \begin{split}
       \widehat{W}(K) &= \int_{-\infty}^{+\infty}  \mathrm{d}s e^{-i K s} W(s) \\ &=  \frac{1}{4 \pi l \sqrt{2} \widetilde{\gamma}} \sum_m \int_{-\infty}^{+\infty}  \mathrm{d}s \bigg[ \frac{e^{-iK s}}{\sqrt{\alpha_m - \cosh(s \sqrt{M}/l-i\epsilon)}}  -  \frac{\zeta e^{-iK s}}{\sqrt{\alpha'_m - \cosh(s \sqrt{M}/l-i\epsilon)}} \bigg] 
       \\ 
       &=   \frac{1}{4 \pi  \sqrt{2} {\gamma}} \sum_m \int_{-\infty}^{+\infty}  \mathrm{d}z \bigg[ \frac{e^{-iK l z / \sqrt{M}}}{\sqrt{\alpha_m - \cosh(z-i\epsilon)}}  -  \frac{\zeta e^{-iK l z / \sqrt{M} }}{\sqrt{\alpha'_m - \cosh(z-i\epsilon)}} \bigg].
    \end{split}
\end{equation} 
with $\widetilde{\gamma}= \gamma / \sqrt{M}$, 
\begin{equation}
     \alpha_m = \frac{ \frac{R^2}{M l^2} \cosh [ 2 \pi m\sqrt{M}]-1}{\frac{R^2}{M l^2}-1 }  = \frac{ \frac{R^2}{M l^2} \cosh [ 2 \pi m\sqrt{M}]-1}{\widetilde{\gamma}^2 }.
\end{equation} and \begin{equation}
     \alpha'_m = \frac{ \frac{R^2}{M l^2} \cosh [ 2 \pi m\sqrt{M}]+1}{\frac{R^2}{M l^2}-1 } = \frac{ \frac{R^2}{M l^2} \cosh [ 2 \pi m\sqrt{M}]+1}{\widetilde{\gamma}^2 } ,
\end{equation} the $z - i \epsilon$ prescription, and $\zeta = -1,0,1$ for Neumann, transparent or Dirichlet boundary conditions.

Note that we have $\alpha_{0} = 1$ so that the term with $\alpha_{0}$ has a pole at $z=0$. 
In this case, we first rewrite using $\cosh(z) - 1 = 2 \sinh^2(z/2)$, and using the Sokhotski formula \begin{equation}
    \frac{1}{\sinh (z-i \epsilon)}=i \pi \delta(z)+\mathrm{PV} \frac{1}{\sinh (z)},
\end{equation} 
We find that the singular part gives rise to the term $1/(4 l \widetilde{\gamma})$.

Following Ref.~\cite{Lifschytz_1994}, the nonsingular parts (including the principal value part of the singular term with $\alpha_0$ above) in the integrals in \cref{eq:w_K_1} can be performed, and give 
\begin{equation}
    \begin{split}
        \widehat{W}(K) &=  \frac{1}{4  {\gamma}} +  \frac{1}{2  {\gamma}} \frac{1}{e^{ K / T_H}+1} \sum_n \bigg[ P_{\frac{i K }{2\pi T_H}-\frac{1}{2}}(\alpha_n) - \zeta  P_{\frac{i K }{2 \pi T_H}-\frac{1}{2}}(\alpha'_n) \bigg]
    \end{split}
\end{equation} with $T_H = \sqrt{M}/(2\pi l)$ the Hawking temperature of the AdS–Rindler horizon with respect to AdS–Rindler time. 
Note that Ref.~\cite{Lifschytz_1994} left out the first term arising from the pole singularity.

\subsection{Analytic expressions for interference term}

For the interference term, we can obtain a similar expression.
We now consider the general situation that comprises both the position- and mass-superposed black holes, with position and mass $R_1,M_1$ and $R_2,M_2$ in the two branches of the superposition, respectively. 
Consider the interference term in the detector transition probability 
\begin{equation}
    \begin{split}
       \mathcal{F}_{12}(E) = \int_{\tau_i}^{\tau_f} d\tau d\tau'  \eta(\tau) \eta(\tau') e^{-i E (\tau -\tau')} W^{(12)}(x(\tau),x(\tau')),
    \end{split} 
\end{equation} with $W^{(12)}$ shorthand for $W^{(12)}_{BTZ}$. Using the AdS time coordinates $\overline{t}=t \sqrt{M_i} =\tau / \widetilde{\gamma_i}$ with $\widetilde{\gamma_i} = \sqrt{R_i^2/(M_i l^2)-1}$ in the branch with BTZ black hole mass $M_i$, we find \begin{equation}
    \begin{split}
       \mathcal{F}_{12}(E) = \widetilde{\gamma_1} \widetilde{\gamma_2} \int_{-\infty}^{+\infty} \int_{-\infty}^{+\infty}  \mathrm{d}\overline{t}  \mathrm{d}\overline{t}'  \eta(\widetilde{\gamma_1} \overline{t}) \eta(\widetilde{\gamma_2} \overline{t}') e^{-i (E \widetilde{\gamma_1} \overline{t} - E \widetilde{\gamma_2} \overline{t}')} W^{(12)}_{BTZ}(\overline{t}-\overline{t}').
    \end{split}
\end{equation} 
We will again try to write $\mathcal{F}_{12}$ in the form of \cref{eq:Fewster_eq_chi_hatW_app}. We start off by changing integration variables with $s= \overline{t}-\overline{t}'$ (we omit the integration limits here, all integrals, also under the changes of variables that occur, run from $-\infty$ to $+\infty$):
\begin{equation}
    \begin{split}
       \mathcal{F}_{12}(E) &=  \widetilde{\gamma_1} \widetilde{\gamma_2} \int \int  \mathrm{d}s  \mathrm{d}\overline{t}' \hspace{0.05cm} \eta(\widetilde{\gamma_1} [s+\overline{t}'] )  \hspace{0.05cm} \eta(\widetilde{\gamma_2} \overline{t}') e^{-i E \widetilde{\gamma_1} s}e^{-iE(\widetilde{\gamma_1}-\widetilde{\gamma_2})\overline{t}'} W^{(12)}_{BTZ}(s) 
       \\ 
       &=  \widetilde{\gamma_1} \widetilde{\gamma_2} \int \int  \mathrm{d}s  \mathrm{d}\overline{t}' \hspace{0.05cm} \eta(\widetilde{\gamma_1} [s+\overline{t}'] )  \hspace{0.05cm} \eta(\widetilde{\gamma_2} \overline{t}') e^{-i E \widetilde{\gamma_1} s}e^{-iE ( \widetilde{\gamma_1}-\widetilde{\gamma_2})\overline{t}'} \int  \frac{\mathrm{d}\omega }{2 \pi} \widehat{W}^{(12)}_{BTZ}(\omega) e^{i \omega s} 
       \\ 
       &= \frac{\widetilde{\gamma_1} \widetilde{\gamma_2}}{2\pi}  \int  \mathrm{d}\omega \int \int  \mathrm{d}s  \mathrm{d}\overline{t}' \hspace{0.05cm}   \eta(\widetilde{\gamma_1} [s+\overline{t}'] )  \hspace{0.05cm} \eta(\widetilde{\gamma_2} \overline{t}') e^{i \omega s}e^{-iE (\widetilde{\gamma_1}-\widetilde{\gamma_2})\overline{t}'} \widehat{W}^{(12)}_{BTZ}(\omega+\widetilde{\gamma_1} E)   
       \\ 
       &= \frac{\widetilde{\gamma_2}}{2\pi}  \int  \mathrm{d}\omega \int \int   \mathrm{d}\overline{t}' \mathrm{d}u \hspace{0.05cm} \eta( u )  e^{i \frac{\omega}{\widetilde{\gamma_1}} u} \eta(\widetilde{\gamma_2} \overline{t}') e^{-i [ \omega + E (\widetilde{\gamma_1}-\widetilde{\gamma_2}) ]\overline{t}'} \widehat{W}^{(12)}_{BTZ}(\omega+\widetilde{\gamma_1} E)  
       \\
       &= \frac{1}{2\pi} \int  \mathrm{d}\omega \int \mathrm{d}u \hspace{0.05cm} \eta( u )  e^{i \frac{\omega}{\widetilde{\gamma_1}} u} \int \mathrm{d} v \eta(v) e^{-i\big[ \frac{\omega}{\widetilde{\gamma_2}} + E \big(\frac{\widetilde{\gamma_1}}{\widetilde{\gamma_2}}-1\big) \big] v} \widehat{W}^{(12)}_{BTZ}(\omega+\widetilde{\gamma_1} E)  
       \\
       &=  \frac{1}{2\pi} \int_{-\infty}^{+ \infty}  \mathrm{d}\omega  \hspace{0.05cm} \widehat{\eta}({\scriptstyle\frac{\omega}{\widetilde{\gamma_1}}})^* \hspace{0.05cm} \widehat{\eta}({\scriptstyle\frac{\omega}{\widetilde{\gamma_2}} + E [\frac{\widetilde{\gamma_1}}{\widetilde{\gamma_2}} -1] } ) \hspace{0.05cm} \widehat{W}^{(12)}_{BTZ}(\omega+\widetilde{\gamma_1} E)
    \end{split}
\end{equation} where we changed variables to $u= \widetilde{\gamma_1} (s + \overline{t}'), v = \widetilde{\gamma_2} \overline{t}'$.

\vspace{0.5cm}

Next, the main calculation of this section, we  calculate $\widehat{W}^{(12)}(K)$: \begin{equation} \label{eq:w_K_1_app}
    \begin{split}
       \widehat{W}^{(12)}(K) &= \int_{-\infty}^{+\infty}  \mathrm{d}s e^{-i K s} W^{(12)}(s) 
       \\ 
       &=  \frac{1}{4 \pi l \sqrt{2 \widetilde{\gamma_1} \widetilde{\gamma_2}} } \frac{1}{\sum_k \Upsilon^{2k}} \sum_{m,n} \int_{-\infty}^{+\infty}  \mathrm{d}s \bigg[ \frac{e^{-iK s}}{\sqrt{\beta_{mn}^{(12)} - \cosh(s /l-i\epsilon)}}  -  \frac{\zeta e^{-iK s}}{\sqrt{\beta_{mn}^{'(12)} - \cosh(s /l-i\epsilon)}} \bigg] 
       \\ 
       &=   \frac{1}{4 \pi \sqrt{2 \widetilde{\gamma_1} \widetilde{\gamma_2}} }  \frac{1}{\sum_k \Upsilon^{2k}}  \sum_{m,n} \int_{-\infty}^{+\infty}  \mathrm{d}z \bigg[ \frac{e^{-iK l z }}{\sqrt{\beta_{mn}^{(12)} - \cosh(z-i\epsilon)}}  -  \frac{\zeta e^{-iK l z }}{\sqrt{\beta_{mn}^{'(12)} - \cosh(z-i\epsilon)}} \bigg].
    \end{split}
\end{equation} 
with 
\begin{equation} \label{eq:alpha_12_m}
     \beta_{mn}^{(12)} = \frac{ \frac{R_1 R_2}{\sqrt{M_1 M_2} l^2} \cosh [ 2 \pi (m\sqrt{M_1}- n \sqrt{M_2})]-1}{\sqrt{\frac{R_1^2}{M_1 l^2}-1}\sqrt{\frac{R_2^2}{M_2 l^2}-1} }
\end{equation} and \begin{equation}
     \beta_{mn}^{'(12)} =  \frac{ \frac{R_1 R_2}{\sqrt{M_1 M_2} l^2} \cosh [ 2 \pi (m\sqrt{M_1}- n \sqrt{M_2})]+1}{\sqrt{\frac{R_1^2}{M_1 l^2}-1}\sqrt{\frac{R_2^2}{M_2 l^2}-1} },
\end{equation} and $\zeta = -1,0,1$ for Neumann, transparent or Dirichlet boundary conditions.
Note that if, for some $m,n$ we have that $\beta_{mn}^{(12)} = 1$, which occurs at rational values of $\sqrt{M_1/M_2}$ in the mass-superposed case (so $M_1 \neq M_2$) while $R_1 /\sqrt{M_1} = R_2 / \sqrt{M_2}$, then the term with $\beta_{mn}^{(12)}$ has a pole at $z=0$. 
In this case, we first rewrite using $\cosh(z) - 1 = 2 \sinh^2(z/2)$, and using the Sokhotski formula \begin{equation}
    \frac{1}{\sinh (z-i \epsilon)}=i \pi \delta(z)+\mathrm{PV} \frac{1}{\sinh (z)},
\end{equation} 
we find the singular part giving rise to the term $1/(4  \sqrt{\widetilde{\gamma_1}\widetilde{\gamma_2}})$.
We have to divide this term by $\sum_k \Upsilon^{2k}$, and it occurs for all $n = \sqrt{M_1/M_2} m$ (and $R_1/\sqrt{M_1}=R_2/\sqrt{M_2}$). This yields a finite contribution which in general might be cut-off dependent, i.e. depending on how the infinite sums are dealt with. 
However, there is a natural way to deal with the sum, arising from the group structure of the groups generated by the identification $\Gamma_M$ that defines the BTZ spacetime as a quotient of AdS--Rindler (cf. Appendix~\ref{sec:app_W_BTZ} and Appendix of Ref.~\cite{foo2022quantum}). 
Namely, let $\Gamma_{M_i}$ denote the identification that defined the BTZ spacetime with mass $M_i$, $\Gamma_{M_i}^n: \varphi \mapsto \varphi + 2\pi n \sqrt{M_i}$. If $\sqrt{M_1/M_2}=p/q \in \mathbb{Q}$ with $\text{gcd}(p,q)=1$, then the groups $G_1,G_2$ generated by $\Gamma_{M_1},\Gamma_{M_2}$ share a subgroup $H$, with indices $[G_1:H]=q, [G_2:H]=p$, suggesting their volumes relate as $\text{Vol}(H)/\sqrt{\text{Vol}(G_1) \text{Vol}(G_2)} = 1/\sqrt{pq}$. 
The singularity is hit at every point of $H$ (assuming also $R_1/ \sqrt{M_1}= R_2/\sqrt{M_2}$), and a suggestive normalization of the BTZ two-point function is then \begin{equation}
    \begin{split} \label{eq:BTZ_corr_appF}
    W_{\mathrm{BTZ}}^{(12)}\left(x, x^{\prime}\right)&=\frac{1}{\sqrt{\text{Vol}(G_1) \text{Vol} (G_2)}}  \sum_{\Gamma_{M_1}^n \in G_1} \sum_{\Gamma_{M_2}^m \in G_2} W_{\mathrm{AdS}}\left(\Gamma_{M_1}^n x, \Gamma_{M_2}^m x^{\prime}\right)
\end{split}
\end{equation} for untwisted fields ($\Upsilon = 1$), arising from the identity for the automorphic field~\cite{foo2022quantum,langlois2005imprintsspacetimetopologyhawkingunruh}: \begin{equation}
    {\phi}_{\text{BTZ},M_i}({x}):=\frac{1}{\sqrt{\text{Vol}(G_i)}} \sum_n {\psi}_{\text{AdS--Rindler}}\left(\Gamma_{M_i}^n {x}\right)
\end{equation} for untwisted fields. As the singular term gives a contribution $1/(4\sqrt{\widetilde{\gamma_1}\widetilde{\gamma_2}})$ at each element of $H$, we get a contribution 
\begin{equation}
    \frac{1}{4\sqrt{\widetilde{\gamma_1}\widetilde{\gamma_2}}} \, \frac{ \text{Vol}(H)}{\sqrt{\text{Vol}(G_1) \text{Vol}(G_2)}} = \frac{1}{4\sqrt{\widetilde{\gamma_1}\widetilde{\gamma_2}}} \, \frac{1}{\sqrt{pq}} 
\end{equation} from the singularities. 
This thus gives the total contribution
\begin{equation} \label{eq:sing.app}
    \text{sing.} = \frac{1}{4  \sqrt{\widetilde{\gamma_1}\widetilde{\gamma_2} p q}}  \quad \text{for } \sqrt{M_1/M_2} = p/q \in \mathbb{Q}^+ \backslash \{1\} \text{ and } R_1/\sqrt{M_1}=R_2/\sqrt{M_2} 
\end{equation} with $\text{gcd}(p,q)=1$, to be inserted as the contribution of the (potential) singular part to $\widehat{W}^{(12)}(K)$. 
Had one taken a square-box regularisation of the sums instead (which is perhaps less well motivated than the above volume argument), i.e., taking $|m|, |n|, |k| \leq N$ such that the double sum becomes $1/(2N+1) \sum_{|m|,|n| \leq N}$ one would get a contribution $1/(4  \sqrt{\widetilde{\gamma_1}\widetilde{\gamma_2}} \max\{q,p\})$ for $\text{sing.}$ above instead.

Regarding all other terms without poles, including the principal value part of the term that was singular before, following Ref.~\cite{Lifschytz_1994}, the integrals in \cref{eq:w_K_1_app} can be performed, giving 
\begin{equation} \label{eq:widehat_interf_app}
    \begin{split}
        \widehat{W}^{(12)}(K) &= \textsf{sing.} + \frac{1}{2  \sqrt{\widetilde{\gamma_1} \widetilde{\gamma_2}}} \frac{1}{\sum_k \Upsilon^{2k}} \hspace{0.07cm} \frac{1}{e^{K/T_H^{AdS}}+1} \sum_{m,n} \bigg[ P_{\frac{i K}{2 \pi T_H^{AdS}}-\frac{1}{2}}\big(\beta_{mn}^{(12)} \big) - \zeta  P_{\frac{i K}{2 \pi T_H^{AdS}}-\frac{1}{2}}\big(\beta_{mn}^{'(12)}\big) \bigg],
    \end{split}
\end{equation} with $T_H^{AdS}$ the Hawking temperature of the AdS--Rindler horizon with respect to AdS--Rindler time, $T_H^{AdS} = 1/(2 \pi l)$.
If $\sqrt{M_1/M_2} \notin \mathbb{Q} \backslash \{1\}$, we have $\textsf{sing.} = 0$, else it gives the contribution as in \cref{eq:sing.app}.

\vspace{0.5cm}

In fact, for the position-superposition, we can remove the double sum in \cref{eq:widehat_interf_app}. Namely, for the position-superposed case we have $M_1 = M_2$ and we can write \begin{equation}
     \begin{split}
        \widehat{W}^{(12)}(K) &= \frac{1}{2  \sqrt{\widetilde{\gamma_1} \widetilde{\gamma_2}}} \hspace{0.07cm} \frac{1}{e^{K/T_H^{AdS}}+1} \sum_{m} \bigg[ P_{\frac{i K}{2 \pi T_H^{AdS}}-\frac{1}{2}}\big(\alpha_{m}^{(12)} \big) - \zeta  P_{\frac{i K}{2 \pi T_H^{AdS}}-\frac{1}{2}}\big(\alpha_{m}^{'(12)}\big) \bigg],
    \end{split}
\end{equation} with \begin{equation}
    \alpha_{m}^{(12)} = \frac{ \frac{R_1 R_2}{M l^2} \cosh [ 2 \pi m \sqrt{M}]-1}{\sqrt{\frac{R_1^2}{M l^2}-1}\sqrt{\frac{R_2^2}{M l^2}-1} }.
\end{equation} 
To be precise, in case $R_1 = R_2$, an additional pole contribution $1/(4 \widetilde{\gamma})$ to $\widehat{W}^{(12)}$ arises, but at $R_1=R_2$ our initial `superposed' state of \cref{eq:initial_state_BH_superp} becomes an unnormalized non-superposed state, and similarly for the interferometric measurement effect, $\ket{+}_B = (\ket{\mathbf{x}_1}_B + \ket{\mathbf{x}_2}_B ) / \sqrt{2}$, as commented upon in \Cref{sec:results}.

\vspace{0.5cm}

\textbf{A note about energy and time coordinates.} 
We briefly comment upon how to compare energy in different frames, choosing different time coordinates, important for example for establishing when the detector's energy gap is small compared to the black hole mass concerning the approximation of neglecting the detector's gravitational backreaction, or for comparing energies across different spacetimes. 
Let $H^{(D)}_i$ be the detector's free Hamiltonian in its rest frame, in branch $i$ of the superposition.
We take the detector in each branch of the superposition to be identical, i.e. $H^{(D)}_i=H^{(D)}$.
Then, acting on the detector energy eigenstate $\ket{E}_{D}$ with energy eigenvalue $E^{(D)}$ we have $ H \ket{E}_{D} = E^{(D)} \ket{E}_{D}$ in the detector frame. 
If we consider the detector in a superposition of locations, or in a superposed spacetime, with branches labelled by $i$, we need to be able to relate this detector frame energy to energy and Hamiltonians with respect to other time parametrizations.
The detector proper time and coordinate time are related by $\tau_i = \gamma_i t_i$ in the branch labelled by $i$. 
In such branch, the Hamiltonians are related by $H^{(D)} = H^{\tau_i} = dt/d\tau H^{t_i} = H^{t_i}/\gamma_i$. 
Therefore, the energies are related by \begin{equation}
    E^{(D)} = E^{\tau,i} = \frac{1}{\gamma_i(R_i)} E^{t_i}.
\end{equation} 
Moving to AdS--Rindler coordinates, we find \begin{equation}
     E^{(D)} = E^{\tau,i} = \frac{1}{\widetilde{\gamma_i}(R_i)} E^{\overline{t}_i}.
\end{equation}

\end{document}